\documentclass[runningheads]{llncs}
\usepackage{graphicx} 
\usepackage{hyperref}
\usepackage{comment}
\usepackage{subcaption}
\usepackage{amsmath,amssymb}
\usepackage{bbm}
\usepackage{xcolor}
\usepackage{booktabs}
\usepackage{tikz}
\usepackage{algpseudocode,algorithm,algorithmicx}
\usepackage{wrapfig}

\newcommand{\ignore}[1]{{}}

\begin{document}

%
\title{On the Verification of the Correctness of a Subgraph Construction Algorithm}
%
\titlerunning{On the Verification of a Subgraph Construction Algorithm}
%
\author{Lucas Böltz, Viorica Sofronie-Stokkermans, Hannes Frey}%
\authorrunning{Lucas Böltz, Viorica Sofronie-Stokkermans, Hannes Frey}
%
\institute{Computer Science Department, University of Koblenz, Germany\\
\email{$\{$boeltz,sofronie,frey$\}$@uni-koblenz.de}}
\maketitle

\begin{abstract}
We automatically verify the crucial steps in the 
original proof of correctness of an algorithm which, 
given a geometric graph
satisfying certain additional properties 
removes edges in a systematic way for producing a connected 
graph in which edges do not (geometrically) intersect. 
The challenge in this case is representing and reasoning about geometric properties of graphs in the Euclidean plane, about their vertices and edges, and about connectivity.
For modelling the geometric aspects, we use an axiomatization of plane geometry; 
for representing the graph structure 
we use additional predicates; for representing certain classes of paths in geometric graphs we use linked lists. 
\end{abstract}

\section{Introduction}

We present an approach for automatically verifying the main steps in
the  correctness proof of an algorithm which, given a geometric graph
satisfying two properties (the redundancy property and the coexistence
property) removes edges in a systematic way for producing a connected 
graph in which edges do not (geometrically) intersect. 
One of the problems to be solved, in this case, is representing and reasoning about 
properties of {\em geometric graphs} with vertices in the Euclidean plane. 
For modelling the geometric aspects, we use an axiomatization of
plane geometry related to Hilbert's and Tarski's axiomatization \cite{hilbert,TarskiG99}.
The vertices and edges are modeled using unary resp.\ binary predicates; 
the properties of geometric graphs we consider refer both to 
geometric conditions and to conditions on the vertices and edges. 
In addition, we use linked lists to
represent certain paths in such geometric graphs.  
We show that this axiomatization can be used for describing the classes
of geometric graphs we are interested in, discuss the reasoning tasks occurring in the verification of
the graph algorithm, and the way we automatically 
proved the crucial steps in the original proof of correctness of the algorithm, using the prover Z3 \cite{z3-2018,z3-2020}. 
The main contributions of this paper can be described as follows: 
\begin{itemize}
\item We present an axiomatization of a part of Euclidean geometry, referring to bounded domains of the Euclidean plane,
which is necessary to describe graph drawings in the Euclidean
plane. We show how this axiomatization can be used to prove theorems
in plane geometry (Axiom of Pasch, transitivity of the inside
predicate, symmetry of the intersection predicate).
\item We extend this axiomatization with additional predicate symbols for graphs and vertices, and formulate properties of graphs also containing geometric conditions in this extended language. 
We show how this axiomatization can be used to prove simple properties (e.g.\ the clique property inside a triangle) in the class of geometric graphs we consider.
\item We formalize constructions used in the correctness proof of a
  planarization algorithm for geometric graphs (the CP-algorithm), and use Z3 to perform
  the corresponding proof tasks and to find counterexamples resp.\ to justify the existence of counterexamples if the hypotheses are weakened.  
\item We extend the axiomatization with a specification of list structures for expressing lists of nodes along the convex hull of certain specified sets of points. 
We use this extended axiomatization for proving the connectedness of the graphs obtained as a result of applying the algorithm we analyze. 
\end{itemize}
The main challenges in this paper were (1) to find a suitable axiomatization of notions from geometric graph theory which would allow us to automatically 
verify the main steps in an existing correctness proof of an algorithm
for geometric graphs, and (2) find a way of expressing connectedness,
in this context, without using higher-order logic.

\smallskip
\noindent {\em Related work:} Existing approaches to the study of graphs and graph algorithms are often 
based on monadic second-order logic or higher-order logic. 
\\In \cite{Courcelle90,Courcelle95,Courcelle06,Courcelle09} (these are only a few of the numerous publications of Courcelle on this topic) monadic second-order logic is used; many of the results are for graphs with
bounded tree-width or bounded clique-width. 
The classes of graphs we study do not necessarily have such properties.  
For the verification of graph algorithms, higher-order theorem provers like Isabelle/HOL and Coq were used (cf. e.g.\  \cite{MehlhornNipkow};  
the possibility of checking graph theoretical properties in Coq is discussed 
in \cite{doczkal2020graph}). 
An overview on checking graph transformation systems is given in \cite{Heckel_2019}.
We are not aware of research that combines reasoning about graphs and
reasoning about geometric properties of graphs. 
In our work we focus on the study of geometric graphs
and identify a situation in which the use of higher-order logic and higher-order logic
systems can be avoided. 
We show that for the type of graphs we consider (geometric graphs with vertices in the Euclidean plane) and for the verification problems we consider we can 
avoid second-order logic also for handling 
connectedness, a property known to not be first-order definable: in our context it will be
sufficient to use certain, uniquely determined paths along the convex
hull of determined sets of points, which we represent using linked
data structures. 

When modelling and reasoning about geometric graphs we need to
consider in addition to the encoding of the topological aspects of
graphs (vertices and edges) also the geometric aspects, and combine 
reasoning about the graphs themselves with reasoning about geometric
properties.
For reasoning in geometry, one can use approaches using coordinates and ultimately analytic geometry or 
axiomatic approaches. 
Among the approaches to automated reasoning in {\em analytic
geometry}, we mention methods based on Gr{\"o}bner bases (cf.\ e.g.\ \cite{CoxLittleOShea}) or approaches based on  quantifier elimination in real closed fields
(cf.\ e.g.\ \cite{SturmWeispfenning96,DolzmannSturmWeispfenning98}). 
Systems of {\em axioms for plane geometry}  have been proposed by 
Tarski \cite{TarskiG99} or Hilbert \cite{hilbert}; such systems were used 
for theorem proving in geometry for instance by Beeson and Wos in \cite{Wos17}. 
We decided to use an axiomatic approach. Compared to
classical axiom systems for plane geometry such as those proposed by 
Tarski \cite{TarskiG99} and Hilbert \cite{hilbert} we only use axioms of order, but did not include axioms for incidence, congruence, parallelity and continuity 
  because our goal is to describe the part of plane Euclidean
geometry necessary to describe geometrically bounded graph drawings in
the Euclidean plane.
Due to technical considerations (which we explain in Section~\ref{sec:geometric_graphs}), 
the axioms we propose are slightly different from the axioms of Hilbert \cite{hilbert}, %
but all necessary conditions to prove correctness of the algorithm are satisfied. 
An advantage of our axiomatic approach to plane geometry is that the
axiom system is simple and only information about the relative
positions of nodes is required; we do not need to 
consider concrete coordinates, lengths and angles between the nodes.
So this axiom system is stable for small changes, where the relative positions of the nodes and therefore the predicates do not change.
The simple axiom system makes it easier for a prover to decide whether a condition for a specific combination of the predicates is satisfiable or not. 
We perform proofs by contradiction; if unsatisfiability cannot be
proved we use the possibilities of generating models with Z3 - we used
this, for instance, to show that the axioms we consider cannot be 
relaxed further. 

\medskip
\noindent {\em Structure of the paper.}  The paper is structured as follows: 
In Section~\ref{sec:Cp_algo} we define geometric graphs and describe the CP-algorithm.
In Section~\ref{sec:geometric_graphs} we describe the representation of geometric graphs we use and define the axiom system we use to model geometric properties, properties of geometric graphs and properties used for proving correctness of the CP-algorithm.
In Section~\ref{sec:autreasoning} we discuss the fragments in which these axioms are, and describe the automated reasoning tools we use and the steps of verifying the correctness proof of the CP-algorithm.
Section~\ref{sec:conclusion} contains conclusions and plans for future work.

\medskip
\noindent This paper is the extended version of
\cite{BoeltzSofronieFrey-vmcai24}: it provides details, especially concerning the
axioms and the tests.

\section{The CP-algorithm}\label{sec:Cp_algo}

We start by presenting the algorithm we will verify.
Before doing so, we introduce the main notions related to geometric graphs 
needed in what follows. 

\medskip
\noindent {\bf Geometric graphs.}
A {\em geometric
graph} is a graph $G = (V, E)$ in which vertices are seen as points
(usually in the Euclidean plane) and an edge $uv$ is represented as
{\em straight line segment} between the vertices $u$ and $v$.
We consider undirected graphs without self-loops,
i.e. graphs $G = (V, E)$ in which $uv \in E$ iff $vu \in E$,  and $uu \not\in E$ for all $u,v \in V$.

\medskip 
\noindent {\bf Plane graphs.} A \emph{plane graph} (or \emph{plane
  drawing}) is a geometric graph on a plane 
without intersecting edges, where edges are seen as line segments.
Plane graphs have applications in the context of wireless networks. 
Given a plane drawing, network algorithms can be implemented, 
distributed across the graph vertices, which rely only on the
local view of the vertices. Locality means that each vertex has only information
about vertices and edges of the graph that can be reached via a path of a
fixed, limited path length. All such local methods achieve a
network-wide objective by vertex-local decisions.

Such local approaches are of particular interest for wireless networked systems because with local methods nodes only need to know their immediate neighborhood and thus the control message overhead can be significantly reduced compared to non-local methods. This is particularly the case with reactive local approaches in which the local view of a node is only created if a network-wide task is currently pending. During the rest of the time, the energy requirements of the network nodes can be reduced to a minimum, which then increases the overall lifespan of a network with battery-operated nodes. In addition, the minimal control message demand also means lower bandwidth requirements of the limited available radio resource.

Local approaches for graph related problems based on plane drawings have been studied for different problem settings and it has been discovered that representing a wireless network as a plane graph is a sufficient condition to ensure the correctness of various types of routing tasks like
unicast routing \cite{ka00,Bose2001,frey10face-tc},
multicast \cite{sa07,fr08},
geocast \cite{st04},
anycast \cite{mi09},
mobicast \cite{hu04},
and broadcast
\cite{se01,st02}, and as well as other algorithmic problems like network void and boundary detection \cite{fa04}, distributed data storage \cite{data_storage}, mobile object tracking \cite{object_tracking} and clustering algorithms \cite{clustering_MCFL}.

Given a global view of an arbitrary geometric graph on a plane, constructing a plane subgraph alone is not difficult, however, ensuring connectedness at the same time can be a conflicting goal. When only a local view on the graph is available, even edge intersections are not always detectable. 

Considering wireless networks, however, the graph topology is not arbitrary. 
Due to limited transmission range, a vertex can only be connected to spatially close by vertices. For this reason, graph models for wireless network graphs usually have this spatial locality inherent, including unit disk graphs being the simplest such model, quasi unit disk graph as a extension thereof \cite{barriere01robust,kuhn03beyond}, or log-normal shadowing graphs, a more practical model \cite{Bettstetter2005,Boehmer2019}.

Local solutions for wireless networks can then be studied in view of such model constraints.
In particular, regarding the question in how far a network graph can be planarized by just local information, the specific models 
\emph{redundancy property} and \emph{coexistence property}
(cf. Definitions~\ref{def-redundancy}~and~\ref{def-coexistence}) had been introduced and studied in the context of wireless network graphs \cite{philip06overlay-globecom,frey07topology,mathews2012llrap-icdcn,Mathews2012a,Neumann2016,Boehmer2019}.
These two models also fundamentally take the above-mentioned locality into account.

A question that was not completely clarified until a few years ago is
in how far graphs with redundancy and coexistence property can be
planarized and if this can be done with local rules.  The first
question was answered to the positive in \cite{Boeltz2019}. Moreover,
in \cite{Boeltz2021} a local algorithm was described which -- as
proved there manually -- guarantees planarity and connectivity provided that the underlying graph satisfies both redundancy and coexistence property.

In this work we take the manual proof from \cite{Boeltz2021} as a blue print and verify each of the proof steps. This requires an axiomatic formalization of the redundancy and coexistence property, as well as a formalization of the considered algorithm.

\begin{figure}[t]
    \centering
   \begin{subfigure}{0.4\textwidth}
    \centering
\definecolor{rvwvcq}{rgb}{0.08235294117647059,0.396078431372549,0.7529411764705882}

\begin{tikzpicture}[line cap=round,line join=round,x=1cm,y=1cm,scale=0.4]
\draw [line width=0.4/0.7pt] (-7,1)-- (-3,1);
\draw [line width=0.4/0.7pt] (-5,2)-- (-5,0);
\draw [line width=0.4/0.7pt] (-7,1)-- (-5,2);
\draw [line width=0.4/0.7pt] (-5,2)-- (-3,1);
\draw [line width=0.4/0.7pt] (-1,1)-- (3,1);
\draw [line width=0.4/0.7pt] (1,2)-- (1,0);
\draw [line width=0.4/0.7pt] (-1,1)-- (1,0);
\draw [line width=0.4/0.7pt] (1,0)-- (3,1);
\draw [line width=0.4/0.7pt] (-5,-2)-- (-5,-4);
\draw [line width=0.4/0.7pt] (-7,-3)-- (-3,-3);
\draw [line width=0.4/0.7pt] (-7,-3)-- (-5,-2);
\draw [line width=0.4/0.7pt] (-7,-3)-- (-5,-4);
\draw [line width=0.4/0.7pt] (-1,-3)-- (3,-3);
\draw [line width=0.4/0.7pt] (1,-2)-- (1,-4);
\draw [line width=0.4/0.7pt] (1,-2)-- (3,-3);
\draw [line width=0.4/0.7pt] (3,-3)-- (1,-4);

\draw [line width=0.4/0.7pt,dashed] (-7,1)-- (-5,0);
\draw [line width=0.4/0.7pt,dashed] (-5,0)-- (-3,1);

\draw [line width=0.4/0.7pt,dashed] (-1,1)-- (1,2);
\draw [line width=0.4/0.7pt,dashed] (1,2)-- (3,1);

\draw [line width=0.4/0.7pt,dashed] (-3,-3)-- (-5,-2);
\draw [line width=0.4/0.7pt,dashed] (-3,-3)-- (-5,-4);

\draw [line width=0.4/0.7pt,dashed] (1,-2)-- (-1,-3);
\draw [line width=0.4/0.7pt,dashed] (-1,-3)-- (1,-4);
\draw [fill=black] (-7,1) circle (2.5pt);
\draw[color=black] (-7.5,1) node {$u$};
\draw [fill=black] (-3,1) circle (2.5pt);
\draw[color=black] (-2.5,1) node {$v$};
\draw [fill=black] (-5,2) circle (2.5pt);
\draw[color=black] (-5,2.4) node {$w$};
\draw [fill=black] (-5,0) circle (2.5pt);
\draw[color=black] (-5,-0.4) node {$x$};
\draw [fill=black] (-1,1) circle (2.5pt);
\draw[color=black] (-1.4,1) node {$u$};
\draw [fill=black] (3,1) circle (2.5pt);
\draw[color=black] (3.4,1) node {$v$};
\draw [fill=black] (1,2) circle (2.5pt);
\draw[color=black] (1,2.4) node {$w$};
\draw [fill=black] (1,0) circle (2.5pt);
\draw[color=black] (1,-0.4) node {$x$};
\draw [fill=black] (-5,-2) circle (2.5pt);
\draw[color=black] (-5,-1.6) node {$w$};
\draw [fill=black] (-5,-4) circle (2.5pt);
\draw[color=black] (-5,-4.4) node {$x$};
\draw [fill=black] (-7,-3) circle (2.5pt);
\draw[color=black] (-7.4,-3) node {$u$};
\draw [fill=black] (-3,-3) circle (2.5pt);
\draw[color=black] (-2.6,-3) node {$v$};
\draw [fill=black] (-1,-3) circle (2.5pt);
\draw[color=black] (-1.4,-3) node {$u$};
\draw [fill=black] (3,-3) circle (2.5pt);
\draw[color=black] (3.4,-3) node {$v$};
\draw [fill=black] (1,-2) circle (2.5pt);
\draw[color=black] (1,-1.6) node {$w$};
\draw [fill=black] (1,-4) circle (2.5pt);
\draw[color=black] (1,-4.4) node {$x$};
\end{tikzpicture}
       \caption{Redundancy property.}
        \label{fig:redundancy}
    \end{subfigure}
     ~~~~~
    \begin{subfigure}[h]{0.45\textwidth}
\vspace{-3.5cm}
\centering
\definecolor{rvwvcq}{rgb}{0.08235294117647059,0.396078431372549,0.7529411764705882}
\begin{tikzpicture}[line cap=round,line join=round,x=1cm,y=1cm,scale=0.4]
\draw [line width=0.4/0.7pt] (-3,-5)-- (5,-5);
\draw [line width=0.4/0.7pt] (5,-5)-- (1,1);
\draw [line width=0.4/0.7pt] (1,1)-- (-3,-5);
\draw [line width=0.4/0.7pt] (-3,-5)-- (1,-3);
\draw [line width=0.4/0.7pt] (1,-3)-- (1,1);
\draw [line width=0.4/0.7pt] (1,-3)-- (5,-5);
\draw [fill=black] (-3,-5) circle (2.5pt);
\draw[color=black] (-3.4,-5.4) node {$u$};
\draw [fill=black] (5,-5) circle (2.5pt);
\draw[color=black] (5.4,-5.4) node {$v$};
\draw [fill=black] (1,1) circle (2.5pt);
\draw[color=black] (1,1.43) node {$w$};

\draw [fill=black] (1,-3) circle (2.5pt);
\draw[color=black] (1.3,-2.57) node {$x$};

\draw[color=white] (0.16,-5.4) node {};
\end{tikzpicture}
        \caption {Coexistence Property.}
        \label{fig:coexsistence}
    \end{subfigure}
    \caption{Redundandcy and Coexistence property \cite{Boeltz2019}.}
    \label{fig:RCG}
\end{figure}
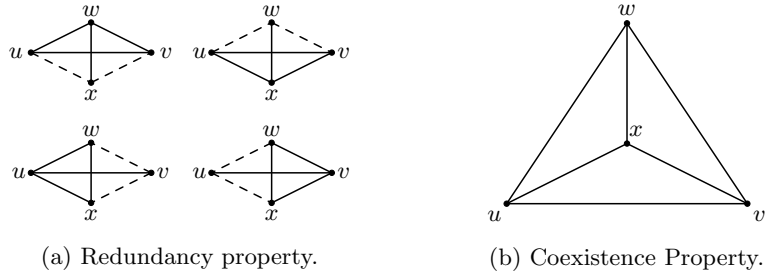

\begin{definition}[Redundancy property]\label{def:redundancy}
We say that a geometric graph $G = (V, E)$ satisfies the \textbf{redundancy property}  if for
every pair of intersecting edges $uv$ and $wx$ of $G$ one of the four
vertices $u,v,w,x$ is connected to the other three (see 
Fig.~\ref{fig:redundancy}). 
The edges $uw, ux, vw, vx$ are called the redundancy edges for the intersection of $uv$ and $wx$.
\label{def-redundancy}
\end{definition}
\begin{definition} [Triangle in a graph; interior of a triangle]\label{def:interior}
Let $G = (V, E)$ be a geometric graph and 
let $u, v, w \in V$ such that the edges $uv,vw,wu$ exist. We say that
$u, v, w$ build a triangle. The interior of the triangle is the set
$Cl(u,v,w) \backslash \{ u, v, w \}$, where 
$Cl(u,v,w)$ is the closed convex set in the Euclidean plane 
delimited by the points associated with the vertices $u,v,w$. 
For three colinear vertices, where $w$ is located between $u$ and $v$, the interior of the triangle $\Delta uvw$ is the open segment between $u$ and $w$, while the interior of the triangle $\Delta vuw$ is the open segment between $w$ and $v$.
\end{definition}
\begin{definition}[Coexistence property]\label{def:coexistence}
Let $G = (V, E)$ be a geometric graph. 
We say that $G$ satisfies the \textbf{coexistence property} if for
every triangle formed by three edges $uv,vw,wu$ of $G$ every vertex $x$
located in
the interior of the triangle  is connected to all three border vertices
of the
 triangle, i.e.\ the edges $ux,vx,wx$ have to exist (see Fig.~\ref{fig:coexsistence}).
\label{def-coexistence}
\end{definition}
\begin{definition}[RCG]
A geometric graph satisfying the redundancy and coexistence property is called redundancy-coexistence graph (RCG).
\end{definition}

\subsection{The CP-algorithm; idea of the correctness proof}
The {\em connected planarization (CP)} algorithm from
\cite{Boeltz2021}, which we verify in this work, is sketched by the following pseudo code. We assume the given graph to be a finite, connected RCG.

\begin{algorithm}[h]
\caption{Global CP-algorithm ({\bf Input:} $G = (V, E)$; {\bf Output:} $G' = (V, F)$)}
\begin{algorithmic}
\State $F \gets \emptyset$ (* initialize the result *)
\State $W \gets E$ (* set up the working set *)
\While {$W \not= \emptyset$}
\State    choose $uv \in W$
\State    remove $uv$ from $W$
\If {for all $w_ix_i{\in} E$ intersecting $uv: ((uw_i {\in} E, ux_i {\in} E)$ or  $(vw_i {\in} E, vx_i {\in} E))$} 
\State add $uv$ to $F$
\State remove all intersecting edges $w_ix_i$ from $W$
\EndIf
\EndWhile
\State $G' \gets (V,F)$
\end{algorithmic}\label{alg:globalCP}
\end{algorithm}

\noindent 
In the initial state, the CP-algorithm has a working set $W$ containing all edges $E$ and an empty result set $F$.
In every iteration step of the algorithm an edge $uv$ of $W$ is chosen 
and removed from $W$, 
and it is decided whether $uv$ is added to $F$ or not. 
This decision depends on the CP-condition.
\begin{definition}[The CP-condition]
The CP-condition is satisfied in a graph $G = (V, E)$ for an edge $uv \in E$ if for all intersecting edges $w_ix_i \in E$ the two redundancy edges $uw_i$ and $ux_i$ or the two redundancy edges $vw_i$ and $vx_i$ exist. (see Fig.~\ref{fig:CP_Input}).
\end{definition}
If the CP-condition holds, $uv$ is added to $F$ and all intersecting edges are removed from $W$.
On the other hand, for every input RCG graph $G = (V, E)$, an edge $uv
\in E$ is not added to $F$ in any iteration
of the CP-algorithm for two reasons:
\begin{itemize}
    \item[(a)] An edge $wx$ intersecting with $uv$ was added to $F$ before $uv$ was considered.
    \item[(b)] The edge $uv$ does not satisfy the CP-condition: there
      is an intersecting edge $wx$ such that ($uw \not\in E$ and $vw
      \not\in E$) or ($ux \not\in E$ and $vx \not\in
      E$).
\end{itemize}
\begin{figure*}[b!]
    \centering
    \begin{subfigure}[t]{0.45\textwidth}
        \centering
\begin{tikzpicture}
\draw [line width=0.5pt,color=black] (-5.012585197384735,1.7)-- (-6,0.3);
\draw [line width=0.5pt,color=black, dashed] (-7,1)-- (-3,1);
\draw [line width=0.5pt,color=black] (-3,1)-- (-6,0.3);
\draw [line width=0.5pt,color=black] (-4,0.3)-- (-7,1);
\draw [line width=0.5pt,color=black] (-5.012585197384735,1.7)-- (-4,0.3);
\draw [line width=0.5pt,color=black] (-4,0.3)-- (-3,1);
\draw [line width=0.5pt,color=black] (-7,1)-- (-6,0.3);
\draw [line width=0.5pt,color=black] (-6,0.3)-- (-4,0.3);
\draw [fill=black] (-5.012585197384735,1.7) circle (1pt);
\draw[color=black] (-5,1.9) node {$w$};
\draw [fill=black] (-6,0.3) circle (1pt);
\draw[color=black] (-6.05,0.1) node {$x$};
\draw [fill=black] (-7,1) circle (1pt);
\draw[color=black] (-7.2,1) node {$u$};
\draw [fill=black] (-3,1) circle (1pt);
\draw[color=black] (-2.8,1) node {$v$};
\draw [fill=black] (-4,0.3) circle (1pt);
\draw[color=black] (-3.95,0.1) node {$y$};
\end{tikzpicture}

        \caption{The edge $uv$ does not satisfy the CP-condition; all other edges satisfy it.}\label{fig:CP_Input}
    \end{subfigure}%
    ~~~ 
    \begin{subfigure}[t]{0.45\textwidth}
        \centering
\begin{tikzpicture}
\draw [line width=1pt,color=black] (-5.012585197384735,1.7)-- (-6,0.3);
\draw [line width=0.5pt,color=black, dashed] (-7,1)-- (-3,1);
\draw [line width=0.5pt,color=black,dashed] (-3,1)-- (-6,0.3);
\draw [line width=0.5pt,color=black,dashed] (-4,0.3)-- (-7,1);
\draw [line width=1pt,color=black] (-5.012585197384735,1.7)-- (-4,0.3);
\draw [line width=1pt,color=black] (-4,0.3)-- (-3,1);
\draw [line width=1pt,color=black] (-7,1)-- (-6,0.3);
\draw [line width=1pt,color=black] (-6,0.3)-- (-4,0.3);
\draw [fill=black] (-5.012585197384735,1.7) circle (1pt);
\draw[color=black] (-5,1.9) node {$w$};
\draw [fill=black] (-6,0.3) circle (1pt);
\draw[color=black] (-6.05,0.1) node {$x$};
\draw [fill=black] (-7,1) circle (1pt);
\draw[color=black] (-7.2,1) node {$u$};
\draw [fill=black] (-3,1) circle (1pt);
\draw[color=black] (-2.8,1) node {$v$};
\draw [fill=black] (-4,0.3) circle (1pt);
\draw[color=black] (-3.95,0.1) node {$y$};
\end{tikzpicture}

        \caption{A possible output graph after the execution of the CP-algorithm.}\label{fig:Result_CP}
    \end{subfigure}%
\caption{Illustration of the CP-condition and the CP-algorithm.} \label{fig:CP_Condition}
\end{figure*}
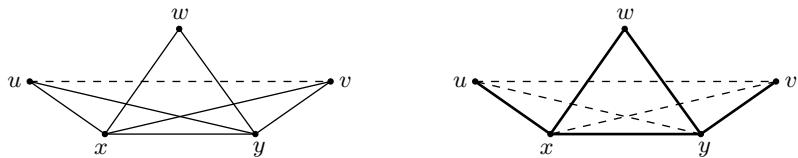
\begin{definition}[Deleting edges]\label{def:deleting_edges}
If there is an edge $wx$ intersecting with $uv$ such that one of the 
properties (a) or (b) holds, then $wx$ deletes $uv$. 
\end{definition}
One can prove (cf.\ \cite{Boeltz2021}) that if an edge $uv$ is not in $F$ because it does not satisfy the CP-condition, then there also exists an edge intersecting with $uv$ that is in $F$. One can therefore assume that every edge that is not in $F$ was not added to $F$ because of condition (a). 
In order to make the terminology even clearer, 
we will in this context consider the order of the vertices $w, x$ when $wx$ deletes $uv$. 
\begin{definition}[Deleting edges (oriented)] 
Assume that $wx$ and $uv$ intersect. 
If {\em $wx$ deletes $uv$} then 
$wu, wv \in E$, and if 
{\em $xw$ deletes $uv$} then $xu, xv \in E$. 
\end{definition}
The order in which the algorithm chooses the edges from the working set $W$ determines which edges are added to $F$ and which not.

\begin{figure*}[h!]
    \centering
    \begin{subfigure}[t]{0.45\textwidth}
        \centering
\begin{tikzpicture}
\draw [line width=0.5pt,color=black, dashed] (-5.012585197384735,1.7)-- (-6,0.3);
\draw [line width=1pt,color=black] (-7,1)-- (-3,1);
\draw [line width=1pt,color=black] (-7,1)-- (-5.012585197384735,1.7);
\draw [line width=1pt,color=black] (-5.012585197384735,1.7)-- (-3,1);
\draw [line width=0.5pt,color=black, dashed] (-5.012585197384735,1.7)-- (-4,0.3);
\draw [line width=1pt,color=black] (-4,0.3)-- (-3,1);
\draw [line width=1pt,color=black] (-7,1)-- (-6,0.3);
\draw [line width=1pt,color=black] (-6,0.3)-- (-4,0.3);
\draw [fill=black] (-5.012585197384735,1.7) circle (1pt);
\draw[color=black] (-5,1.9) node {$w$};
\draw [fill=black] (-6,0.3) circle (1pt);
\draw[color=black] (-6.05,0.1) node {$x$};
\draw [fill=black] (-7,1) circle (1pt);
\draw[color=black] (-7.2,1) node {$u$};
\draw [fill=black] (-3,1) circle (1pt);
\draw[color=black] (-2.8,1) node {$v$};
\draw [fill=black] (-4,0.3) circle (1pt);
\draw[color=black] (-3.95,0.1) node {$y$};
\end{tikzpicture}

        \caption{$uv$ is added to $F$ and deletes $wx$ and $wy$, therefore $wx$ and $wy$ are not in $F$.}\label{fig:Deleting1}
    \end{subfigure}%
    ~~~ 
    \begin{subfigure}[t]{0.45\textwidth}
        \centering
\begin{tikzpicture}
\draw [line width=1pt,color=black] (-5.012585197384735,1.7)-- (-6,0.3);
\draw [line width=0.5pt,color=black, dashed] (-7,1)-- (-3,1);
\draw [line width=1pt,color=black] (-7,1)-- (-5.012585197384735,1.7);
\draw [line width=1pt,color=black] (-5.012585197384735,1.7)-- (-3,1);
\draw [line width=1pt,color=black] (-5.012585197384735,1.7)-- (-4,0.3);
\draw [line width=1pt,color=black] (-4,0.3)-- (-3,1);
\draw [line width=1pt,color=black] (-7,1)-- (-6,0.3);
\draw [line width=1pt,color=black] (-6,0.3)-- (-4,0.3);
\draw [fill=black] (-5.012585197384735,1.7) circle (1pt);
\draw[color=black] (-5,1.9) node {$w$};
\draw [fill=black] (-6,0.3) circle (1pt);
\draw[color=black] (-6.05,0.1) node {$x$};
\draw [fill=black] (-7,1) circle (1pt);
\draw[color=black] (-7.2,1) node {$u$};
\draw [fill=black] (-3,1) circle (1pt);
\draw[color=black] (-2.8,1) node {$v$};
\draw [fill=black] (-4,0.3) circle (1pt);
\draw[color=black] (-3.95,0.1) node {$y$};
\end{tikzpicture}

        \caption{One of the edges $wx$ and $wy$ is added to $F$ and
          deletes $uv$, therefore $uv$ is not in $F$. The other of the
          edge $wx$, $wy$ is added to $F$ later.}\label{fig:Deleting2}
    \end{subfigure}%
\caption{Two possible graphs after the execution of the CP-algorithm.} \label{fig:Deleting_Example}
\end{figure*}
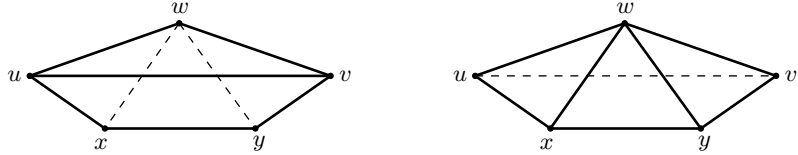

\medskip
\noindent {\bf Termination.} Since the graph $G$ is finite and in every iteration of the while loop at least one edge is removed from the finite set $W$ of edges, it is easy to see that the algorithm terminates.

\medskip
\noindent {\bf Idea 
of the correctness proof.}
For the proof of the correctness of the algorithm we need to show that 
if the input graph $G = (V, E)$ is connected and satisfies
the redundancy and the coexistence property, then the  
graph $G' = (V, F)$ constructed by the CP-algorithm is connected and plane.

\smallskip
\noindent {\bf Planarity} is easy to prove: if an edge $uv$ is added to $F$, then all intersecting edges are removed from the working set $W$.
Therefore, at any time $W$ only contains edges which are not intersecting with edges so far added to $F$. This means that only edges can be added to $F$, which are not intersecting with edges that are already in $F$, which leads to a plane graph as output.

\smallskip
\noindent {\bf Connectivity.} The proof is given by contradiction. 
Assume that after the termination of the algorithm, the resulting graph $G'=(V,F)$ has $k \geq 2$ components which are not connected.
Let $X$ be one of the components and $Y$ be the induced subgraph of $G'\setminus X$, which contains the other $k-1$ components. One can prove that a path with edges in $F$ connecting a vertex in $X$ and a vertex in $Y$ exists, which is a contradiction.
For 
proving this, 
the following lemmas are needed.
The correctness of the lemmas is verified in  Section~\ref{sec:verification_cp} (Lemma \ref{lemma1a}), Section~\ref{sec:part3} (Lemma \ref{lemma2a}) and Section~\ref{sec:convexhull} (Lemma \ref{lemma3a}).

\begin{lemma}\label{lemma1a}
For every edge $uv$ that is not in $F$ there exists an edge $wx$ in $F$ that intersects with $uv$.
\end{lemma}

\begin{lemma}\label{lemma2a}
There exists an edge $uv$ in $E$ with $u \in X$ and $v \in Y$, such that all edges in $F$ intersecting with $uv$ are located in $Y$ (Example: the edge $u_4 v_1$ in Fig.~\ref{fig:connectivity1} which is resulting from edge sequence construction $u_1 v_1,~ u_2 v_1,~ \dots,~ u_4 v_1$ described in Section~\ref{sec:part3}). 
\end{lemma}

\begin{lemma}\label{lemma3a}
For an edge $u v$ and the intersecting edge $w x \in F$ with the redundancy edges $w u$ and $w v$, and intersection point $q$ closest to $u$ among all intersecting edges in $F$, the finest path $P$ from $u$ to $w$ on the convex hull of the triangle $\Delta u q w$ contains only edges in $F$ (Example: the path $i_0 i_1 \dots i_6$ from $u_1$ to $w_1$ in Fig.\ref{fig:connectivity2}, constructed for the edge $u_1 v_1$ and the intersecting edge $w_1 x_1 \in F$).
\end{lemma}

 \definecolor{uuuuuu}{rgb}{0.26666666666666666,0.26666666666666666,0.26666666666666666}
\definecolor{ududff}{rgb}{0.30196078431372547,0.30196078431372547,1}
\definecolor{xdxdff}{rgb}{0.49019607843137253,0.49019607843137253,1}


\begin{figure}[t]
\centering
\begin{subfigure}{0.48\columnwidth}
    \centering
 \begin{tikzpicture}[line cap=round,line join=round,x=1cm,y=1cm, scale=0.65]
\draw [line width=0.5pt] (-3,0)-- (3,0);
\draw [line width=0.5pt] (-0.98,2.96)-- (-0.54,0.98);
\draw [line width=1pt] (-0.54,0.98)-- (0,3);
\draw [line width=0.5pt] (-0.54,0.98)-- (0.24,1.36);
\draw [line width=1pt] (0.24,1.36)-- (0.2,0.24);
\draw [line width=0.5pt] (-3,0)-- (-0.98,2.96);
\draw [line width=0.5pt] (-0.98,2.96)-- (3,0);
\draw [line width=0.5pt] (-0.54,0.98)-- (3,0);
\draw [line width=0.5pt] (0.24,1.36)-- (3,0);
\draw [line width=1pt] (-0.98,2.96)-- (-1.52,-0.66);
\draw [rotate around={49.88520256811728:(-0.97,1.27)},line width=0.5pt,dotted] (-0.97,1.27) ellipse (2.893905241691998cm and 2.3699973729716293cm);
\draw [rotate around={-68.74949449286684:(3.21,0.08)},line width=0.5pt,dotted] (3.21,0.08) ellipse (1.7791563512119308cm and 1.1565886572406587cm);
\draw [fill=xdxdff] (-3,0) circle (2.5pt);
\draw[color=black] (-3.1006291967757003,-0.23) node {$u_1$};
\draw [fill=ududff] (3,0) circle (2.5pt);
\draw[color=black] (3.0975289053404507,-0.23) node {$v_1$};
\draw [fill=ududff] (-0.98,2.96) circle (2.5pt);
\draw[color=black] (-1.023130220720148,3.25) node {$u_2$};
\draw [fill=ududff] (-0.54,0.98) circle (2.5pt);
\draw[color=black] (-0.5767254820635832,0.72) node {$u_3$};
\draw [fill=ududff] (0,3) circle (2.5pt);
\draw [fill=ududff] (0.24,1.36) circle (2.5pt);
\draw[color=black] (0.2645757561737891,1.6) node {$u_4$};
\draw [fill=ududff] (0.2,0.24) circle (2.5pt);
\draw[color=black] (-2.8,2.072175883387477) node {$X$};
\draw[color=black] (3.6,0.97) node {$Y$};
\draw [fill=ududff] (-1.52,-0.66) circle (2.5pt);
\draw [color=uuuuuu] (-1.4215469613259668,0)-- ++(-3pt,-3pt) -- ++(6pt,6pt) ++(-6pt,0) -- ++(6pt,-6pt);
\draw [color=uuuuuu] (-0.17144635816559461,2.3586636231583316)-- ++(-3pt,-3pt) -- ++(6pt,6pt) ++(-6pt,0) -- ++(6pt,-6pt);
\draw [color=uuuuuu] (0.2189250749250749,0.7699020979020978)-- ++(-3pt,-3pt) -- ++(6pt,6pt) ++(-6pt,0) -- ++(6pt,-6pt);
 \end{tikzpicture}
    \caption{Illustration of a sequence of edges connecting $X$ and $Y$.}
    \label{fig:connectivity1}
    \end{subfigure}
  ~~~
    \begin{subfigure}{0.48\columnwidth}
        \centering
\begin{tikzpicture}[line cap=round,line join=round,x=1cm,y=1cm,scale=0.65]
\draw [line width=0.5pt] (-2.02,2.68)-- (3.14,-1.04);
\draw [line width=0.5pt] (2.68,3.06)-- (1.86,-1.1);
\draw [line width=1pt] (2.68,3.06)-- (2.24,2.52);
\draw [line width=1pt] (2.24,2.52)-- (1.8,2.22);
\draw [line width=1pt] (1.8,2.22)-- (1.14,2.08);
\draw [line width=1pt] (1.14,2.08)-- (-0.6,2.2);
\draw [line width=1pt] (-0.6,2.2)-- (-1.36,2.36);
\draw [line width=1pt] (-1.36,2.36)-- (-2.02,2.68);
\draw [line width=0.5pt] (-2.02,2.68)-- (2.68,3.06);
\draw [rotate around={76.38145833855903:(-1.95,1.26)},line width=0.5pt,dotted] (-1.95,1.26) ellipse (2.6166467063676033cm and 1.6918451424242191cm);
\draw [rotate around={-81.66255556115324:(2.18,1.1)},line width=0.5pt,dotted] (2.18,1.1) ellipse (3.0030668490072943cm and 1.8762756992528036cm);
\draw [fill=ududff] (-2.02,2.68) circle (2.5pt);
\draw[color=black] (-2.214774346418534,3.0430504103437546) node {$u_1 = i_0$};
\draw [fill=ududff] (3.14,-1.04) circle (2.5pt);
\draw[color=black] (3.2387830268745597,-1.3) node {$v_1$};
\draw [fill=ududff] (2.68,3.06) circle (2.5pt);
\draw[color=black] (2.7430050838479145,3.456198696199293) node {$w_1=i_6$};
\draw [fill=ududff] (1.86,-1.1) circle (2.5pt);
\draw[color=black] (1.6853454720577388,-1.32) node {$x_1$};
\draw [fill=ududff] (2.24,2.52) circle (2.5pt);
\draw[color=black] (2.395960523729263,2.22) node {$i_5$};
\draw [fill=ududff] (1.8,2.22) circle (2.5pt);
\draw[color=black] (1.9332344435710611,1.9) node {$i_4$};
\draw [fill=ududff] (1.14,2.08) circle (2.5pt);
\draw[color=black] (1.2060934604653153,1.72) node {$i_3$};
\draw [fill=ududff] (-1.36,2.36) circle (2.5pt);
\draw[color=black] (-1.438055569010124,2.0) node {$i_1$};
\draw [fill=ududff] (-0.6,2.2) circle (2.5pt);
\draw[color=black] (-0.5613367916017137,1.852338935685133) node {$i_2$};
\draw [color=ududff] (-0.9918398453942183,2.4941076437324816) circle (1.5pt);
\draw[color=black] (-3.206330232471824,1.7705536899086962) node {$X$};
\draw[color=black] (3.5031979298221034,2.2002279071984563) node {$Y$};
\draw [color=ududff] (0.6990318293022216,2.5471843330727286) circle (1.5pt);
\draw [color=ududff] (1.4374364508651991,2.7830976566371226) circle (1.5pt);
\draw [color=ududff] (-0.7264563986929834,2.4561957227751625) circle (1.5pt);
\draw [color=ududff] (-0.6885444777356641,2.6760848643276143) circle (1.5pt);
\draw [color=ududff] (1.0283813706863434,2.7830976566371226) circle (1.5pt);
\draw [color=ududff] (1.2752930278534746,2.448613338583699) circle (1.5pt);
\draw [color=ududff] (-1.1813994501808147,2.6230081749873673) circle (1.5pt);
\draw [color=ududff] (1.8667189947876555,2.842897316539819) circle (1.5pt);
\draw [color=ududff] (-0.37008434169418214,2.532019564689801) circle (1.5pt);
\draw [color=ududff] (0.7105729964710454,2.240956638235068) circle (1.5pt);
\draw [color=ududff] (1.7721983847202283,2.5491704606299925) circle (1.5pt);
\draw [color=ududff] (2.2002731380465117,2.857384283024917) circle (1.5pt);
\draw [color=uuuuuu] (2.03,-0.24)-- ++(-3pt,-3pt) -- ++(6pt,6pt) ++(-6pt,0) -- ++(6pt,-6pt);
\draw[color=black] (1.75,-0.44) node {$q$};
\end{tikzpicture}
 \caption{Illustration of the path on the convex hull of the triangle $\Delta u_1qw_1$ connecting $X$ and $Y$.}
    \label{fig:connectivity2}
    %
\end{subfigure}
\caption{Connecting paths from $X$ to $Y$.}
\label{fig:Connection}
\end{figure}
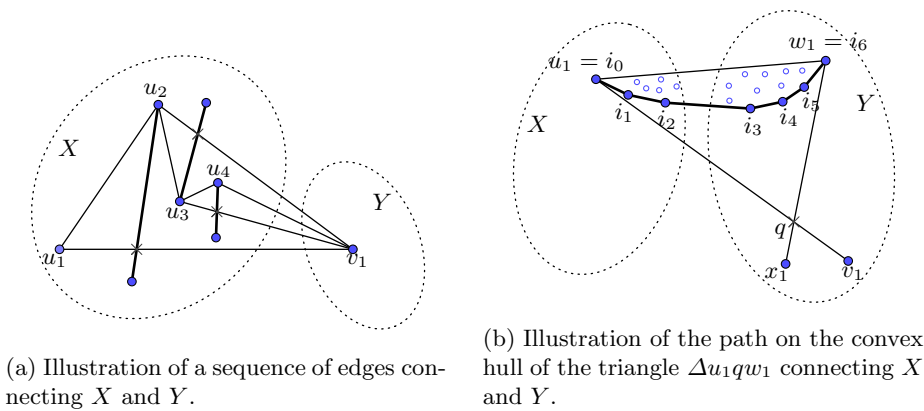

Combining the results of the Lemmas~\ref{lemma1a}--\ref{lemma3a} leads to the connectivity of $G'$ with the following argumentation:
Since the original graph $G=(V,E)$ before applying the CP-algorithm was connected, there has to exist an edge 
$u_1v_1$ in $E$ but not in $F$ with $u_1 \in X$ and $v_1 \in Y$.
Lemma~\ref{lemma1a} ensures that $u_1v_1$ is intersected by an edge $w_1x_1$ in $F$ and by Lemma~\ref{lemma2a} we can choose $u_1v_1$ such that  all intersecting edges in $F$ are located in $Y$.
Lemma~\ref{lemma3a} ensures now -- for the intersecting edge $w_1x_1 \in F$ (located in $Y$) with redundancy edges $w_1u_1$ and $w_1v_1$ and intersection point $q$ closest to $u_1$ -- the existence of a path $P$ from $u_1$ to $w_1$ with all edges in $F$ (see Fig.~\ref{fig:connectivity2}). This path is connecting $u_1$ and $w_1$ and therefore connects $X$ and $Y$ in $G'$.
This is a contradiction to the assumption that $G'$ is not connected.
Therefore the following theorem holds.

\begin{theorem}\label{theorem1}
The edges of F form a connected plane spanning subgraph of G after the termination of the
CP-algorithm.
\end{theorem}

\

\noindent {\bf A local CP-algorithm.} In \cite{Boeltz2021} a local version of
the CP-algorithm is proposed.
In the local variant, each vertex $u$ takes its own decision to keep or remove an edge based on the 2-hop-neighborhood $N(u)$ (i.e.\ the set of vertices connected to $u$ in at most 2 steps).

\begin{algorithm}
\caption{Distributed version of the algorithm from the view of a vertex $u$ \label{alg:sample-triangle-uncorrelated}}
\begin{algorithmic}[1]

\State $F \gets \emptyset$ (* initialize the result for the local view of $u$ *)
\State $W  \gets$ Edges connecting $u$ with its 2-hop neighborhood (* set up working set *)
\For {$v \in N(u)$}
\State $c(uv)=true$ (condition for adding the edge $uv$)
\For {$w \in N(u)$}
\For {$x \in N(w)$}
\If {$wx$ intersects $uv$ and $( vx \notin E) \land (ux \notin E)$}
\State $c(uv)=false$
\EndIf
\EndFor
\EndFor
\If {$c(uv)=true$} 
\State add $uv$ to $F$
\State inform $w$ and $x$ of all edges $wx \in W$ intersecting $uv$ to remove $wx$
\Else
\State inform all neighbors to remove $uv$
\EndIf
\EndFor
\State $G'(u) \gets (N(u),F)$

\end{algorithmic}
\label{alg:localedge}
\end{algorithm}

The redundancy property ensures that each intersection $wx$ that can
remove the edge $uv$ can be detected with two hop information from
$u$. The remaining 
\begin{wrapfigure}[11]{r}{0.4\linewidth}
    \centering
\vspace{-0.7cm}    
\begin{tikzpicture}[line cap=round,line join=round,x=1cm,y=1cm]

\draw [line width=0.5pt] (1,0)-- (1,1);
\draw [-,line width=0.5pt,red,solid] (0,1)-- (2,1);

\draw [-,line width=0.5pt,red,solid] (1,2)-- (1,0);

\draw [-,line width=0.5pt,black,solid] (0,1)-- (1,0);

\draw [-,line width=0.5pt,black,solid] (2,1)-- (1,0);

\draw [-,line width=0.5pt,black,solid] (2,1)-- (1,2);

\draw [fill=black] (0,1) circle (0.5pt);
\draw[color=black] (0,1.2) node {$u$};

\draw [fill=black] (1,0) circle (0.5pt);
\draw[color=black] (1,-0.2) node {$w$};

\draw [fill=black] (2,1) circle (0.5pt);
\draw[color=black] (2,1.2) node {$v$};

\draw
[fill=black] (1,2) circle (0.5pt);
\draw[color=black] (1,2.2) node {$x$};
\end{tikzpicture}
    
   \caption{Illustration of a conflict \\
   in the local approach.}
    \label{fig:local_conflict}
\end{wrapfigure}
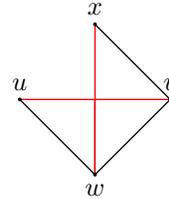
intersections are the ones, where only the redundancy
edges $vw$ and $vx$ exist and therefore the intersection of $wx$ does
not contradict the CP-condition for the intersection with $uv$.
Furthermore, with 2-hop information it can be checked if the CP-condition is satisfied.
The only problem can occur if for two intersecting edges, $uv$ and
$wx$, two nodes might take their decision of adding these edges to $F$
at the same time.  
In this case a global ordering is required, which decides whether $uv$ or $wx$ can be added to $F$.
Because intersections may no be recognized when using information about the set $W$ only, already removed edges can still be considered for the decision if the currently regarded edge is kept or not.
The only additional constraint for the distributed algorithm is therefore that the decisions of all intersecting edges $w_ix_i$ with higher priority have to be available at the moment when a vertex $u$ takes the decision for the edge $uv$.

Since in the global version of the CP-algorithm described in 
Algorithm~\ref{alg:globalCP} the order in which the edges are chosen from the working set $W$ is arbitrary, the correctness proof of
Theorem \ref{theorem1} can be adapted for the local version without problems.

\setcounter{lemma}{0}
\setcounter{theorem}{0}

\section{Modeling geometric graphs}\label{sec:geometric_graphs}

In order to automatically verify the main steps of the correctness proof in the 
CP-algorithm, we have to find an efficient way of representing geometric graphs and their properties, in particular the redundancy and coexistence property.
For expressing the redundancy property we need to be 
able to express the fact that two edges geometrically 
intersect when regarded as segments.
For expressing the coexistence property we need to be 
able to describe the interior of a triangle according to Definition~\ref{def:interior}. 

Expressing such conditions in analytic geometry using coordinates and distances may lead to a high computational effort.
One reason is that for a line passing through two points $A$ and $B$ first an equation of the form $l(x, y) = ax+by+c=0$ has to be computed and then for other points $C, D$ with coordinates $(x_C, y_C)$ resp.\ $(x_D, y_D)$, 
the values $l(x_C, y_C)$ and $l(x_D, y_D)$
have to be computed to decide if $C$ and $D$ are on the same side of the line $AB$ or not.
An important problem in this context is to correctly cover the limit cases such as: "Is a point located on the line passing through two other points?" or "Is a point located on the border of a triangle formed by three other points?", 
and to obtain the desired conditions for describing the 
interior of a triangle according to Definition~\ref{def:interior}. 
In order to minimize the computational effort, alternative approaches like the axiomatization of Hilbert \cite{hilbert} and Tarski \cite{TarskiG99} can be considered.
The advantage of these approaches is that for each point instead of its concrete position only the relative position to other points is considered.
This might be enough to distinguish if two lines intersect or if a point is located inside the triangle formed by three other points.

\smallskip
\noindent 
We show that this axiomatic approach is sufficient for automatically checking the correctness proof of the 
CP-algorithm and its distributed version.
We start with axioms describing geometric aspects, 
then add axioms describing graph properties, 
and finally notions and constructions needed for 
describing the CP-algorithm and for checking the 
main steps in its correctness proof.

\subsection{Geometric axioms}
\label{axioms-geometry}
The geometric axioms are structured in a hierarchical way,
starting with an axiomatization of the properties of 
a predicate $\textbf{left}$ of arity 3. Using this predicate 
additional notions are axiomatized (\textbf{intersection}, \textbf{inside}). 

\medskip
\noindent {\bf Axioms for} $\textbf{left}$. 
The intended meaning for $\textsf{left}(u,v,w)$ is 
{\em "$w$ is on the left side of the (oriented) line through $u$ and $v$ or on the ray starting in $u$ and passing through $v$, but $u \neq w$"} see Fig.~\ref{fig:def_left}.
This includes the case that $v=w$ (if $u\neq v)$.
Note that $\textsf{left}(u,v,w)$ is false for $u=v$ and $u=w$.
If $u \neq v$ then for every point $w$ at least one of of the predicates $\textsf{left}(u,v,w)$ or $\textsf{left}(v,u,w)$ is true.
If both $\textsf{left}(u,v,w)$ and $\textsf{left}(v,u,w)$ are true then $w$ is located strictly between $u$ and $v$.

\noindent 
We proposed a set of axioms for the predicate \textbf{left}, cf.\ also Appendix~\ref{sec:Appendix}. 
When analyzing these axioms 
we noticed that there is a minimal choice of 6 axioms {\bf A1}--{\bf A6} from which the other axioms can be 
derived. These axioms are given below (the variables $u, v, w, x, y,
z$ are universally quantified), see also Fig.~\ref{fig:Axioms_left}. 

\medskip
\noindent $\begin{array}{ll}
\textbf{A1}: & 
(\textsf{left}(u,v,w) \land \textsf{left}(v,u,w) \rightarrow \lnot \textsf{left}(w,u,v))\\
\textbf{A2}: &  
(u \neq w \land v \neq w \land \lnot \textsf{left}(w,u,v) \land \lnot \textsf{left}(w,v,u) \rightarrow \textsf{left}(u,v,w))\\
    
    
\textbf{A3}: & 
(\textsf{left}(u,v,w) \land \textsf{left}(v,u,w) \land \textsf{left}(u,x,w) \rightarrow  \textsf{left}(u,x,v))\\
     
     \textbf{A4}: & 
     (\textsf{left}(u,v,w) \land \textsf{left}(v,u,w) \land \textsf{left}(u,x,v) \rightarrow  \textsf{left}(u,x,w))\\
     
     \textbf{A5}: & 
     (\textsf{left}(u,v,w) \land \textsf{left}(v,w,u) \land \textsf{left}(w,u,v) \\ 
     & \rightarrow  \textsf{left}(u,v,x) \lor \textsf{left}(v,w,x) \lor \textsf{left}(w,u,x))\\
     
     \textbf{A6}: & 
     (\textsf{left}(u,v,z) \land \textsf{left}(v,w,z) \land \textsf{left}(w,u,z) \land \\ & 
     \textsf{left}(x,y,u) \land  \textsf{left}(x,y,v) \land  \textsf{left}(x,y,w) \rightarrow  \textsf{left}(x,y,z))\\
\end{array}$

\medskip
\noindent 
The other axioms, which are derived from {\bf A1, A2, A3, A4, A5, A6} are listed in Appendix~\ref{sec:Appendix}.
Axiom $A5$ can be derived from the other axioms if there exist at
least 5 different points.

\begin{figure}[t]
    \centering
    \begin{minipage}{.5\textwidth}
    \centering
    \begin{tikzpicture}[line cap=round,line join=round,x=1cm,y=1cm]

\draw [-,line width=0.5pt,blue,dashed] (0,0)-- (2,0);

\draw [-,line width=0.5pt,red,dashed] (-2,0)-- (0,0);

\draw [-,line width=0.5pt,green,dashed] (2,0)-- (4,0);

\draw [fill=black] (0,0) circle (0.5pt);
\draw[color=black] (0,0.2) node {$u$};

\draw [color=green, fill=green] (1.5,0.5) circle (0.5pt);
\draw[color=green] (1.5,0.7) node {$w_1$};

\draw [color=green, fill=green] (3,0) circle (0.5pt);
\draw[color=green] (3,0.2) node {$w_2$};

\draw [color=blue, fill=blue] (1,0) circle (0.5pt);
\draw[color=blue] (1,0.2) node {$w_3$};

\draw [color=red, fill=red] (0.5,-0.5) circle (0.5pt);
\draw[color=red] (0.5,-0.3) node {$w_4$};

\draw [color=red, fill=red] (-1,0) circle (0.5pt);
\draw[color=red] (-1,0.2) node {$w_5$};

\draw [fill=black] (2,0) circle (0.5pt);
\draw[color=black] (2,0.2) node {$v$};
\end{tikzpicture}
    
    \captionof{figure}{$w_1$, $w_2$ and $w_3$ are left of $uv$, \\
    while $w_3$, $w_4$ and $w_5$ are left of $vu$.}
    \label{fig:def_left}
\end{minipage}%
\begin{minipage}{.5\textwidth}
    \centering
    \begin{tikzpicture}[line cap=round,line join=round,x=1cm,y=1cm]

\draw [-,line width=0.5pt,blue,dashed] (0,0)-- (2,0);

\draw [-,line width=0.5pt,red,dashed] (1,0)-- (0.5,0.5);

\draw [-,line width=0.5pt,green,dashed] (2,0)-- (0.5,0.5);

\draw [-,line width=0.5pt,black,dashed] (0,0)-- (1,1);

\draw [fill=black] (0,0) circle (0.5pt);
\draw[color=black] (0,0.2) node {$u$};

\draw [fill=black] (1,0) circle (0.5pt);
\draw[color=black] (1,0.2) node {$w$};

\draw [fill=black] (2,0) circle (0.5pt);
\draw[color=black] (2,0.2) node {$v$};

\draw
[fill=black] (0.5,0.5) circle (0.5pt);
\draw[color=black] (0.5,0.7) node {$x$};
\end{tikzpicture}
    
    \captionof{figure}{Illustration of the axioms \\
    {\bf A1}, {\bf A2}, {\bf A3} and {\bf A4}.}
    \label{fig:Axioms_left}
    \end{minipage}
\end{figure}
\medskip
\noindent 
Based on the {\bf left} predicate, the axiomatization of two additional predicates, $\textbf{intersection}$ and 
$\textbf{inside}$, can be given.

\smallskip
\noindent {\bf Axioms for intersection:} The intended meaning for $\textsf{intersection}(u,v,w,x)$ 
for the segments $uv$ and $wx$ is {\em "the segments $uv$ and $wx$ intersect"}. 
This is the case iff one of points $w$ or $x$ is located left of $uv$ and one left of $vu$ and one of points $u$ or $v$ is located left of $wx$ and one left of $xw$.

\smallskip
\noindent With this definition two segments intersect if they have a point in common, which is not an end vertex of both segments.
Moreover, a segment can not intersect with itself, therefore  $\textsf{intersection}(u,v,u,v)$ and $\textsf{intersection}(u,v,v,u)$ are false.
 The axioms for the intersection predicate
  (names starting with \textbf{I}) are given below 
(the variables $u,v,w,x$ are universally quantified).\footnote{Since the clause form of
some of the axioms contain more than one formula, and for doing the
tests we used axioms in clause form, in such cases a range of numbers is used for the axioms.} 

\noindent 
\smallskip
\noindent $\begin{array}{ll}
\textbf{I1-2}: &  \textsf{intersection}(u,v,w,x)  \rightarrow (u \neq w \lor v \neq x) \land (u \neq x \lor v \neq w)
\\
\textbf{I3}: &  \textsf{intersection}(u,v,w,x)  \rightarrow \textsf{left}(u,v,w) \lor \textsf{left}(u,v,x)
\\
\textbf{I4}: &  \textsf{intersection}(u,v,w,x)  \rightarrow \textsf{left}(v,u,w) \lor \textsf{left}(v,u,x)
\\
\textbf{I5}:&  \textsf{intersection}(u,v,w,x)  \rightarrow \textsf{left}(w,x,u) \lor
\textsf{left}(w,x,v)
\\
\textbf{I6}:&  
\textsf{intersection}(u,v,w,x)  \rightarrow \textsf{left}(x,w,u) \lor \textsf{left}(x,w,v)
\\
\textbf{I11-12}: & 
\textsf{left}(u,v,w) \land \textsf{left}(v,u,x) \land \textsf{left}(w,x,u) \land \textsf{left}(x,w,v) \rightarrow \\
& ~~~~~~~~~~~~~~~~~~~~~~~~~~~~~~~~~~~~\textsf{intersection}(u,v,w,x) \lor (u=x \land v=w)
\end{array}$\\[-1ex]
\noindent $\begin{array}{ll}
\textbf{I13}: & 
\textsf{left}(u,v,w) \land \textsf{left}(v,u,x) \land \textsf{left}(w,x,v) \land \textsf{left}(x,w,u) \rightarrow \\
& ~~~~~~~~~~~~~~~~~~~~~~~~~~~~~~~~~~~~\textsf{intersection}(u,v,w,x)
\\
\textbf{I14}: & 
\textsf{left}(u,v,x) \land \textsf{left}(v,u,w) \land \textsf{left}(w,x,u) \land \textsf{left}(x,w,v) \rightarrow \\
& ~~~~~~~~~~~~~~~~~~~~~~~~~~~~~~~~~~~~\textsf{intersection}(u,v,w,x)
\\
\textbf{I15-16}: & 
\textsf{left}(u,v,x) \land \textsf{left}(v,u,w) \land \textsf{left}(w,x,v) \land \textsf{left}(x,w,u) \rightarrow \\
& ~~~~~~~~~~~~~~~~~~~~~~~~~~~~~~~~~~~~\textsf{intersection}(u,v,w,x) \lor (u=w \land v=x)
\end{array}$

\smallskip
\noindent {\bf Axioms for inside:} The intended meaning for $\textsf{inside}(u,v,w,x)$ is {\em ``the point $x$ is located inside or on the boundary the triangle $\Delta uvw$, but $x$ is not equal to $u$, $v$ or $w$''}.
We can therefore define $\textsf{inside}$ by the following axioms (the variables $u,v,w,x$ are universally quantified): 

\smallskip 
\noindent $\begin{array}{ll}
\textbf{T1-6}: & 
(\textsf{inside}(u,v,w,x) \leftrightarrow \textsf{left}(u,v,x) \land \textsf{left}(v,w,x) \land \textsf{left}(w,u,x)) 
\end{array}$

\smallskip
\noindent The triangles we consider are oriented, so the order in which 
the vertices are written is important: 
In the (oriented) triangle $\Delta uvw$, 
$\textsf{left}(u,v,w)$, $\textsf{left}(v,w,u)$ and $\textsf{left}(w,u,v)$ are all true.
A point $x$ is located inside the triangle $\Delta uvw$ iff $\textsf{left}(u,v,x)$, $\textsf{left}(v,w,x)$ and $\textsf{left}(w,u,x)$ are all true; 
$x$ is located inside the triangle $\Delta uwv$ iff $\textsf{left}(u,w,x)$, $\textsf{left}(w,v,x)$ and $\textsf{left}(v,u,x)$ are all true.
The triangle $\Delta uvw$ is equal to the triangles $\Delta vwu$ and $\Delta wuv$.
We could prove:
\begin{itemize}
\item The inside predicate is transitive, i.e.\ if $y$ is located inside the triangle $\Delta uvx$ and $x$ is located inside the triangle $\Delta uvw$, then $y$ is also located inside the triangle $\Delta uvw$ ({\bf T21}-{\bf T23}).
    
\item If $x$ and $y$ are located inside a triangle $\Delta uvw$, where $x \neq y$, then $y$ is also located in at least one of the triangles $\Delta uvx$, $\Delta vwx$ or $\Delta wux$ ({\bf T24}).
\end{itemize} 

\noindent In what follows we will refer to the union of the axioms for
{\bf left}, {\bf intersection} and {\bf inside} by ${\sf AxGeom}$. A
list of all the axioms we considered can be found in Appendix~\ref{sec:Appendix}--\ref{sec:ax-inside}. 

\medskip
\noindent {\bf Link to other axiom systems for plane geometry.}
Tarski's \cite{TarskiG99} and Hilbert's \cite{hilbert} axiom systems for plane geometry use both a predicate "between". 
Tarski's betweenness notion is non-strict: a point $c$ is between $a$
and $b$ iff $c$ is on the closed segment $[ab]$, while Hilbert's
notion is strict: $c$ is between $a$ and $b$ iff $c$ is on the open segment $(ab)$.
The relation "$c$ is between $a$ and $b$" can be expressed in our axiomatization by "$\textsf{left}(a,b,c) \land \textsf{left}(b,a,c)$" and is strict, so Hilbert's axiomatization is closer to ours than Tarski's.
Since we only consider line segments the axioms of Hilbert's group 1,3 and 4 (cf.\ \cite{hilbert}) are not important for us; 
the related axioms can be found in group 2 (cf.\ \cite{hilbert}).
Hilbert's axioms 2.1, 2.3 and 2.4 can be derived from our axioms (cf. also the remarks in Section~\ref{geom-ar} on Pasch's axiom).
Since we only consider a finite set of points (related to the vertices
of the graphs we consider or to intersections of the edges in the
graphs we consider), Hilbert's axiom 2.2 (a $\forall
\exists$-sentence) was not needed for the correctness proof, so no
equivalent was included. 
%
Pasch's axiom could also be added in our axiom set and replace the
axioms {\bf A4} and {\bf A6}, because they can be derived by the remaining axioms of ${\sf AxGeom}$ and Pasch's axiom.
But we decided to use the axioms in the proposed way, because the axioms for \textsf{left} can be introduced independently first and afterwards the axioms for \textsf{intersection} and \textsf{inside} can be formulated based on the axioms for \textsf{left}.

\subsection{Axiomatizing graph properties}
In the next step conditions for vertices and edges are defined. We use
a unary predicate ${\sf V}$ for representing the vertices\footnote{
The predicate ${\sf V}$ for the vertices is listed here to distinguish between the axioms for geometry, which hold for arbitrary points, and properties for edges between vertices in a graph.} and a binary
predicate ${\sf E}$ for representing the edges. Axioms for the edge
predicate (denoted by \textbf{E}) are given below (the variables $u,v$ are universally quantified):

\medskip
\noindent $\begin{array}{lll}
\textbf{E1:}~~ & 
\neg {\sf E}(u, u) & \quad (\text{self-loops are not allowed})\\
\textbf{E2:}~~ & 
{\sf E}(u, v) \rightarrow {\sf E}(v, u)
&\quad  (E \text{ is symmetric})\\
\textbf{E3:}~~ & 
{\sf E}(u, v) \rightarrow {\sf V}(u)
\wedge {\sf V}(v) & \quad (\text{edges exist only between vertices})
\end{array}$

\medskip
\noindent Using the predicates  {\sf intersection}, {\sf inside}, {\sf
  V} and {\sf E}, the conditions for the redundancy property and for
the coexistence property can be described by formulae.

\smallskip
\noindent The redundancy property expresses the fact that if two edges $uv$ and $wx$ intersect, then one of the vertices $u$, $v$, $w$ or $x$ is connected to the other 3 ones. This is expressed with the following axiom (the variables $u,v,w, x$ are universally quantified):

\medskip
\noindent {\bf R1:}~~ $
{\sf E}(u,v) \wedge {\sf E}(w, x)
\wedge {\sf intersection}(u,v,w,x) \rightarrow {\sf E}(u,w) \vee {\sf E}(v,x)$

\medskip
\noindent 
The coexistence property expresses the fact that if $\textsf{inside}(u,v,w,x)$ is true and the edges $uv$, $vw$ and $wu$ exist, then also the edges $ux$, $vx$ and $wx$ exist. By the rotation of the triangle it is sufficient that only one of the edges has to exist.
Axiom {\bf C1} (the variables $u,v,w,x$ are universally quantified) expresses this: 

\medskip
\noindent {\bf C1:}~~ $
{\sf inside}(u,v,w,x) \wedge {\sf E}(u, v) \wedge {\sf E}(v, w) \wedge {\sf E}(w, u) \wedge {\sf V}(x) \rightarrow {\sf E}(u, x)$

\subsection{Axioms for notions used in the CP-algorithm}
For graphs satisfying the redundancy and coexistence property the CP
algorithm constructs a connected intersection-free subgraph $G' = (V, F)$. 
The properties of $F$ are described by a set of axioms 
\textbf{F1}--\textbf{F6} (in these formulae the variables $u,v,w,x$ are universally quantified).
Axioms \textbf{F1}--\textbf{F3} express the fact that the edges in $F$ are a
subset of $E$ and $F$ is symmetric and intersection-free. 
Axiom $\textbf{F4}$ expresses the fact that no vertex can be located on an edge in $F$.

\medskip
\noindent \textbf{F1}:~~ $\textsf{F}(u,v) \rightarrow \textsf{F}(v,u)$
\\
\textbf{F2}:~~ $\textsf{F}(u,v) \rightarrow \textsf{E}(u,v)$
\\
\textbf{F3}:~~ $\textsf{F}(u,v) \land \textsf{F}(w,x) \rightarrow \lnot \textsf{intersection}(u,v,w,x)$
\\
\noindent In addition, no vertex can be located on an edge in $F$:\\
\noindent \textbf{F4}:~~ $\textsf{left}(u,v,w) \land \textsf{left}(v,u,w) \land \textsf{V}(w) \rightarrow \lnot \textsf{F}(u,v)$\\ 

\noindent 
Furthermore, the edges in $F$ have to satisfy the CP-condition,
i.e. for each edge $wx \in E$ intersecting with an edge $uv$ in
$F$, both edges $uw$ and $ux$ or both edges $vw$ and $vx$ have to
exist in $E$. This can be expressed (after removing the implications already entailed by {\bf R1} and {\bf F2}) by the following axioms:

\medskip
\noindent \textbf{F5}:~~ $\textsf{F}(u,v) \land \textsf{E}(w,x) \land
\textsf{intersection}(u,v,w,x) \rightarrow \textsf{E}(u,w) \lor
\textsf{E}(v,w)$ (CP-condition)

\noindent \textbf{F6}:~~ $\textsf{F}(u,v) \land \textsf{E}(w,x) \land \textsf{intersection}(u,v,w,x) \rightarrow \textsf{E}(u,x) \lor \textsf{E}(v,x)$ (CP-condition)\\
(only one of the conditions {\bf F5} or {\bf F6} is needed due to symmetry)

\subsection{Axioms for notions used in the correctness proof}

For the proof of correctness a further predicate \textbf{deleting} of arity 4 is defined, and notions such as path along the 
convex hull of points in a certain triangle are formalized. 
We describe the axiomatizations we used for this below.

\smallskip
\noindent {\bf Axioms for deleting:} The intended meaning for $\textsf{deleting}(u,v,w,x)$ is 
{\em "$uv$ and $wx$ are intersecting edges and the edge $wx$ prohibits that $uv$ is in $F$"}.

\smallskip
\noindent Note that if $\textsf{deleting}(u,v,w,x)$ is true then the two redundancy edges $wu$ and $wv$ exist, while if $\textsf{deleting}(u,v,x,w)$ is true then the two redundancy edges $xu$ and $xv$ exist.
The edge $wx$ can prohibit that the edge $uv$ is in $F$ if $wx$ is in $F$ or if only the redundancy edges $wu$ and $wv$, but neither $xu$ nor $xv$ exist for the intersection of $uv$ and $wx$.
Thus,  $\textsf{deleting}(u,v,w,x)$ should hold iff $\textsf{E}(u,v) \land \textsf{E}(w,x) \land \textsf{intersection}(u,v,w,x) \land \textsf{E}(u,w) \land \textsf{E}(v,w) \land \big(( \lnot \textsf{E}(u,x) \land \lnot \textsf{E}(v,x)) \lor \textsf{F}(w,x)\big)$ holds.
The axioms for the deleting predicate are denoted by \textbf{D} and represent the condition above
(the variables $u,v,w,x$ are universally quantified).\\
\\
\textbf{D1-3}: $\textsf{deleting}(u,v,w,x) \rightarrow \textsf{E}(u,v) \land \textsf{E}(w,x) \land \textsf{intersection}(u,v,w,x)$
\\
\textbf{D4-5}: $\textsf{deleting}(u,v,w,x) \rightarrow  \textsf{E}(u,w) \land \textsf{E}(v,w)$
\\
\textbf{D11}: $\textsf{E}(u,x) \land \textsf{deleting}(u,v,w,x) \rightarrow \textsf{F}(w,x)$
\\
\textbf{D12}: $\textsf{E}(v,x) \land \textsf{deleting}(u,v,w,x) \rightarrow \textsf{F}(w,x)$
\\
\textbf{D13}: $\textsf{E}(u,v) \land \textsf{E}(w,x) \land
\textsf{intersection}(u,v,w,x) \land \textsf{E}(u,w) \land \textsf{E}(v,w)$ \\
$~~~~~~~~~~~~~~~~~~~~\rightarrow \textsf{deleting}(u,v,w,x) \lor \textsf{E}(u,x) \lor \textsf{E}(v,x)$
\\
\textbf{D14}: $\textsf{E}(u,v) \land \textsf{F}(w,x) \land
\textsf{intersection}(u,v,w,x) \land \textsf{E}(u,w) \land
\textsf{E}(v,w)$ \\
$~~~~~~~~~~~~~~~~~~~~\rightarrow \textsf{deleting}(u,v,w,x)$

\medskip
\noindent In what follows, we will refer to the axioms describing properties of $E$ and $F$, and of the {\sf deleting} predicate
(axioms {\bf E1}--{\bf E3}, {\bf F1}--{\bf F6}, and {\bf D1}--{\bf D14}) by ${\sf AxGraphs}$.

\medskip
\noindent {\bf Path along the convex hull of a set of points.} 
Consider the triangle $\Delta u_1 q w_1$, where $q$ is the intersection of edges $u_1 v_1$ and $w_1 x_1$.
For representing a path from $u_1$ to $w_1$ along the convex hull of all points located inside of this triangle we use a list structure, with a function symbol 
${\sf next}$, where for every point $i$, ${\sf next}(i)$ is the next vertex on this path if $i$ is a vertex on the convex hull that has a successor, or {\sf nil} otherwise.
Axioms {\bf Y0}--{\bf Y4} express the properties of the convex hull. In these axioms we explicitly write the universal quantifiers.

\smallskip 
\noindent \textbf{Y0}: $\textsf{next}(w_1)=\textsf{nil}$
\\
\textbf{Y1}: $\forall y, i ~(\textsf{V}(y) \land \textsf{left}(u_1,v_1,y) \land \textsf{left}(x_1,w_1,y) \rightarrow  i=\textsf{nil} \lor \textsf{next}(i)=\textsf{nil} \lor  \\
~~~~~~~~~~~~~~~~y=i \lor  y=\textsf{next}(i) \lor \textsf{left}(i,\textsf{next}(i),y) \lor \textsf{left}(y,\textsf{next}(i),i))$\\
\textbf{Y2}: $\forall y, i ~ (\textsf{V}(y) \land \textsf{left}(u_1,v_1,y) \land \textsf{left}(x_1,w_1,y) \rightarrow  i=\textsf{nil} \lor \textsf{next}(i)=\textsf{nil} \lor \\
~~~~~~~~~~~~~~~~y=i \lor y=\textsf{next}(i) \lor \textsf{left}(i,\textsf{next}(i),y) \lor \textsf{left}(\textsf{next}(i),y,i))$\\
\textbf{Y3}: $\forall y, i ~ (\textsf{V}(y) \land \textsf{left}(u_1,v_1,y) \land \textsf{left}(x_1,w_1,y) \land \textsf{left}(\textsf{next}(i),i,y) \rightarrow \\
~~~~~~~~~~~~~~~~i=\textsf{nil} \lor \textsf{next}(i)=\textsf{nil} \lor y=i \lor y=\textsf{next}(i) \lor \textsf{left}(y,\textsf{next}(i),i))$\\
\textbf{Y4}: $\forall y, i ~ (\textsf{V}(y) \land \textsf{left}(u_1,v_1,y) \land \textsf{left}(x_1,w_1,y) \land \textsf{left}(\textsf{next}(i),i,y) \rightarrow  \\
~~~~~~~~~~~~~~~~i=\textsf{nil} \lor \textsf{next}(i)=\textsf{nil}  \lor y=i \lor y=\textsf{next}(i) \lor \textsf{left}(\textsf{next}(i),y,i))$

\noindent Axioms {\bf Y11}--{\bf Y13} express the fact that all the vertices on the path are in $\Delta u_1 q w_1$
and that there is an edge between the vertices of the convex hull.\\[2ex]
\noindent \textbf{Y11}: $\forall i ~(i=\textsf{nil} \lor \textsf{next}(i)=\textsf{nil} \lor \textsf{inside}(u_1,v_1,w_1,\textsf{next}(i)) \lor \textsf{next}(i)=w_1)$\\
\textbf{Y12}: $\forall i ~(i=\textsf{nil} \lor \textsf{next}(i)=\textsf{nil} \lor \textsf{inside}(u_1,x_1,w_1,i) \lor i=u_1)$\\
\textbf{Y13}: $\forall i ~(\textsf{E}(i,\textsf{next}(i)) \lor i=\textsf{nil} \lor \textsf{next}(i)=\textsf{nil})$\\[2ex]
Any vertex $i$ on the path is located in $\Delta u_1v_1{\sf next}(i)$ and $\Delta u_1x_1{\sf next}(i)$ or equal to $u_1$; ${\sf next}(i)$ is located in $\Delta v_1w_1i$ and $\Delta x_1
w_1i$ or equal to $w_1$.\\[2ex]
\textbf{Y14}: $\forall i ~(i=\textsf{nil} \lor \textsf{next}(i)=\textsf{nil} \lor \textsf{inside}(u_1,v_1,\textsf{next}(i),i) \lor i=u_1)$\\
\textbf{Y15}: $\forall i ~(i=\textsf{nil} \lor \textsf{next}(i)=\textsf{nil} \lor \textsf{inside}(u_1,x_1,\textsf{next}(i),i) \lor i=u_1)$\\
\textbf{Y16}: $\forall i ~(i=\textsf{nil} \lor \textsf{next}(i)=\textsf{nil} \lor \textsf{inside}(v_1,w_1,i,\textsf{next}(i)) \lor \textsf{next}(i)=w_1)$\\
\textbf{Y17}: $\forall i ~(i=\textsf{nil} \lor \textsf{next}(i)=\textsf{nil} \lor \textsf{inside}(x_1,w_1,i,\textsf{next}(i)) \lor \textsf{next}(i)=w_1)$

\section{Automated Reasoning}\label{sec:autreasoning}

We used the axioms above for automating reasoning about geometric graphs 
and for proving correctness of results used in the proof of the CP-algorithm. 

\subsection{Analyzing the axioms; Decidability}
If we analyze the form of the axioms used in the proposed 
axiomatization (with the exception of the axioms describing the path along the convex hull), we notice that they are all universally quantified 
formulae over a signature containing only constants and predicate symbols (thus no function symbols with arity $\geq 1$), i.e.\ 
all the verification tasks can be formulated as satisfiability tasks for formulae in the Bernays-Sch{\"o}nfinkel class.
\begin{definition}[The Bernays-Sch{\"o}nfinkel class]
Let $\Sigma = \{ c_1, \dots c_n \}$ be a signature consisting only of 
constants (it does not contain function symbols with arity $\geq 1$).
The Bernays-Sch{\"o}nfinkel class consists of all formulae 
of the form $\exists x_1 \dots x_n \forall y_1 \dots \forall y_m
F(x_1, \dots, x_n, y_1, \dots, y_m)$, where $n, m \geq 0$ and
$F$ is a quantifier-free formula containing only variables $x_1, \dots, x_n, y_1, \dots, y_m$. 
\end{definition}
Satisfiability of formulae in the Bernays-Sch{\"o}nfinkel class is decidable.

\medskip
\noindent {\bf Description of paths along the convex hull of points in a triangle.} In \cite{necula-mcpeak-2005}, McPeak and Necula investigate 
reasoning in pointer data structures. 
The language used has sorts ${\sf p}$ (pointer) and 
${\sf s}$ (scalar). Sets $\Sigma_p$ and $\Sigma_s$ of pointer 
resp.\ scalar fields are modeled by functions of sort 
${\sf p} \rightarrow {\sf p}$
and ${\sf p} \rightarrow {\sf s}$, respectively. A constant 
${\sf nil}$ of sort ${\sf p}$ exists.
The only predicate of sort ${\sf p}$ is 
equality; predicates of 
sort ${\sf s}$ can have any arity.  
The axioms considered in \cite{necula-mcpeak-2005} are of the form 
\begin{eqnarray}
\forall p ~~ {\cal E} \vee {\cal C} \label{loc-ax}
\end{eqnarray}
\noindent where ${\cal E}$ contains disjunctions of pointer equalities and 
${\cal C}$ contains scalar constraints 
(sets of both positive and negative literals). It is assumed that 
for all terms $f_1(f_2(\dots f_n(p)))$ occurring in 
the body of an axiom, the axiom also contains 
the disjunction $p = {\sf nil} \vee f_n(p) = {\sf nil} \vee \dots \vee 
f_2(\dots f_n(p)) = {\sf nil}$.\footnote{This nullable subterm property has the role 
of excluding null pointer errors.} 

\noindent In \cite{necula-mcpeak-2005} a decision procedure for conjunctions of axioms of type~(\ref{loc-ax}) is proposed; in \cite{sofronie-tacas-08} a decision procedure based on instantiation of the universally quantified variables of sort ${\sf p}$ in axioms of type~(\ref{loc-ax}) is proposed. 

\medskip
\noindent The axioms {\bf Y1}--{\bf Y17} for the convex hull are all of type~(\ref{loc-ax}) and satisfy the nullable subterm condition above. 
The axioms needed for the verification conditions can therefore be represented as ${\cal T}_0 \subseteq {\cal T}_1 = {\cal T}_0 \cup \{ {\bf Y1}, \dots, {\bf Y4}, {\bf Y11}, \dots {\bf Y17}\}$
where the theory ${\cal T}_0$ axiomatized by ${\sf AxGeom} \cup {\sf AxGraphs} \cup \{ {\bf R1, C1 } \}$ can be regarded as a theory of scalars. Therefore, satisfiability of ground instances w.r.t.\ ${\cal T}_1$ is decidable, and decision procedures based on instantiation exist.

\subsection{Choosing the proof systems} 
For solving the proof tasks, we used SPASS \cite{SPASS} and Z3 \cite{z3}. 
The tests reported here were made with Z3 because 
the axioms we used were (with the exception of the modelling of the 
paths along the convex hull) in the Bernays-Sch{\"o}nfinkel class, a class for which 
Z3 can be used as a decision procedure 
\cite{PiskacMB10}, whereas SPASS might not terminate. 
In addition, instantiation-based decision procedures exist for the type of 
list structures used for representing the paths along the convex hulls of points we consider
\cite{necula-mcpeak-2005}. Since Z3 is an instantiation-based SMT-prover, it performs the 
necessary instances during the proof process.

We also considered using 
methods for complete hierarchical reasoning in local theory extensions,  proposed in  \cite{Sofronie-cade2005,sofronie-tacas-08,ihlemann-sofronie-cade-2010}, 
since when defining the properties of geometric graphs we proceeded hierarchically -- 
the properties of new predicate symbols used symbols already introduced. 
This was not possible because the current implementation of the hierarchical reduction in our system H-PILoT \cite{hpilot} can only be used for chains of theory extensions in which for every extension in this chain the free variables occur below extension symbols, which is not 
the case in the axiomatization we propose here. 

\medskip
\noindent The tests can be found under 
  \url{https://github.com/sofronie/tests-vmcai-2024.git} and 
 \url{https://userpages.uni-koblenz.de/~boeltz/CP-Algorithm-Verification/}.

\noindent In what follows, when mentioning only {\em file or folder names}, we refer
to files/folders in
\url{https://userpages.uni-koblenz.de/~boeltz/CP-Algorithm-Verification/};
the files or folder names are linked and can be seen by clicking on
them.
We did not include links for folder names in
\url{https://github.com/sofronie/tests-vmcai-2024.git} we mention, but
the file structure is the same.

\subsection{Automated proof of properties of geometric graphs}
\label{geom-ar}

In a first step, we analyzed the axiomatization of plane geometry we propose and the axioms of geometric graphs. 
We automatically proved the following results in plane geometry: 
\begin{description}
    \item[Dependencies between axioms:] We identified axioms of the left predicate which can be derived from the other ones. 
    We proved that some axioms of the inside predicate follow from the
    others. A discussion can be found in
    Appendix~\ref{sec:Appendix}. The proofs for the dependencies of
    axioms are listed in the folder
    \href{https://userpages.uni-koblenz.de/~boeltz/CP-Algorithm-Verification/Axioms/}{\color{blue}Axioms\color{black}}
    and also under
    \url{https://github.com/sofronie/tests-vmcai-2024.git} in the
    directory {\sf CP-Algorithm-Verification/Axioms/}.
    \item[Pasch's axiom:] We proved Pasch's Axiom (stating that if a
      line intersects one side of a triangle then it intersects
      exactly one other side) using the axioms {\bf A}, {\bf I} and
      {\bf T} (cf.\ the files 
      \href{https://userpages.uni-koblenz.de/~boeltz/CP-Algorithm-Verification/Axioms/Basic}{\color{blue}z3-test\_pasch\color{black}}
      and also the tests available under 
      \url{https://github.com/sofronie/tests-vmcai-2024.git}, 
    directory {\sf CP-Algorithm-Verification/Axioms/Basic-Properties-for-Geometry-and-RCG}). 
\end{description}
We then automatically proved the following properties of geometric graphs $G = (V, E)$ 
which are useful in the correctness proof of the CP-algorithm: 
\begin{description}
    \item[For graphs with the coexistence property:] We proved that the coexistence property implies the clique property for vertices in the interior of a triangle defined by edges, i.e.\ the property that for every triangle $\Delta uvw$, where $u, v, w$ are vertices of $G$ and $uv, vw, wu \in E$, every two different vertices $x, y$ located in the interior of the triangle 
    according to Definition~\ref{def:interior} are connected by the
    edge $xy \in E$ (cf.\ the files 
    \href{https://userpages.uni-koblenz.de/~boeltz/CP-Algorithm-Verification/Axioms/Basic}{\color{blue}z3-test\_clique\color{black}}
     and also the tests available in 
     \url{https://github.com/sofronie/tests-vmcai-2024.git}, 
    directory {\sf CP-Algorithm-Verification/Axioms/Basic-Properties-for-Geometry-and-RCG}). 
    \item[For graphs with the redundancy and coexistence property:] For three colinear vertices $u$, $v$ and $w$, where $w$ is located between $u$ and $v$, it can be proven that if the edge $uv$ exists, then also the edges $uw$ and $wv$ exist.
\end{description}

\subsection{Correctness of the CP-algorithm: Verifying main steps}
\label{sec:verification_cp}

In order to prove the correctness of the CP-algorithm which is
executed on an input graph $G=(V,E)$, that satisfies the redundancy and the
coexistence property, it has to be proven that the resulting graph $G'
= (V, F)$ is connected and intersection-free.
%
As mentioned before, the graph $G'$ is guaranteed to be
intersection-free. This follows from the fact that if an edge is added to $F$ then all intersecting edges are removed from the working set $W$.

Therefore it remains to prove the connectivity of $F$.
We assume that $F$ is not connected after the termination of the algorithm, and we derive a contradiction.
For this, the following lemma is proven:

\begin{lemma}\label{lemma1}
For every edge $uv$ that is not in $F$ there exists an edge $wx$ in $F$ that intersects with $uv$.
\end{lemma}
The proof of Lemma~\ref{lemma1} is split in two parts: 
\begin{description}
    \item[Part 1:] First, it is proven that for an edge $e_0 = u_1v_1$, which is not in $F$ there exists always a sequence of other edges $e_1, ..., e_n$ in $E$ that intersect with $e_0$.
    
The sequence is constructed in the following way:
We start with an edge $d_1 = e_1=w_1x_1$ that deletes, and therefore also intersects $u_1v_1$. If $e_1 \in F$ we are done. 
Assume that we built a sequence $e_1, \dots, e_{i-1}$ of edges which are not in $F$ and intersect $e_0$. Then an edge $d_i$ can be chosen, that deletes $e_{i-1}$.  Let $e_i$ be a redundancy edge of the intersection between $e_{i-1}$ and $d_i$, such that $e_i$ intersects with $e_0$. 

\item[Part 2:] We then show that such a sequence can not be cyclic, by showing that the maximal length of a shortest cycle is 2 and such a cycle can not exist.
This proves the statement of Lemma \ref{lemma1}.
\end{description}
For the construction of the shortest cycle the following cases have to be considered.
If $e_2$ also deletes $u_1v_1$, then a sequence of shorter length can be obtained by choosing $e_2$ instead of $e_1$.
The other situation where a sequence of shorter length can be obtained is if an edge $d_a$ formed by vertices of the edges $d_1,...,d_i$ deletes an edge $e_j$ for $j<i-1$.
In this case $d_a$ can be chosen instead of $d_{j+1}$ to obtain a sequence of shorter length.

\medskip
\noindent {\bf Automated verification of Part 1.} We prove\footnote{The tests can be found in {\scriptsize \url{https://github.com/sofronie/tests-vmcai-2024.git}} (folder Proof) and {\scriptsize \url{https://userpages.uni-koblenz.de/~boeltz/CP-Algorithm-Verification/Proof}}.} Part 1 in two steps. 
\begin{description}
    \item[Step 1] (see Fig.~\ref{fig:part1_case1}). We prove the following statement: 
Let $u_1v_1$ be an edge which is deleted by an edge $w_1x_1$, i.e.\ $u_1v_1$ and $w_1x_1$ intersect and the edges $w_1u_1$ and $w_1v_1$ exist. Assume that $w_1x_1$ is not in $F$ 
because an intersecting edge $w_2x_2$ deletes $w_1x_1$.
Then at least one of the edges $w_1w_2$ or $x_1w_2$ exists and intersects with $u_1v_1$.
     
For this, we prove that the axioms ${\sf AxGeom} \cup {\sf AxGraphs} \cup \{ {\bf R1, C1} \}$ entail: 
\begin{align*}
    &\textsf{deleting}(u_1,v_1,w_1,x_1) 
    \land \textsf{deleting}(w_1,x_1,w_2,x_2) 
    \\ \rightarrow &\textsf{deleting}(u_1,v_1,w_2,x_1) 
     \lor \textsf{Intersection}(u_1,v_1,w_1,w_2) 
\end{align*}

The case that $x_2w_2$ deletes $w_1x_1$ is analogous, with an intersection of $u_1v_1$ and $x_2w_1$ or an intersection of $u_1v_1$ and $x_2x_1$. 
The test can be found under \href{https://userpages.uni-koblenz.de/~boeltz/CP-Algorithm-Verification/Proof/}{\color{blue}z3-test\_proof\_part1\_step1 \color{black}}. 
If the coexistence property is dropped, the implication of Step 1 does
not hold anymore. We show that the counterexample in
Fig.~\ref{fig:part1_counterexample} is consistent with the
axiomatization from which the coexistence property is removed (cf.\
\href{https://userpages.uni-koblenz.de/~boeltz/CP-Algorithm-Verification/Counterexamples/}{\color{blue}z3-test\_proof\_part1\_step1\_without\_coexistence\color{black}} or
\url{https://github.com/sofronie/tests-vmcai-2024.git} folder {\sf Counterexamples}).

\begin{figure*}[t]
    \centering
    \begin{subfigure}[t]{0.45\textwidth}
        \centering
\begin{tikzpicture}
\draw [line width=0.5pt,color=black] (-5.012585197384735,1.7)-- (-5,0.3);
\draw [line width=1pt,color=black] (-7,1)-- (-3,1);
\draw [line width=0.5pt,color=black] (-7,1)-- (-5.012585197384735,1.7);
\draw [line width=0.5pt,color=black] (-5.012585197384735,1.7)-- (-3,1);
\draw [line width=1pt,color=black] (-5.012585197384735,1.7)-- (-4,0.7);
\draw [line width=0.5pt,color=black] (-6,0.7)-- (-4,0.7);
\draw [line width=0.5pt,color=black] (-5,0.3)-- (-4,0.7);
\draw [fill=black] (-5.012585197384735,1.7) circle (1pt);
\draw[color=black] (-5,1.9) node {$w_1$};
\draw [fill=black] (-5,0.3) circle (1pt);
\draw[color=black] (-5.05,0.1) node {$x_1$};
\draw [fill=black] (-7,1) circle (1pt);
\draw[color=black] (-7.2,1) node {$u_1$};
\draw [fill=black] (-3,1) circle (1pt);
\draw[color=black] (-2.8,1) node {$v_1$};
\draw [fill=black] (-4,0.7) circle (1pt);
\draw[color=black] (-3.95,0.5) node {$w_2$};
\draw [fill=black] (-6,0.7) circle (1pt);
\draw[color=black] (-5.95,0.5) node {$x_2$};
\end{tikzpicture}

        \caption{The edges $u_1v_1$ and $w_1w_2$ intersect.}\label{fig:fig:part1_case1a}
    \end{subfigure}%
    ~~~ 
  \begin{subfigure}[t]{0.45\textwidth}
        \centering
\begin{tikzpicture}
\draw [line width=0.5pt,color=black] (-5.012585197384735,1.7)-- (-5,0.3);
\draw [line width=1pt,color=black] (-7,1)-- (-3,1);
\draw [line width=0.5pt,color=black] (-7,1)-- (-5.012585197384735,1.7);
\draw [line width=0.5pt,color=black] (-5.012585197384735,1.7)-- (-3,1);
\draw [line width=0.5pt,color=black] (-5.012585197384735,1.7)-- (-4.5,1.3);
\draw [line width=0.5pt,color=black] (-5.5,1.3)-- (-4.5,1.3);
\draw [line width=1pt,color=black] (-5,0.3)-- (-4.5,1.3);
\draw [fill=black] (-5.012585197384735,1.7) circle (1pt);
\draw[color=black] (-5,1.9) node {$w_1$};
\draw [fill=black] (-5,0.3) circle (1pt);
\draw[color=black] (-5.05,0.1) node {$x_1$};
\draw [fill=black] (-7,1) circle (1pt);
\draw[color=black] (-7.2,1) node {$u_1$};
\draw [fill=black] (-3,1) circle (1pt);
\draw[color=black] (-2.8,1) node {$v_1$};
\draw [fill=black] (-4.5,1.3) circle (1pt);
\draw[color=black] (-4.35,1.1) node {$w_2$};
\draw [fill=black] (-5.5,1.3) circle (1pt);
\draw[color=black] (-5.45,1.1) node {$x_2$};
\end{tikzpicture}

        \caption{The edges $u_1v_1$ and $x_1w_2$ intersect.}\label{fig:part1_case1b}
    \end{subfigure}%
\caption{One of the edges $w_1w_2$ or $x_1w_2$ intersects with $u_1v_1$.} \label{fig:part1_case1}
\end{figure*}
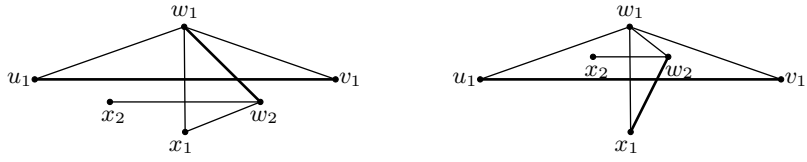

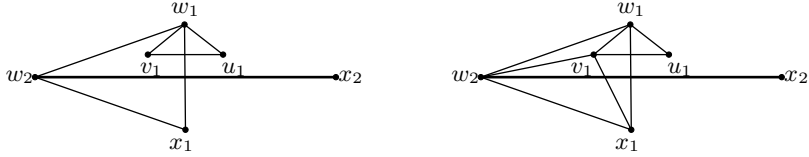
\begin{figure*}[t]
    \centering
     \begin{subfigure}[t]{0.45\textwidth}
        \centering
\begin{tikzpicture}
\draw [line width=0.5pt,color=black] (-5.012585197384735,1.7)-- (-5,0.3);
\draw [line width=1pt,color=black] (-7,1)-- (-3,1);
\draw [line width=0.5pt,color=black] (-7,1)-- (-5.012585197384735,1.7);
\draw [line width=0.5pt,color=black] (-5.012585197384735,1.7)-- (-5.5,1.3);
\draw [line width=0.5pt,color=black] (-5.012585197384735,1.7)-- (-4.5,1.3);
\draw [line width=0.5pt,color=black] (-5.5,1.3)-- (-4.5,1.3);
\draw [line width=0.5pt,color=black] (-5,0.3)-- (-7,1);
\draw [fill=black] (-5.012585197384735,1.7) circle (1pt);
\draw[color=black] (-5,1.9) node {$w_1$};
\draw [fill=black] (-5,0.3) circle (1pt);
\draw[color=black] (-5.05,0.1) node {$x_1$};
\draw [fill=black] (-7,1) circle (1pt);
\draw[color=black] (-7.2,1) node {$w_2$};
\draw [fill=black] (-3,1) circle (1pt);
\draw[color=black] (-2.8,1) node {$x_2$};
\draw [fill=black] (-4.5,1.3) circle (1pt);
\draw[color=black] (-4.35,1.1) node {$u_1$};
\draw [fill=black] (-5.5,1.3) circle (1pt);
\draw[color=black] (-5.45,1.1) node {$v_1$};
\end{tikzpicture}

        \caption{Counterexample for Step 1 without coexistence property.}\label{fig:fig:part1_case1coexistence}
    \end{subfigure}%
    ~~~ 
    \begin{subfigure}[t]{0.45\textwidth}
        \centering
\begin{tikzpicture}
\draw [line width=0.5pt,color=black] (-5.012585197384735,1.7)-- (-5,0.3);
\draw [line width=1pt,color=black] (-7,1)-- (-3,1);
\draw [line width=0.5pt,color=black] (-7,1)-- (-5.012585197384735,1.7);
\draw [line width=0.5pt,color=black] (-5.012585197384735,1.7)-- (-5.5,1.3);
\draw [line width=0.5pt,color=black] (-5.012585197384735,1.7)-- (-4.5,1.3);
\draw [line width=0.5pt,color=black] (-5.5,1.3)-- (-4.5,1.3);
\draw [line width=0.5pt,color=black] (-5,0.3)-- (-7,1);
\draw [line width=0.5pt,color=black] (-5.5,1.3)-- (-7,1);
\draw [line width=0.5pt,color=black] (-5.5,1.3)-- (-5,0.3);
\draw [fill=black] (-5.012585197384735,1.7) circle (1pt);
\draw[color=black] (-5,1.9) node {$w_1$};
\draw [fill=black] (-5,0.3) circle (1pt);
\draw[color=black] (-5.05,0.1) node {$x_1$};
\draw [fill=black] (-7,1) circle (1pt);
\draw[color=black] (-7.2,1) node {$w_2$};
\draw [fill=black] (-3,1) circle (1pt);
\draw[color=black] (-2.8,1) node {$x_2$};
\draw [fill=black] (-4.5,1.3) circle (1pt);
\draw[color=black] (-4.35,1.1) node {$u_1$};
\draw [fill=black] (-5.5,1.3) circle (1pt);
\draw[color=black] (-5.65,1.1) node {$v_1$};
\end{tikzpicture}

        \caption{How the graph would look like if the coexistence property is satisfied.}\label{fig:part1_case1counterexample}
    \end{subfigure}%
\caption{$v_1$ is located inside the triangle $\Delta w_2x_1w_1$.} \label{fig:part1_counterexample}
\end{figure*}
\end{description}

\begin{description}
\item[Step 2:] In the previous step it was proven that
if $w_1 x_1$ deletes $u_1 v_1$ and $w_2 x_2$ deletes $w_1 x_1$ then either $w_1w_2$ or $x_1w_2$ intersects with $u_1v_1$.
The case that $x_1w_2$ (resp. $x_1x_2$) intersects with $u_1v_1$ corresponds to  Step 1, with $w_2$ (resp. $x_2$) instead of $w_1$.
So in Step 2 
$w_1w_2$ is the edge intersecting with $u_1v_1$.
Furthermore $w_3x_3$ deletes $w_1w_2$.
We distinguish the following cases: 

{\bf Case 2a:} Only the edge $w_3x_3$, but not $x_3w_3$ deletes the edge $w_1w_2$.

{\bf Case 2b:} Both of the edges $w_3x_3$ and $x_3w_3$ delete the edge $w_1w_2$.

\smallskip 
In {\bf Step 2a} we prove that in Case 2a another edge deletes $w_1x_1$ (consider Step 1 with this edge instead of $w_2x_2$)
or one of the edges $w_1w_3$ or $w_2w_3$ is 
intersecting with $u_1v_1$.
(cf.\ \href{https://userpages.uni-koblenz.de/~boeltz/CP-Algorithm-Verification/Proof/}{\color{blue}z3-test\_proof\_part1\_step2a\color{black}}).

    \item[Step 2a]: We prove that the axioms ${\sf AxGeom} \cup {\sf AxGraphs} \cup \{ {\bf R1}, {\bf C1} \}$ entail: 
\begin{align*}
    &\textsf{deleting}(u_1,v_1,w_1,x_1) 
    \land 
    \textsf{deleting}(w_1,x_1,w_2,x_2) 
    \land 
    \\ &
    \textsf{intersection}(u_1,v_1,w_1,w_2) 
    \land 
    \textsf{deleting}(w_1,w_2,w_3,x_3) \\ 
    \rightarrow 
    &\textsf{deleting}(w_1,x_1,w_3,x_2) 
     \lor 
    \textsf{deleting}(w_1,x_1,w_3,x_3) 
    \lor \\
    &\textsf{deleting}(w_1,x_1,x_3,w_3)  
    \lor 
    \textsf{deleting}(w_1,x_1,x_3,x_2) 
    \lor 
    \textsf{deleting}(w_1,w_2,x_3,w_3) 
    \lor \\
    &\textsf{intersection}(u_1,v_1,w_1,w_3) 
    \lor \textsf{intersection}(u_1,v_1,w_2,w_3) 
\end{align*}
In {\bf Step 2b} we prove that in Case 2b  another edge deletes $w_1x_1$\footnote{Then $d_2=w_3x_3$ or $d_2=x_3w_3$ or $d_2 = w_3x_2$ or $d_2 = x_3x_2$.} (consider Step 1 with this edge instead of $w_2x_2$) or one of the edges $w_1w_3$, $w_2w_3$, $w_1x_3$, $w_2x_3$ is 
intersecting with $u_1v_1$\footnote{Depending on the edge intersecting with $u_1v_1$, $d_3$ is either $w_3x_3$ or $x_3w_3$.} (cf. \href{https://userpages.uni-koblenz.de/~boeltz/CP-Algorithm-Verification/Proof/}{\color{blue}z3-test\_proof\_part1\_step2b\color{black}}).

\smallskip
%
 \item[Step 2b] We prove  that ${\sf AxGeom} \cup {\sf AxGraphs} \cup \{ {\bf R1, C1} \}$ entails: 
\begin{align*}
    &\textsf{deleting}(u_1,v_1,w_1,x_1) 
    \land 
    \textsf{deleting}(w_1,x_1,w_2,x_2) 
    \land \\ &
    \textsf{intersection}(u_1,v_1,w_1,w_2) 
    \land \\
    & 
    \textsf{deleting}(w_1,w_2,w_3,x_3) 
    \land \textsf{deleting}(w_1,w_2,x_3,w_3) 
    \\
    \rightarrow 
    & \textsf{deleting}(w_1,x_1,w_3,x_2) 
    \lor 
    \textsf{deleting}(w_1,x_1,w_3,x_3) 
    \lor \\
    &\textsf{deleting}(w_1,x_1,x_3,w_3)  
    \lor 
    \textsf{deleting}(w_1,x_1,x_3,x_2) 
    \lor \\
    & \textsf{intersection}(u_1,v_1,w_1,w_3) 
    \lor \textsf{intersection}(u_1,v_1,w_2,w_3) 
    \lor \\
    &\textsf{intersection}(u_1,v_1,w_1,x_3) 
    \lor \textsf{intersection}(u_1,v_1,w_2,x_3) 
\end{align*}

\end{description} 

\medskip
\noindent {\bf Automated verification of Part 2.}
It remains to prove that a sequence of edges which intersect with $u_1v_1$ exists, which is not cyclic, i.e all edges in the sequence differ.
To prove this, it is enough to show that if a sequence of intersecting edges with $uv$ is greater than 2, then also a shorter sequence exists.

This part of the proof is split in the Steps 3--7.
\begin{description}
\smallskip
\item[Step 3:] In this step it is proven that if an edge $w_1x_1$ is deleted by an edge $w_2x_2$, the edge $w_1w_2$ is deleted by an edge $w_3x_3$ and $w_1w_3$ intersects $w_2x_2$ then there exists another edge between two of the vertices $w_2,x_2,w_3,x_3$ which is deleting $w_1x_1$, but is not deleted by $w_2x_2$ (cf.\ \href{https://userpages.uni-koblenz.de/~boeltz/CP-Algorithm-Verification/Proof/}{\color{blue}z3-test\_proof\_part2\_step3 \color{black}}).
For this, we prove  that ${\sf AxGeom} \cup {\sf AxGraphs} \cup \{ {\bf R1, C1 } \}$ entails the formula below 
(some of the atoms are displayed in light gray and commented out in the tests, because they are not needed for proving the conclusion):  
\begin{align*}
    & \color{lightgray} \textsf{deleting}(u_1,v_1,w_1,x_1) 
    \land 
   \color{black}
    \textsf{deleting}(w_1,x_1,w_2,x_2) 
    \land 
    \color{lightgray} \textsf{intersection}(u_1,v_1,w_1,w_2) 
    \land \\& 
    \textsf{deleting}(w_1,w_2,w_3,x_3) 
    \land \textsf{intersection}(w_1,w_3,w_2,x_2) \\
    \rightarrow 
     &\textsf{deleting}(w_1,x_1,w_3,x_2) 
      \lor 
    \textsf{deleting}(w_1,x_1,w_3,x_3) 
    \lor \textsf{deleting}(w_1,x_1,x_2,w_2)\lor\\ 
      & \textsf{deleting}(w_1,x_1,x_3,w_3) \lor 
      \textsf{deleting}(w_1,x_1,x_3,x_2)  
\end{align*}

For the further steps of the proof we therefore assume that the edges inside the triangle $\Delta w_1x_1w_2$ considered in Step 5 as well as the edge $w_1w_3$ considered in Step 6, do not intersect with $w_2x_2$.

\smallskip
\item[Step 4:] In this step it is proven that if an edge $w_1x_1$ is deleted by an edge $w_2x_2$ and $w_1w_2$ is deleted by $w_3x_3$, then there exits another edge between two of the vertices $w_2,x_2,w_3,x_3$ deleting $w_1x_1$, one of the vertices $w_3$ or $x_3$ is located inside the triangle $\Delta w_1x_1w_2$ resp. $\Delta x_1w_1w_2$ (considered in \textbf{Step 5}) or the edge $w_2x_2$ is in $F$ (considered in \textbf{Step 6}).
The special case of $w_3=x_1$ is considered in \textbf{Step 4a} (cf.\ \href{https://userpages.uni-koblenz.de/~boeltz/CP-Algorithm-Verification/Proof/}{\color{blue}z3-test\_proof\_part2\_step4a\color{black}}). 

We prove that ${\sf AxGeom} \cup {\sf AxGraphs} \cup \{ {\bf R1, C1} \}$
entails the following formula 
(the atoms displayed in light gray are commented out in the tests, because they are not needed for proving the conclusion): 
\begin{align*}
    & \color{lightgray} \textsf{deleting}(u_1,v_1,w_1,x_1) 
    \land 
   \color{black}
    \textsf{deleting}(w_1,x_1,w_2,x_2) 
    \land \\
    &
    \color{lightgray} \textsf{intersection}(u_1,v_1,w_1,w_2) 
    \land \color{black} \textsf{deleting}(w_1,w_2,w_3,x_3) 
    \\ 
    \rightarrow 
    &\textsf{deleting}(w_1,x_1,w_3,x_2) 
    \lor
      \textsf{deleting}(w_1,x_1,w_3,x_3) 
    \lor
    \textsf{deleting}(w_1,x_1,x_3,w_3) \lor\\  
     & \textsf{deleting}(w_1,x_1,x_3,x_2) 
    \lor 
    \textsf{inside}(w_1,x_1,w_2,w_3)  
    \lor 
    \textsf{inside}(w_1,x_1,w_2,x_3)  
    \lor \\
    &
    \textsf{inside}(w_1,w_2,x_1,w_3)  
    \lor 
    \textsf{inside}(w_1,w_2,x_1,x_3)  
    \lor 
    \textsf{F}(w_2,x_2) 
\end{align*}

\

\item[Step 5:] It is then proven that if an edge $w_1x_1$ is deleted by an edge $w_2x_2$
and an edge $yw_2$ \textbf{(Step 5a)}, $yw_1$ \textbf{(Step 5b)} or $yz$ \textbf{(Step 5c)} inside the triangle $\Delta w_1x_1w_2$\footnote{In \textbf{Step5d-f} it is proven that the considered edges do not intersect with $w_2x_2$.} is deleted by an edge $w_3x_3$ with both vertices located outside the triangle $\Delta w_1x_1w_2$, then there exists another edge between two of the vertices $w_2,x_2,w_3,x_3$ deleting $w_1x_1$ or another edge between two of the vertices $x_2,w_3,x_3$ deleting $w_1w_2$.
(cf. \href{https://userpages.uni-koblenz.de/~boeltz/CP-Algorithm-Verification/Proof/}{\color{blue}z3-test\_proof\_part2\_step5a\color{black}} to \href{https://userpages.uni-koblenz.de/~boeltz/CP-Algorithm-Verification/Proof/}{\color{blue}z3-test\_proof\_part2\_step5f\color{black}}). Since the formulae are relatively large, we included them in Appendix~\ref{app:proofs-formulae}. 

\

\item[Step 6:] As in the previous steps an edge $w_1x_1$ is deleted by an edge $w_2x_2$ and $w_1w_2$ is deleted by $w_3x_3$.
In addition in the tests of this step the edge $w_1w_3$ (\textbf{Step 6a} and \textbf{Step 6b}) resp. $w_2w_3$ (\textbf{Step 6c}, \textbf{Step 6d} and \textbf{Step 6e}) is deleted by an edge $w_4x_4$ with both vertices outside the triangle $\Delta w_1w_2w_3$.
It is proven that then there exists another edge between two of the vertices $w_2,x_2,w_3,x_3,w_4,x_4$ deleting $w_1x_1$ or another edge  between two of the vertices $x_2,w_3,x_3,w_4,x_4$ deleting $w_1w_2$.

In the special case of $w_4=w_1$ (\textbf{Step 6e}) it is proven that there exists another edge deleting $w_1x_1$.
Analogously in the special case of $w_4=x_1$\textbf{(Step 6b} resp. \textbf{Step 6d)} it is proven that there exists another edge deleting $w_1w_2$.

In the test (\textbf{Step 6g}) it is proven that if $w_4$ is located inside the triangle $\Delta w_1x_1w_2$ and $w_1w_4$\footnote{A similar result is proven in tests \textbf{Step 6h-j} for the edges $w_2w_4$ and $w_3w_4$.} is deleted by $w_5x_5$, then there exists another edge deleting $w_1w_2$ or $w_1w_3$, one of the edges $w_1w_3$, $w_1w_4$ or $w_1w_5$ intersects with $w_2x_2$ or $w_5$ or $x_5$ are located inside a triangle formed by $w_1$, $w_3$ and $w_4$.
The tests are available under  \href{https://userpages.uni-koblenz.de/~boeltz/CP-Algorithm-Verification/Proof/}{\color{blue}z3-test\_proof\_part2\_step6a\color{black}} to  \href{https://userpages.uni-koblenz.de/~boeltz/CP-Algorithm-Verification/Proof/}{\color{blue}z3-test\_proof\_part2\_step6j\color{black}}. Since the formulae are relatively large, we included them in Appendix~\ref{app:proofs-formulae}.

\

\item[Step 7:]Finally it is proven that if $w_1x_1$ is deleted by $w_2x_2$ and $x_2w_2$, $w_1w_2$ is deleted by $x_1w_3$ and $w_1x_2$ is deleted by $x_1w_4$ then a contradiction is obtained.
For this we prove that ${\sf AxGeom} \cup {\sf AxGraph} \cup \{ {\bf R1, C1} \}$ entails:  
\begin{align*}
    &\textsf{deleting}(w_1,x_1,w_2,x_2) 
    \land
    \textsf{deleting}(w_1,x_1,x_2,w_2)\\ 
    \land
    &\textsf{deleting}(w_1,w_2,x_1,w_3) \land 
    \textsf{deleting}(w_1,x_2,x_1,w_4) 
    \rightarrow \bot
\end{align*}
\end{description}
\ignore{
\begin{figure}
\centering

\begin{subfigure}{0.41\columnwidth}
\centering
\begin{tikzpicture}[line cap=round,line join=round,x=1cm,y=1cm,scale=0.5]

\draw [line width=0.4pt] (3,1)-- (9,1);
\draw [line width=0.4pt] (6,-3)-- (9,-2);
\draw [line width=0.4pt] (6,-3)-- (6,3);
\draw [line width=0.4pt] (4,-2)-- (9,-2);
\draw [line width=0.4pt] (9,-2)-- (6,3);
\draw [line width=0.4pt] (7.6,-1.2)-- (9.1,-1.2);
\draw [line width=0.4pt] (7.6,-1.2)-- (6.6,0.7);
\draw [line width=0.4pt,dotted] (6.6,0.7)-- (8,0.5);
\draw [line width=0.4pt] (8,0.5)-- (6,-3);
\draw [line width=0.4pt,dotted] (8,0.5)-- (6,3);
\draw [line width=0.4pt,dotted] (8,0.5)-- (7.6,-1.2);
\draw [line width=0.4pt,dotted] (8,0.5)-- (9,-2);
\draw [line width=0.8pt] (4,-2)-- (6,-3);
\draw [line width=0.8pt] (4,-2)-- (6,3);

\draw [fill=black] (3,1) circle (1.5pt);
\draw[color=black] (2.66,1.19) node {$u$};
\draw [fill=black] (9,1) circle (1.5pt);
\draw[color=black] (9.32,1.19) node {$v$};
\draw [fill=black] (6,3) circle (1.5pt);
\draw[color=black] (6.34,3.25) node {$w_1$};
\draw [fill=black] (6,-3) circle (1.5pt);
\draw[color=black] (6.88,-3.31) node {$x_1$};
\draw [fill=black] (9,-2) circle (1.5pt);
\draw[color=black] (9.42,-2.35) node {$w_2$};
\draw [fill=black] (4,-2) circle (1.5pt);
\draw[color=black] (3.8,-2.35) node {$x_2$};
\draw [fill=black] (9.1,-1.2) circle (1.5pt);

\draw [fill=black] (7.6,-1.2) circle (1.5pt);
\draw[color=black] (7.77,-1.69) node {$y$};
\draw [fill=black] (6.6,0.7) circle (1.5pt);
\draw[color=black] (6.2,0.5) node {$z$};
\draw [fill=black] (8,0.5) circle (1.5pt);
\draw[color=black] (8.32,0.69) node {$x_3$};

\end{tikzpicture}
    \caption{Example for an edge $yz$ inside the triangle $\Delta w_1x_1w_2$ deleted by the edge $x_1x_3$}
    \label{fig:triangle_yz}
    \end{subfigure}
    ~~~
    \begin{subfigure}{0.41\columnwidth}
    \centering
\begin{tikzpicture}[line cap=round,line join=round,x=1cm,y=1cm,scale=0.5]

\draw [line width=0.4pt] (3,0)-- (9,0);

\draw [line width=0.4pt] (6,1)-- (4,-2);
\draw [line width=0.4pt] (4,-2)-- (8,-0.5);
\draw [line width=0.4pt] (6,1)-- (8,-2);
\draw [line width=0.4pt] (8,-2)-- (4,-0.5);
\draw [line width=0.4pt] (4,-2)-- (8,-2);
\draw [line width=0.4pt] (4,-0.5)-- (6,-3);
\draw [line width=0.4pt] (8,-0.5)-- (6,-3);
\draw [line width=0.4pt] (6,1)-- (6,-3);
\draw [line width=0.4pt] (8,-2)-- (6,-3);
\draw [line width=0.4pt] (4,-2)-- (6,-3);
\draw [line width=0.4pt,dotted] (4,-0.5)-- (6,1);
\draw [line width=0.4pt,dotted] (8,-0.5)-- (6,1);
\draw [line width=0.4pt,dotted] (4,-2)-- (4,-0.5);
\draw [line width=0.4pt,dotted] (8,-0.5)-- (8,-2);

\draw [fill=black] (3,0) circle (1.5pt);
\draw[color=black] (2.7,0.19) node {$u$};
\draw [fill=black] (9,0) circle (1.5pt);
\draw[color=black] (9.32,0.19) node {$v$};
\draw [fill=black] (6,-3) circle (1.5pt);
\draw[color=black] (6.6,-3.21) node {$x_1$};
\draw [fill=black] (4,-0.5) circle (1.5pt);
\draw[color=black] (3.5,-0.81) node {$w_4$};
\draw [fill=black] (8,-0.5) circle (1.5pt);
\draw[color=black] (8.5,-0.81) node {$w_3$};
\draw [fill=black] (6,1) circle (1.5pt);
\draw[color=black] (6.5,1.09) node {$w_1$};
\draw [fill=black] (4,-2) circle (1.5pt);
\draw[color=black] (3.7,-2.37) node {$x_2$};
\draw [fill=black] (8,-2) circle (1.5pt);
\draw[color=black] (8.5,-2.21) node {$w_2$};

\end{tikzpicture}
    \caption{Dotted edges are assumed not to exist, therefore the intersection of $w_2w_4$ and $x_2w_3$ does not satisfy the redundancy property. The remaining redundancy and coexistence edges are omitted in the drawing for simplification}
    \label{fig:case6}
    \end{subfigure}%
~~~
\caption{Illustration of two cases considered in part2 of the proof}
\label{fig:Counterexample_Step2}
\end{figure}
}

\subsection{Connection between the components $X$ and $Y$}
\label{sec:part3}

In this section the following Lemma is proven.

\begin{lemma}\label{lemma2}
There exists an edge $uv$ in $E$ with $u \in X$ and $v \in Y$, such that all edges in $F$ intersecting with $uv$ are located in $Y$.
\end{lemma}
We consider the resulting graph $G'=(V,F)$ obtained after the
termination of the CP-algorithm with the components $X$ and
$Y=G'\setminus X$. In the input graph $G=(V,E)$ (which was assumed to
be connected) there was an edge $u_1v_1$ with $u_1 \in X$ and $v_1 \in Y$.
Now a sequence of edges connecting the components $X$ and $Y$ is constructed.
The following tests show that a sequence of edges intersecting with
edges connecting the components $X$ and $Y$ can not be cyclic (see
Fig.~\ref{fig:connectivity1}). 

For all tests\footnote{The tests can be found in {\scriptsize \url{https://github.com/sofronie/tests-vmcai-2024.git}},
  folder {\sf Proof} and in {\scriptsize \url{https://userpages.uni-koblenz.de/~boeltz/CP-Algorithm-Verification/Proof}}.} the edge $u_1v_1$ is deleted by an edge $w_1x_1$ in $F$ and the redundancy edge for this intersection, $v_1w_1$, is deleted by the edge $w_2x_2$ in $F$.
\begin{description}
\item[Step 8:] If $w_2x_2$ intersects $v_1w_1$, then $w_2$ or $x_2$ are located inside one of the triangles formed by $u_1,v_1$ and $w_1$ ($\Delta u_1v_1w_1$ or $\Delta v_1u_1w_1$) (considered in Step 9) or $u_1v_1$ and $w_2x_2$ intersect \textbf{(Step 8a)}.
In case that $u_1v_1$ and $w_2x_2$ intersect, the edge $w_2x_2$ can be considered instead of the edge $w_1x_1$.
The special case of $w_2=u_1$ is considered in \textbf{Step 8b} and proves that no cycle of length 2 can occur  (the tests are described in \href{https://userpages.uni-koblenz.de/~boeltz/CP-Algorithm-Verification/Proof/}{\color{blue}z3-test\_proof\_part3\_step8a \color{black}} and \href{https://userpages.uni-koblenz.de/~boeltz/CP-Algorithm-Verification/Proof/}{\color{blue}z3-test\_proof\_part3\_step8b\color{black}}). 

{\bf Step 8a:} We prove that ${\sf AxGeom} \cup {\sf AxGraphs} \cup \{ {\bf R1, C1} \}$ entails: 
\begin{align*}
    &\textsf{deleting}(u_1,v_1,w_1,x_1) \land
    \textsf{F}(w_1,x_1) \land
    \textsf{deleting}(v_1,w_1,w_2,x_2)  \land
    \textsf{F}(w_2,x_2)\\ 
    \rightarrow& \textsf{inside}(u_1,v_1,w_1,w_2) \lor
    \textsf{inside}(v_1,u_1,w_1,w_2) \lor \\ &\textsf{inside}(u_1,v_1,w_1,x_2) \lor
    \textsf{inside}(v_1,u_1,w_1,x_2) \lor \\
    &\textsf{deleting}(u_1,v_1,w_2,x_2) \lor
    \textsf{deleting}(u_1,v_1,x_2,w_2)
\end{align*}

{\bf Step 8b:}  We prove that ${\sf AxGeom} \cup {\sf AxGraphs} \cup \{ {\bf R1, C1} \}$ entails: 

\smallskip 
$\textsf{deleting}(u_1,v_1,w_1,x_1) \land
    \textsf{F}(w_1,x_1)\land
    \textsf{deleting}(v_1,w_1,u_1,x_2)
    \land
    \textsf{F}(u_1,x_2)
    \rightarrow \bot$

\smallskip
\item[Step 9:] If $w_2$ is inside the triangle $\Delta u_1v_1w_1$ and the edge in F $w_3x_3$ deletes $v_1w_2$, then $w_3$ or $x_3$ are located inside the triangle $\Delta v_1w_1w_2$ or the edge $w_3x_3$ is intersecting with the edge $u_1v_1$ or $v_1w_1$ \textbf{(Step 9a)}.
If the edge $w_3x_3$ intersects with $u_1v_1$, then $w_3x_3$ can be chosen instead of $w_1x_1$.
If the edge $w_3x_3$ intersects with $v_1w_1$, then $w_3x_3$ can be chosen instead of $w_2x_2$.
The special case of $w_3=u_1$ is considered in \textbf{Step 9b} and proves that no cycle of length 3 can occur. 
The tests are described in 
\href{https://userpages.uni-koblenz.de/~boeltz/CP-Algorithm-Verification/Proof/}{\color{blue}z3-test\_proof\_part3\_step9a \color{black}} and \href{https://userpages.uni-koblenz.de/~boeltz/CP-Algorithm-Verification/Proof/}{\color{blue}z3-test\_proof\_part3\_step9b\color{black}}.

\smallskip
{\bf Step 9a:} We prove that ${\sf AxGeom} \cup {\sf AxGraphs} \cup \{ {\bf R1, C1} \}$ entails: 
\begin{align*}
    &\textsf{deleting}(u_1,v_1,w_1,x_1) \land
    \textsf{F}(w_1,x_1) \land
    \textsf{deleting}(v_1,w_1,w_2,x_2)
    \land \textsf{F}(w_2,x_2) \land\\
    & \textsf{inside}(u_1,v_1,w_1,w_2) \land
    \textsf{deleting}(v_1,w_2,w_3,x_3)
    \land
    \textsf{F}(w_3,x_3)\\
    \rightarrow& \textsf{inside}(u_1,v_1,w_1,x_1) \lor
    \textsf{inside}(v_1,w_1,w_2,w_3) \lor \textsf{inside}(v_1,w_1,w_2,x_3)
    \lor \\
    &\textsf{inside}(w_1,v_1,w_2,w_3) \lor \textsf{inside}(w_1,v_1,w_2,x_3) 
    \lor
    \textsf{deleting}(u_1,v_1,w_3,x_3)
    \lor\\
    &\textsf{deleting}(u_1,v_1,x_3,w_3)
    \lor
    \textsf{deleting}(v_1,w_1,w_3,x_3)
    \lor
    \textsf{deleting}(v_1,w_1,x_3,w_3)
\end{align*}

{\bf Step 9b:} We prove that ${\sf AxGeom} \cup {\sf AxGraphs} \cup \{ {\bf R1, C1} \}$ entails:
\begin{align*}
    &\textsf{deleting}(u_1,v_1,w_1,x_1) \land
    \textsf{F}(w_1,x_1) \land
    \textsf{deleting}(v_1,w_1,w_2,x_2) 
    \land 
    \textsf{F}(w_2,x_2) \land \\
    &\textsf{inside}(u_1,v_1,w_1,w_2) 
   \land
   \textsf{deleting}(v_1,w_2,u_1,x_3) 
    \land 
    \textsf{F}(u_1,x_3) 
    \rightarrow \bot
\end{align*}

\end{description}
\ignore{
The contradiction obtained by Step8a is given by: 
\begin{align*}
    \textsf{deleting}(u_1,v_1,w_1,x_1) \land
    \textsf{F}(w_1,x_1)& \land\\ 
    \textsf{deleting}(v_1,w_1,u_1,x_2) \land
    \textsf{F}(u_1,x_2)&
    \rightarrow \bot
\end{align*} 
}

\subsection{Proof of properties for the convex hull}
\label{sec:convexhull}

After proving Lemma \ref{lemma1} and \ref{lemma2a}, the following
Lemma is proven.
\begin{lemma}\label{lem:convexhull}
For an edge $u_1v_1$ and an intersecting edge $w_1x_1 \in F$, with intersection point $q$ closest to $u_1$ among all intersecting edges in $F$ there exists a path $P$ on the convex hull of the triangle $\Delta u_1qw_1$ with all edges in $F$ connecting $u_1$ and $w_1$ (see Fig.~\ref{fig:connectivity2}). 
\end{lemma}

The concept used for the proof of Lemma \ref{lemma3a} is the construction of paths on convex hulls.
We consider {\em the finest path} along the convex hull: 
If vertices $u, v, w$ are on the convex hull, $v$ is located on the segment $uw$ and we have edges $uv$ and $vw \in F$, then $u, v, w$ are all on the finest path of the convex hull.
We prove that a path on the convex hull of a triangle $\Delta u_1qw_1$ formed by an edge $u_1v_1$ and an intersecting edge $w_1x_1 \in F$ with intersection point $q$ closest to $u_1$ is not intersected by edges of $F$.
For this, we prove 
(cf. also \href{https://userpages.uni-koblenz.de/~boeltz/CP-Algorithm-Verification/Proof/}{\color{blue}z3-convexhull \color{black}} for the tests) 
that for an edge $u_1v_1$ and an edge $w_1x_1$ with intersection point $q$ closest to $u_1$, no edge $ab \in F$ can intersect with an edge of the convex hull in $\Delta u_1qw_1$.
From Lemma \ref{lemma1} it now follows that the edges of the convex hull are in $F$. 
Therefore there exists a path with all edges in $F$ that connects $u_1$ with $w_1$.
This path connects the components $X$ and $Y$ and therefore the following Theorem holds.

\begin{theorem}
The edges of F form a connected plane spanning subgraph of G after the termination of the
CP-algorithm.
\end{theorem}

\subsection{Counterexamples in the absence of the coexistence property}
\label{sec:counterexample}
We analyzed whether the proof still can be carried out if the
coexistence property is dropped, and showed the existence of an 
example of a geometric graph satisfying axiom ${\bf R1}$ but in 
which axiom {\bf C1} does not hold, for which the application of 
the CP-algorithm does not yield a connected, intersection free subgraph.
We also considered axiomatizations in which 
the coexistence property (formula {\bf C1}) was removed or was 
replaced with a weaker form, {\bf Cw}, in which only two of the coexistence 
edges are required to exist, and not all three (the variables $u, v,
w, x$ are universally quantified):

\medskip
\noindent $\begin{array}{ll} {\bf Cw}: & ({\sf E}(u,v) \wedge {\sf E}(v,w)
\wedge {\sf E}(w,u) \wedge {\sf V}(x) \wedge {\sf inside}(u,v,w,x) \\
& \rightarrow ({\sf E}(u,x) {\wedge} {\sf E}(v,x)) \vee ({\sf E}(u,x)
{\wedge} {\sf  E}(w,x)) \vee ({\sf E}(v,x)
{\wedge} {\sf E}(w,x))
\end{array}$

\medskip
\noindent A problem when trying to solve this task automatically, is that a
formula for the connectedness of the input graph is required; however,
this notion is not first-order definable. 
To solve this problem, we analyzed the existence of certain types of counterexamples.
We could consider, for instance, connectivity by paths of bounded length.
However, these axioms are in the $\forall \exists$-fragment, a fragment for which
the termination of Z3 is not guaranteed.

\smallskip 
\noindent In order to avoid this type of problems, we attempted 
\begin{itemize}
\item[(1)] to construct
counterexamples with an upper bound on the number of vertices, and on the
number of edges which connect them, 
and for which the disconnectedness of the resulting graph is simple to
express, 
and 
\item[(2)] to prove that the existence of certain redundancy cycles leading to
disconnectedness is consistent with the considered axiomatization.
\end{itemize}

\smallskip
\noindent {\bf (1) Generating counterexamples with given number of
  vertices.} We checked whether in the absence of the coexistence
property, or if the coexistence property is replaced with the weaker
version {\bf Cw} it is possible that after applying the CP-algorithm to 
a graph in which all vertices are connected
by paths of length at most 3, we can obtain a graph in which one
vertex is not connected to any other vertex. 
We proved that this is not possible for graphs with 2, 3, 4, 5, and 6
vertices, but it is possible for graphs with 7 vertices. Using Z3, we prove that
this is the case and can construct an example of a graph with this property. 
\begin{figure}[t]  
  \centering
    \begin{tikzpicture}[line cap=round,line join=round,x=1cm,y=1cm,scale=0.7]
\draw [line width=0.4pt] (0,0)-- (-4,-2);    
\draw [line width=0.4pt] (0,0)-- (4,-2); 
\draw [line width=0.4pt] (0,0)-- (0,3); 
\draw [line width=0.4pt] (0,3)-- (-4,-2);    
\draw [line width=0.4pt] (-4,-2)-- (4,-2); 
\draw [line width=0.4pt] (4,-2)-- (0,3); 
\draw [line width=0.4pt] (2,-0.8)-- (-4,-2);    
\draw [line width=0.4pt] (-0.4,1.5)-- (4,-2); 
\draw [line width=0.4pt] (-2,-1.2)-- (0,3);
\draw [line width=0.4pt] (2,-0.8)-- (4,-2);    
\draw [line width=0.4pt] (-0.4,1.5)-- (0,3); 
\draw [line width=0.4pt] (-2,-1.2)-- (-4,-2);
\draw [line width=0.4pt] (2,-0.8)-- (0,3);    
\draw [line width=0.4pt] (-0.4,1.5)-- (-4,-2); 
\draw [line width=0.4pt] (-2,-1.2)-- (4,-2);

\draw [fill=black] (0,0) circle (2.5pt);
\draw[color=black] (0.2,-0.2) node {$z$}; 
\draw [fill=black] (-4,-2) circle (2.5pt);
\draw[color=black] (-3.6,-2.2) node {$x_1$};
\draw [fill=black] (4,-2) circle (2.5pt);
\draw[color=black] (4.4,-2.2) node {$x_2$};
\draw [fill=black] (0,3) circle (2.5pt);
\draw[color=black] (0.4,2.8) node {$x_3$};
\draw [fill=black] (2,-0.8) circle (2.5pt);
\draw[color=black] (2.4,-1) node {$y_1$};
\draw [fill=black] (-0.4,1.5) circle (2.5pt);
\draw[color=black] (0,1.3) node {$y_2$};
\draw [fill=black] (-2,-1.2) circle (2.5pt);
\draw[color=black] (-1.6,-1.4) node {$y_3$};
\end{tikzpicture}

    \caption{Example of a graph with 7 vertices which satisfies {\bf
        R1} and {\bf Cw} and becomes disconnected after applying the CP-algorithm.}
    \label{fig:redundancy-cycle-7}
\end{figure}
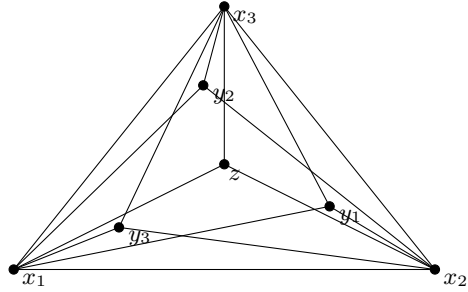
For instance, the graph in Figure~\ref{fig:redundancy-cycle-7} was constructed by Z3; 
it is a graph $G = (V, E)$ with 7 vertices, all connected by  
paths of length at most 3, and which satisfies {\bf R1} and the weaker  
form of the coexistence property {\bf Cw} with the property that after  
applying the CP-algorithm with input $G$ we obtain a disconnected  
graph. 
 Note that the tests which yielded the counterexample in 
Figure~\ref{fig:redundancy-cycle-7} 
were made under the assumption that Lemma~1 holds 
(i.e.\ if $G = (V, E)$ is the input graph and $G' = (V, F)$ is the
graph obtained after applying the CP-algorithm then 
every edge that is not in $F$ is intersected by an edge in $F$). 
Since Lemma~1 is not guaranted to hold in the absence of the 
coexistence property, other counterexamples can be obtained if 
we do not include a formulation of Lemma~1 in the premises.

We experimented with various types of axiomatizations: Since Z3 took 
rela\-tively long when considering a larger number of vertices if we used $\forall
\exists$ sentences, we also created versions of the tests in which 
the universal quantifiers were instantiated with the (finite number
of) vertices we considered. 
The tests are
described in \url{https://github.com/sofronie/tests-vmcai-2024.git},
in the folder {\sf Counterexamples/N-Vertices/}.

\

\noindent {\bf (2) Checking the existence of counterexamples containing
  redundancy cycles.} 
We considered also less complicated examples. 
We showed, for instance, that we can obtain a counterexample for the case of a redundancy cycle of 
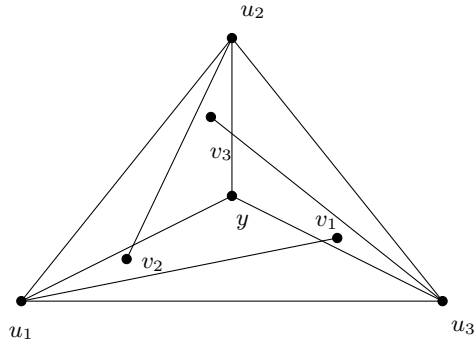
\begin{figure}[t]
    \centering
    \begin{tikzpicture}[line cap=round,line join=round,x=1cm,y=1cm,scale=0.7]
\draw [line width=0.4pt] (0,0)-- (-4,-2);    
\draw [line width=0.4pt] (0,0)-- (4,-2); 
\draw [line width=0.4pt] (0,0)-- (0,3); 
\draw [line width=0.4pt] (0,3)-- (-4,-2);    
\draw [line width=0.4pt] (-4,-2)-- (4,-2); 
\draw [line width=0.4pt] (4,-2)-- (0,3); 
\draw [line width=0.4pt] (2,-0.8)-- (-4,-2);    
\draw [line width=0.4pt] (-0.4,1.5)-- (4,-2); 
\draw [line width=0.4pt] (-2,-1.2)-- (0,3);
    
\draw [fill=black] (0,0) circle (2.5pt);
\draw[color=black] (0.2,-0.5) node {$y$}; 
\draw [fill=black] (-4,-2) circle (2.5pt);
\draw[color=black] (-4,-2.6) node {$u_1$};
\draw [fill=black] (4,-2) circle (2.5pt);
\draw[color=black] (4.4,-2.5) node {$u_3$};
\draw [fill=black] (0,3) circle (2.5pt);
\draw[color=black] (0.4,3.5) node {$u_2$};
\draw [fill=black] (2,-0.8) circle (2.5pt);
\draw[color=black] (1.8,-0.5) node {$v_1$};
\draw [fill=black] (-0.4,1.5) circle (2.5pt);
\draw[color=black] (-0.2,0.8) node {$v_3$};
\draw [fill=black] (-2,-1.2) circle (2.5pt);
\draw[color=black] (-1.5,-1.3) node {$v_2$};
\end{tikzpicture}

    \caption{Example of a redundancy cycle.}
    \label{fig:redundancy_cycle}
\end{figure}
length 3. 
The cycle can be constructed by using the predicate {\sf deleting}: 
\begin{itemize}
\item An edge $u_1y$ is deleted by an edge $u_2v_2$. 
\item The redundancy edge $u_2y$ exists, but $u_2y$ is deleted by
  $u_3v_3$. 
\item This leads to the existence of the redundancy edge $u_3y$. 
\item But $u_3y$ is deleted by $u_1v_1$.
\end{itemize}
This means that the edges $u_1y$, $u_2y$ and $u_3y$ form a redundancy
cycle and the graph containing the vertices $u_1$, $u_2$, $u_3$,
$v_1$, $v_2$, $v_3$ and $y$ is connected. We showed that the existence
of a redundancy cycle of length 3 is consistent with the axioms from
which the weak coexistence property was omitted. 
A similar test was made for a redundancy cycle of length 4.

Since the models for the redundancy cycles of length 3 and 4 using the axiomatization with the weak coexistence property, also hold for the axiomatization where the coexistence property is completly omitted, we only present the test with weak coexistence.

The tests are in the file 
\href{https://userpages.uni-koblenz.de/~boeltz/CP-Algorithm-Verification/Counterexamples/}{\color{blue}z3-test\_counterexample\_length3\color{black}}
(for a cycle of length 3) 
and in the file
{\color{blue}z3-test\_counterexample\_length4\color{black}} (for a
cycle of length 4),
cf. also \url{https://github.com/sofronie/tests-vmcai-2024.git} folder {\sf Counterexamples}.
%

\

\subsection{Tests: Details}

The tests \footnote{The series of tests can be found under 
{\scriptsize \url{https://github.com/sofronie/tests-vmcai-2024.git}}.}  were made using the SMT-solver Z3, first with Z3-4.4.1 on an
Intel(R) Core(TM) i7-3770 CPU @ 3.40GHz, 8192K cache, then
with Z3-4.4.1 and  Z3-4.8.15 on a  
12th Gen Intel(R) Core(TM) i5-12400 CPU @ max 5.6GHz, 18MiB cache.
Below we include some statistics with the tests made on the second
computer, which is considerably faster.
Since the two versions of Z3 yield somewhat different
counterexamples, when describing the tests for finding counterexamples
we include the times with both versions of Z3.

\

\noindent 
{\bf 1. Dependencies between axioms.} We identified axioms of the left predicate which can be derived from the other ones. 
    We proved that some axioms of the inside predicate follow from the
    others. A discussion can be found in
    Appendix~\ref{sec:Appendix}. The proofs for the dependencies of
    axioms are listed in the folder
    \href{https://userpages.uni-koblenz.de/~boeltz/CP-Algorithm-Verification/Axioms/}{\color{blue}Axioms\color{black}}
    and also under
    \url{https://github.com/sofronie/tests-vmcai-2024.git} in the
    directory {\sf CP-Algorithm-Verification/Axioms/}.

\

\noindent
{\bf 2. Proof of properties of geometric graphs.} We present statistics on the tests described in Section~\ref{geom-ar}: 
\begin{itemize}
\item Task {\sf pasch}: Proving that Pasch's axiom can be derived from axiom sets {\bf A}, {\bf I} and {\bf T}.  
\item Task {\sf clique}: Proving that the coexistence property implies the clique
  property for vertices in the interior of a triangle. Proving the clique property only requires the axioms of the sets {\bf A} and {\bf T}, but not the axioms for {\bf I}. 
\end{itemize}
In the table below, the axioms include {\bf A}, {\bf I} and {\bf T}
for task {\sf pasch} and the axioms  {\bf A} and {\bf T} for task {\sf
  clique}; the query consists of the clause form of the negation of 
the property to be proved.  

\begin{center}
\begin{tabular}{@{}||c |c |c |c |c||} 
 \hline
 ~Task~ &~Duration ~& ~\# constants~ &  ~\# Clauses ~& ~\# Clauses~\\ 
 & (s) & ~in the clauses~ & (axioms) & (query) \\
 \hline
 pasch & 0.49 & 5 & 22 & 12 \\ 
 \hline
 clique  & 0.34 & 5 & 14 & 9 \\
 \hline
 \end{tabular}
 \end{center}

 \
 
 \noindent
{\bf 3. Steps in the checking of the correctness proof.} We present statistics on the tests for checking steps 1-9 in the correctness proof described in Section~\ref{sec:verification_cp}; as well as the tests for the connectedness arguments using the convex hull in 
Section~\ref{sec:convexhull}.
In all these tests the axioms used are the same: there are 41 axioms, the size of the query depends on the proof tasks
considered. The query consists of the clause form of the negation of
the formula to be proven.
 
\begin{center}
\begin{tabular}{@{}||c | l | c | c | c | c||} 
 \hline
 Task &~~Step & Duration~ &~ \# constants~ &  \# Clauses & \# Clauses\\ 
 \phantom{{\bf Convex hull}} & & (s) & in the clauses & (axioms) & (query) \\  
 \hline
\hline 
{\bf Part 1}  & ~ Step 1  & 2.47 & 6 & 41 & 4\\ 
 \hline
 & ~ Step 2a  & 62.56 & 8 & 41 & 11 \\
 \hline
 & ~ Step 2b & 33.63 & 8 & 41 & 11\\
 \hline
%
 \hline
 {\bf Part 2} & 
 ~ Step 3 & 5.01 & 6 & 41 & 8\\
 \hline
 & ~ Step 4 & 5.28 & 6 & 41 & 11\\
 \hline
& ~ Step 4a & 0.80 & 5 & 41 & 4 \\
 \hline
& ~ Step 5a & 12.59 & 7 & 41 & 14 \\
 \hline
& ~ Step 5b & 9.18 & 7 & 41 & 14 \\
 \hline
& ~ Step 5c & 73.78 & 8 & 41 & 16 \\
 \hline
& ~ Step 5d & 13.20 & 7 & 41 & 13 \\
 \hline
& ~ Step 5e & 8.59 & 7 & 41 & 13\\
 \hline
& ~ Step 5f & 69.71 & 8 & 41 & 13\\
 \hline
& ~ Step 6a & 68.09 & 8 & 41 & 23\\
 \hline
& ~ Step 6b & 18.02 & 7 & 41 & 11\\
 \hline
& ~ Step 6c & 61.98 & 8 & 41 & 23\\
 \hline
& ~ Step 6d & 17.07 & 7 & 41 & 11\\
 \hline
& ~ Step 6e & 8.22 & 7 & 41 & 12\\
 \hline
& ~ Step 6g & 423.69 & 10 & 41 & 21\\
 \hline
& ~ Step 6h & 428.55 & 10 & 41 & 21\\
 \hline
& ~ Step 6i & 406.41 & 10 & 41 & 21 \\
 \hline
& ~ Step 6j & 428.77 & 10 & 41 & 22\\
 \hline
& ~ Step 7 & 2.83 & 6 & 41 & 4\\
 \hline
\hline 
 {\bf Part 3} & 
 ~ Step 8a & 2.72 & 6 & 41 & 10\\
 \hline
 & ~ Step 8b & 0.12 & 5 & 41 & 4\\
 \hline
 & ~ Step 9a & 18.40 & 8 & 41 & 16\\
 \hline
 & ~ Step 9b & 3.51 & 7 & 41 & 7\\
 \hline
\hline 
{\bf ~Convex hull~} & 
 ~ convexhull ~ & 30.71 & 8 & 53 & 15\\
 \hline
 \end{tabular}
\end{center}
 
\

\noindent
{\bf 4. Constructing counterexamples.} We proved that counterexamples
exist in the absence of the coexistence property
(cf. Sections~\ref{sec:verification_cp}
and~\ref{sec:counterexample}). 
\begin{itemize}
\item Task {\sf counterexample}: We show that if the coexistence
  property is removed, the implication of Step 1 in
  Section~\ref{sec:verification_cp} does
not hold anymore. For this we show that the counterexample in
Fig.~\ref{fig:part1_counterexample} is consistent with the
axiomatization from which the coexistence property is omitted
(this axiomatization contains 40 instead of 41 clauses). 
\item Tasks $N${\sf vertices\_noC}, $N \in \{ 2, 3, 4, 5, 6, 7 \}$: 
\begin{itemize}
\item Task $N${\sf vertices\_noC} ($N \in \{2, 3, 4, 5, 6 \}$): 
We show that one cannot construct a graph with $N$ vertices, in which 
every two vertices are  
connected by a path of length at most 3, which does not satisfy 
the weak coexistence property, and in the graph obtained from $G$ by applying 
the CP-algorithm one point is not connected to the other points. 
\item Task $7${\sf vertices\_noC}: 
We show that there exists a connected 
  graph $G$ with 7 vertices not satisfying the weak coexistence property, 
  such that in the graph obtained from $G$ by applying 
  the CP-algorithm one point is not connected to the other points (Z3 
  returns an example of such a graph). 
\end{itemize}
\item Tasks $N${\sf vertices\_weakC}, $N \in \{ 2, 3, 4, 5, 6, 7
  \}$: 
\begin{itemize}
\item Task $N${\sf vertices\_weakC} ($N \in \{ 2, 3, 4, 5, 6 \}$): 
We show that one cannot construct a graph with $N$ vertices, in which 
every two vertices are  
connected by a path of length at most 3, which satisfies 
the weak coexistence property, and in the graph obtained from $G$ by applying 
the CP-algorithm one point is not connected to the other points. 
\item Task $7${\sf vertices\_weakC}: 
We show that there exists a connected
  graph $G$ with 7 vertices satisfying the weak coexistence property, 
  such that in the graph obtained from $G$ by applying
  the CP-algorithm one point is not connected to the other points (Z3
  returns an example of such a graph). 
\end{itemize}
\item Tasks {\sf  cycle3} and {\sf cycle4}: We check if the existence
  of a redundancy
  cycle of length 3 resp.\ 4 is consistent with the axiomatization
  from which the coexistence property was removed and a weaker form of
  the coexistence property was added (therefore this axiomatization
  contains 41 clauses).
\end{itemize}
The counterexamples returned by Z3 are descriptions of input graphs 
$G = (V, E)$ with the property that after the execution of the
CP-algorithm the resulting graph $G' = (V, F)$ is
not connected. 

In what follows we present statistics on these tests.
Since the two versions of Z3 yield somewhat different
counterexamples, when describing the tests for finding counterexamples
we include the times with both versions of Z3.
%


\begin{center}
\begin{tabular}{@{}|| c |c |c |c |c|c|c||} 
 \hline
 Task &Duration & Duration & \#constants~ &  \#Clauses & \#Clauses & sat/\\ 
 & Z3-4.4.1 (s) & Z3-4.8.15 (s) & in the clauses & (axioms) & (query)
 & unsat \\
 \hline
\hline 
 counterexample & 2.58& 1.36 & 6 & 40 & 4 & sat\\
\hline 
\hline 
 2vertices\_noC~~ & 0.00 & 0.01 & 2 & 43 & 2 & unsat \\
 \hline
 3vertices\_noC~~ & 0.00 & 0.01 & 3 & 43 & 5 & unsat \\
  \hline
 4vertices\_noC~~ & 0.35 & 0.31 & 4 & 43 & 9 & unsat \\
 \hline
 5vertices\_noC~~ & 8.71 & 6.86 & 5 & 43 & 14 & unsat \\ 
 \hline
 6vertices\_noC~~ & 206.56 & 145.50 & 6 & 43 & 20 & unsat \\
 \hline
 7vertices\_noC~~ & 218.60 & 91.00 & 7 & 43 & 27& sat \\
\hline 
\hline 
~2vertices\_weakC & 0.00 & 0.01 & 2 & 44 & 2 & unsat \\
 \hline
 ~3vertices\_weakC & 0.00 & 0.01 & 3 & 44 & 5 & unsat \\
  \hline
 ~4vertices\_weakC & 0.41 & 0.37 & 4 & 44 & 9 & unsat \\
 \hline
 ~5vertices\_weakC & 8.78 & 7.83 & 5 & 44 & 14 & unsat \\ 
 \hline
 ~6vertices\_weakC & 187.35 & 145.62 & 6 & 44 & 20 & unsat \\
 \hline
 ~7vertices\_weakC & 220.02 & 226.23 & 7 & 44 & 27 & sat \\
\hline 
\hline
 cycle3 & 5.82& 13.24 & 7 & 41 & 6 & sat \\
 \hline
 cycle4 & 80.89& 61.42 & 9 & 41 & 8 & sat \\[1ex] 
 \hline
\end{tabular}
\end{center}

\ignore{\marginpar{LB: The rest of the subsection can be removed.} Under redundancy property intersecting edges can always be detected with 2-hop information from at least one of the two endvertices. Thus, with three hop information a distributed variant of the algorithm described before follows immediately.
However, according the previously described algorithm an edge $uv$ is kept if for all intersecting edges $wx$ always one of the vertices $u$ or $v$ is connected to $w$ and $x$. 
If $u$ can not detect any of the intersections $wx$ with edge $uv$ with 2-hop information, since neither $w$ nor $x$ are in the neighborhood of $u$, 
then $uv$ will keep that edge. However, if $u$ is not connected to $w$ and $x$, then the edges $vw$ and $vx$ have to exist. Therefore $uv$ satisfies the CP-condition for the intersection with $wx$ and $v$ will keep the edge $vu$ as well.

Thus, two hop information is sufficient to implement the algorithm in a distributed way as depicted by Alg.~\ref{alg:localedge}.
In order to avoid conflicting distributed decisions about edge removals, however, the algorithm requires an underlying distributed serialization of edge decisions. Such serialization can be based on a total ordering defined on the edges.}

\ignore{\section{Relation to Other Axiom Systems}\label{sec:axiom_hilbert_tarski}

\emph{Tarskis} \cite{TarskiG99} and \emph{Hilberts} \cite{hilbert} axiom systems for plane geometry use both a predicate "between", but formalize different notions of betweenness. 
Tarski's notion of betweenness is non-strict: a point $c$ is between $a$ and $b$ iff $c$ on the closed segment $[ab]$, while Hilbert's notion is strict: a point $c$ is between $a$ and $b$ iff $c$ on the open segment $(ab)$.
Therefore Hilbert's axiomatization is closer to ours than Tarski's.
Since we only consider line segments and not a whole (infinite) line the axioms of Hilberts group 1 (cf.\ \cite{hilbert}) are not important for us.
The related axioms can be found in group 2 (cf.\ \cite{hilbert}).

The relation "$c$ is between $a$ and $b$" can be expressed in our axiomatization by "$\textsf{left}(a,b,c) \land \textsf{left}(b,a,c)$". 
We consider here "strict" betweenness, since $\textsf{left}(b,a,b)$ and $\textsf{left}(a,b,a)$ are always false.
With this definition, \textbf{Hilberts axiom 2.1} can be derived from our axioms axiomatisation, since $c$ has to be left of both $ab$ and $ba$.
Since we only consider a finite set of points, \textbf{Hilberts axiom 2.2} is not needed in our axiomatisation.
\textbf{Hilberts axiom 2.3} is a direct consequence of our \textbf{axiom A1}, since $\lnot \textsf{left(c,a,b)}$ implies that $b$ is not located between $c$ and $a$ and $\lnot \textsf{left(c,b,a)}$ implies that a is not located between $c$ and $b$.
\textbf{Hilberts Axiom 2.4}, which is also known as Paschs axiom can be proven with our axiomatisation with a short test (left predicate test pasch.smt).}

\ignore{
\noindent {\bf Embedding in the Euclidean Plane.}
The axioms of \textbf{left} are defined in a way that a concrete embedding in the Euclidean plane can be obtained.
\marginpar{VSS: Did you prove this?}
\marginpar{What is an assignment of the predicates?}
For each model of the predicates infinitely many embeddings in the Euclidean plane can be found.
This is due to the invariance of the predicates models against translation, scaling and rotation.
Therefore the computational effort to find such an embedding can be quite high.

The following inequality has to hold if $\textsf{left}(u,v,w)$ is true\\
$(X(v)-X(u))\cdot(Y(w)-Y(u))  \geq (Y(v)-Y(u))\cdot(X(w)-X(u)) \\
(X(v)-X(u))\cdot(Y(w)-Y(u))  >  (Y(v)-Y(u))\cdot(X(w)-X(u)) \lor\\ (X(v)-X(u))\cdot(X(w)-X(u)) > 0 \lor (Y(v)-Y(u))\cdot(Y(w)-Y(u)) > 0$
}
\section{Conclusion}\label{sec:conclusion}
We defined an axiom system for geometric graphs, and extended it with possibilities of reasoning about paths along certain convex hulls.
We used this axiom system to check the correctness of two versions of the CP-algorithm (a global and a local version)
for graphs satisfying the redundancy and coexistence property.

Another verification possibility we would like to investigate is proving that (with the notation in Alg.~1) the connectedness of $W \cup F$ is an inductive invariant. However, the form of the redundancy property changes when edges are removed. We would like to investigate if ideas from \cite{PeuterSofronie-cade19} could be used in this context for invariant strengthening. 
Also as future work, we plan to use our axiom system for the verification of other algorithms and transformations in geometric graphs and extend it to a version that covers also the 3-dimensional space. 

We would also like to investigate the existence of possible links
between (1) conditions under which local algorithms in graph theory can be proved
correct, (2) conditions guaranteeing the correctness of methods for local reasoning
for global graph properties as studied e.g.\ in \cite{Wies2020}, and
(3) methods for verification of safety property for families of interacting
similar systems, for which locality properties of the associated
theories could be established
and used to prove small model properties
(cf.\ e.g.\ \cite{ifm2010,DammHorbachSofronie2015}). 

\medskip
\noindent {\bf Acknowledgments.} We thank the VMCAI 2024 reviewers for their helpful comments.
\bibliographystyle{splncs04}
\bibliography{references}

\section{Appendix}
\subsection{Axioms for left}
\label{sec:Appendix}

{\small 
\textbf{A1}: ($\textsf{left}(u,v,w) \land \textsf{left}(v,u,w) \rightarrow \lnot \textsf{left}(w,u,v)$)\\
\textbf{A2}: ($u \neq w \land v \neq w \land \lnot \textsf{left}(w,u,v) \land \lnot \textsf{left}(w,v,u) \rightarrow \textsf{left}(u,v,w)$)\\
\textbf{A3}: ($\textsf{left}(u,v,w) \land \textsf{left}(v,u,w) \land \textsf{left}(u,x,w) \rightarrow  \textsf{left}(u,x,v)$)\\
\textbf{A4}: ($\textsf{left}(u,v,w) \land \textsf{left}(v,u,w) \land \textsf{left}(u,x,v) \rightarrow  \textsf{left}(u,x,w)$)\\
\textbf{A5}: ($\textsf{left}(u,v,w) \land \textsf{left}(v,w,u) \land \textsf{left}(w,u,v) \rightarrow  \textsf{left}(u,v,x) \lor \textsf{left}(v,w,x) \lor \textsf{left}(w,u,x)$)\\[1ex]
(for 5 different points {\bf A5} is implied by {\bf A1-4} and {\bf A6})\\[1ex]
\textbf{Axiom for a triangle $\Delta uvw$ left of $xy$ and $z$ inside $\Delta uvw$}\\[1ex]
\textbf{A6}: $ \textsf{left}(u,v,z) \land \textsf{left}(v,w,z) \land \textsf{left}(w,u,z) \land  \textsf{left}(x,y,u)\land \textsf{left}(x,y,v)\land \textsf{left}(x,y,w) \\
~~~~~~~~~~~ \rightarrow \textsf{left}(x,y,z)$}\\
\\
\textbf{Axioms implied by A1-3}\\
{\small \textsf{B1}: $\lnot \textsf{left}(u,u,v)$\\
\textsf{B2}: $\lnot \textsf{left}(u,v,u)$\\
\textsf{B3}: ($u \neq v \rightarrow \textsf{left}(u,v,v)$)\\[1ex]
\textsf{B11}: ($u \neq v \rightarrow \textsf{left}(u,v,w) \lor \textsf{left}(v,u,w)$)\\
\textsf{B12}: ($u \neq w \rightarrow \textsf{left}(u,v,w) \lor \textsf{left}(w,v,u)$)\\[1ex]
\textsf{B21}: ($\textsf{left}(u,v,w) \land \textsf{left}(v,u,w) \rightarrow \textsf{left}(u,w,v)$)\\
\textsf{B22}: ($\textsf{left}(u,v,w) \land \textsf{left}(v,u,w) \rightarrow \textsf{left}(v,w,u)$)\\
\textsf{B23}: ($\textsf{left}(u,v,w) \land \textsf{left}(v,u,w) \rightarrow \lnot \textsf{left}(w,u,v)$)\\
\textsf{B24}: ($\textsf{left}(u,v,w) \land \textsf{left}(v,u,w) \rightarrow \lnot \textsf{left}(w,v,u)$)\\[1ex]
\textsf{B31}: ($\textsf{left}(u,v,w) \land \textsf{left}(w,v,u) \rightarrow \textsf{left}(u,w,v)$)\\
\textsf{B32}: ($\textsf{left}(u,v,w) \land \textsf{left}(w,v,u) \rightarrow \textsf{left}(w,u,v)$)\\
\textsf{B33}: ($\textsf{left}(u,v,w) \land \textsf{left}(w,v,u) \rightarrow \lnot \textsf{left}(v,u,w)$)\\
\textsf{B34}: ($\textsf{left}(u,v,w) \land \textsf{left}(w,v,u) \rightarrow \lnot \textsf{left}(v,w,u)$)}\\[1ex]
\textbf{Axioms implied by A1-5}\\
{\small \textsf{B41}: ($\textsf{left}(u,v,w) \land \textsf{left}(u,w,v) \rightarrow \textsf{left}(v,u,w) \lor \textsf{left}(w,v,u) \lor v=w$)\\
\textsf{B42}: ($\textsf{left}(u,v,w) \land \textsf{left}(u,w,v) \rightarrow \textsf{left}(v,w,u) \lor \textsf{left}(w,u,v) \lor v=w$)\\[1ex]
\textsf{B51}: ($\textsf{left}(u,v,w) \land \textsf{left}(w,u,v) \rightarrow \lnot \textsf{left}(v,u,w)$)\\
\textsf{B52}: ($\textsf{left}(u,v,w) \land \textsf{left}(u,w,v) \rightarrow \textsf{left}(u,w,v) \lor \textsf{left}(v,w,u)$)\\
\textsf{B53}: ($\textsf{left}(u,v,w) \land \textsf{left}(u,w,v) \rightarrow \textsf{left}(w,v,u) \lor \textsf{left}(v,w,u)$)\\[1ex]
\textsf{B61}: ($\textsf{left}(u,v,w) \land \textsf{left}(v,w,u) \rightarrow \lnot \textsf{left}(w,v,u)$)\\
\textsf{B62}: ($\textsf{left}(u,v,w) \land \textsf{left}(v,w,u) \rightarrow \textsf{left}(v,u,w) \lor \textsf{left}(w,u,v)$)\\
\textsf{B63}: ($\textsf{left}(u,v,w) \land \textsf{left}(v,w,u) \rightarrow \textsf{left}(u,w,v) \lor \textsf{left}(w,u,v)$)}\\[1ex]
\textbf{Axioms implied by A1-6}\\
{\small \textsf{B71}: ($\textsf{left}(u,v,w) \land \textsf{left}(v,u,w) \land \textsf{left}(w,u,x) \rightarrow  \textsf{left}(v,u,x)$)\\
\textsf{B72}: ($\textsf{left}(u,v,w) \land \textsf{left}(v,u,w) \land \textsf{left}(w,x,u) \rightarrow  \textsf{left}(v,x,u)$)\\
\textsf{B73}: ($\textsf{left}(u,v,w) \land \textsf{left}(v,u,w) \land \textsf{left}(x,v,u) \rightarrow  \textsf{left}(x,v,w)$)\\
\textsf{B74}: ($\textsf{left}(u,v,w) \land \textsf{left}(v,u,w) \land \textsf{left}(x,v,u) \rightarrow  \textsf{left}(x,w,u)$)\\
\textsf{B75}: ($\textsf{left}(u,v,w) \land \textsf{left}(v,u,w) \land \textsf{left}(u,w,x) \rightarrow  \textsf{left}(u,v,x)$)\\
\textsf{B76}: ($\textsf{left}(u,v,w) \land \textsf{left}(v,u,w) \land \textsf{left}(u,v,x) \rightarrow  \textsf{left}(u,w,x)$)\\
\textsf{B77}: ($\textsf{left}(u,v,w) \land \textsf{left}(v,u,w) \land \textsf{left}(w,u,x) \rightarrow \lnot \textsf{left}(w,v,x)$)\\
\textsf{B78}: ($\textsf{left}(u,v,w) \land \textsf{left}(v,u,w) \land \textsf{left}(w,x,u) \rightarrow \lnot \textsf{left}(w,x,v)$)}

\subsection{Axioms for intersection}
\textbf{intersection $\rightarrow$}\\[1ex]
{\small \textbf{I1-2}: $\textsf{intersection}(u,v,w,x)  \rightarrow (u \neq w \lor v \neq x) \land (u \neq x \lor v \neq w)$
\\
\textbf{I3}: $\textsf{intersection}(u,v,w,x)  \rightarrow \textsf{left}(u,v,w) \lor \textsf{left}(u,v,x)$
\\
\textbf{I4}: $\textsf{intersection}(u,v,w,x)  \rightarrow \textsf{left}(v,u,w) \lor \textsf{left}(v,u,x)$
\\
\textbf{I5}: $\textsf{intersection}(u,v,w,x)  \rightarrow \textsf{left}(w,x,u) \lor
\textsf{left}(w,x,v)$
\\
\textbf{I6}: $\textsf{intersection}(u,v,w,x)  \rightarrow \textsf{left}(x,w,u) \lor \textsf{left}(x,w,v)$
\\[1ex]
\textbf{intersection $\leftarrow$}\\[1ex]
\textbf{I11-12}: $\textsf{left}(u,v,w) \land \textsf{left}(v,u,x) \land \textsf{left}(w,x,u) \land \textsf{left}(x,w,v) \\
~~~~~~~~~~~~~~~\rightarrow \textsf{intersection}(u,v,w,x) \lor (u=x \land v=w)$
\\
\textbf{I13}: $\textsf{left}(u,v,w) \land \textsf{left}(v,u,x) \land \textsf{left}(w,x,v) \land \textsf{left}(x,w,u) \rightarrow \textsf{intersection}(u,v,w,x)$
\\
\textbf{I14}: $\textsf{left}(u,v,x) \land \textsf{left}(v,u,w) \land \textsf{left}(w,x,u) \land \textsf{left}(x,w,v) \rightarrow \textsf{intersection}(u,v,w,x)$
\\
\textbf{I15-16}: $\textsf{left}(u,v,x) \land \textsf{left}(v,u,w) \land \textsf{left}(w,x,v) \land \textsf{left}(x,w,u) \\
~~~~~~~~~~~~~~~\rightarrow \textsf{intersection}(u,v,w,x) \lor (u=w \land v=x)$ 

\subsection{Axioms for inside}
\label{sec:ax-inside}
\textbf{Axioms defining inside}\\[1ex]
\textbf{T1-T3}: $\textsf{inside}(u,v,w,x)  \rightarrow \textsf{left}(u,v,x)\land \textsf{left}(v,w,x) \land  \textsf{left}(w,u,x)$
\\[1ex]
\textbf{inside $\rightarrow$}\\[1ex]
\textbf{T4}: $ \textsf{left}(u,v,x)\land \textsf{left}(v,w,x) \land  \textsf{left}(w,u,x) \rightarrow \textsf{inside}(u,v,w,x)$
\\
{\bf T5}: $ \textsf{left}(u,v,x)\land \textsf{left}(v,w,x) \land  \textsf{left}(w,u,x) \rightarrow \textsf{inside}(v,w,u,x)$
\\
{\bf T6}: $ \textsf{left}(u,v,x)\land \textsf{left}(v,w,x) \land  \textsf{left}(w,u,x) \rightarrow \textsf{inside}(w,u,v,x)$
\\[1ex]
\textbf{transitivity of inside derived by the other axioms}\\[1ex]
T21-23: $(\textsf{inside}(u,v,x,y) \lor \textsf{inside}(v,w,x,y) \lor \textsf{inside}(w,u,x,y)) \land \textsf{intersection}(u,v,w,x)\\ 
~~~~~~~~~~~~~~~\rightarrow \textsf{inside}(u,v,w,y)$
\\
T24: $\textsf{inside}(u,v,w,y) \land \textsf{intersection}(u,v,w,x) \land x \neq y \\
~~~~~~~~~~~~~~~\rightarrow  (\textsf{inside}(u,v,x,y) \lor \textsf{inside}(v,w,x,y) \lor \textsf{inside}(w,u,x,y))$
}

\subsection{Axioms for edges $E$ in the input graph of the algorithm}

\textbf{E1}: $\lnot \textsf{E}(u,u)$ (no self-loops)
\\
\textbf{E2}: $\textsf{E}(u,v) \rightarrow \textsf{E}(v,u)$ (symmetry of edges)
\\
\textbf{E3}: $\textsf{E}(u,v) \rightarrow \textsf{V}(u)$
(symmetry implies also that $v$ is a vertex)\\ 
\\
\textbf{R1}: $\textsf{E}(u,v) \land \textsf{E}(w,x) \land \textsf{intersection}(u,v,w,x) \rightarrow (\textsf{E}(u,w) \lor \textsf{E}(v,x))$\\
R2: $\textsf{E}(u,v) \land \textsf{E}(w,x) \land \textsf{intersection}(u,v,w,x) \rightarrow (\textsf{E}(u,x) \lor \textsf{E}(v,w))$\\
(symmetry implies R2)\\ 
\textbf{C1}: $\textsf{intersection}(u,v,w,x) \land \textsf{E}(u,v) \land \textsf{E}(v,w) \land \textsf{E}(w,u) \land \textsf{V}(x) \rightarrow \textsf{E}(u,x)$\\
C2: $\textsf{intersection}(u,v,w,x) \land \textsf{E}(u,v) \land \textsf{E}(v,w) \land \textsf{E}(w,u) \land \textsf{V}(x) \rightarrow \textsf{E}(v,x)$\\
C3: $\textsf{intersection}(u,v,w,x) \land \textsf{E}(u,v) \land \textsf{E}(v,w) \land \textsf{E}(w,u) \land \textsf{V}(x) \rightarrow \textsf{E}(w,x)$\\
(symmetry implies C2 and C3) 

\vspace{-2mm}
\subsection{Axioms for edges $F$ in the output graph of the algorithm}
\textbf{Axioms for the definition of $F$}\\[1ex]
\textbf{F1}: $\textsf{F}(u,v) \rightarrow \textsf{F}(v,u)$
\\
\textbf{F2}: $\textsf{F}(u,v) \rightarrow \textsf{E}(u,v)$
\\
\textbf{F3}: $\textsf{F}(u,v) \land \textsf{F}(w,x) \rightarrow \lnot \textsf{intersection}(u,v,w,x)$\\
\textbf{F4}: $\textsf{left}(u,v,w) \land \textsf{left}(v,u,w) \land \textsf{V}(w) \rightarrow \lnot \textsf{F}(u,v)$ \\(no vertex can be located on an edge in $F$)\\
\textbf{F5}: $\textsf{F}(u,v) \land \textsf{E}(w,x) \land \textsf{intersection}(u,v,w,x) \rightarrow \textsf{E}(u,w) \lor \textsf{E}(v,w)$(CP-condition)
\\
F6: $\textsf{F}(u,v) \land \textsf{E}(w,x) \land \textsf{intersection}(u,v,w,x) \rightarrow \textsf{E}(u,x) \lor \textsf{E}(v,x)$(CP-condition)
\\
(F5 and F6 are symmetric)\\[1ex]
\textbf{deleting $\rightarrow$}\\[1ex]
\textbf{D1-3}: $\textsf{deleting}(u,v,w,x) \rightarrow \textsf{E}(u,v) \land \textsf{E}(w,x) \land \textsf{intersection}(u,v,w,x)$
\\
\textbf{D4-5}: $\textsf{deleting}(u,v,w,x) \rightarrow  \textsf{E}(u,w) \land \textsf{E}(v,w)$
\\
\textsf{D6}: $\textsf{deleting}(u,v,w,x) \rightarrow \lnot \textsf{F}(u,v)$
\\[1ex]
\textbf{$E$,deleting $\rightarrow$}\\[1ex]
\textbf{D11}: $\textsf{E}(u,x) \land \textsf{deleting}(u,v,w,x) \rightarrow \textsf{F}(w,x)$
\\
\textbf{D12}: $\textsf{E}(v,x) \land \textsf{deleting}(u,v,w,x) \rightarrow \textsf{F}(w,x)$
\\[1ex]
\textbf{ deleting $\leftarrow$}\\[1ex]
\textbf{D13}: $\textsf{E}(u,v) \land \textsf{E}(w,x) \land \textsf{intersection}(u,v,w,x) \\
~~~~~~~~~~~~~~~~\rightarrow \textsf{deleting}(u,v,w,x) \lor \textsf{E}(u,x) \lor \textsf{E}(v,x)$
\\
\textbf{D14}: $\textsf{E}(u,v) \land \textsf{F}(w,x) \land \textsf{intersection}(u,v,w,x) \land \textsf{E}(u,w) \land \textsf{E}(v,w)\rightarrow \textsf{deleting}(u,v,w,x)$
\\
\textbf{implications derived by the other axioms}\\
D23: $\textsf{E}(u,v) \land \textsf{E}(w,x) \land \textsf{intersection}(u,v,w,x) \\
~~~~~~~~~~~~~~~~\rightarrow \textsf{deleting}(u,v,x,w) \lor \textsf{E}(u,w) \lor \textsf{E}(v,w)$
\\
(equivalent to D13 by replacing $w$ and $x$)\\
D24: $\textsf{E}(u,v) \land \textsf{F}(w,x) \land \textsf{intersection}(u,v,w,x) \land \textsf{E}(u,x) \land \textsf{E}(v,x)\rightarrow \textsf{deleting}(u,v,x,w)$\\
(equivalent to D14 by replacing $w$ and $x$)

\vspace{-2mm}
\subsection{Axioms for the convex hull of the vertices in $\Delta u_1p_1w_1$}

Below, $u_1, v_1, w_1$ and $x_1$ are considered to be Skolem constants. $p_1$ is the intersection of the segments $u_1 v_1$ and $w_1 x_1$.\\[1ex]
\noindent 1) The segments $u_1 v_1$ and $w_1 x_1$ intersect \\[1ex]
\textbf{W1}: $\textsf{left}(u_1,v_1,w_1)$\\
\textbf{W2}: $\textsf{left}(v_1,u_1,x_1)$\\
\textbf{W3}: $\textsf{left}(x_1,w_1,u_1)$\\
\textbf{W4}: $\textsf{left}(w_1,x_1,v_1)$\\[1ex]
\noindent 2) There is an edge from $u_1$ to $v_1$ and an edge in $F$ from $w_1$ to $x_1$: 
\\[1ex]
\textbf{W5}: $\textsf{E}(u_1,v_1)$\\
\textbf{W6}: $\textsf{F}(w_1,x_1)$\\[1ex]
\noindent 3) The edge $w_1x_1$ is the edge in $F$ with intersection point $p_1$ closest to $u_1$.\\[2ex]
\textbf{X1}: $\forall y, z~ (\textsf{F}(y,z) \land \textsf{intersection}(u_1,v_1,y,z) \rightarrow \lnot \textsf{intersection}(u_1,w_1,y,z))$\\
\textbf{X2}: $\forall y, z~ ( \textsf{F}(y,z) \land \textsf{intersection}(u_1,v_1,y,z) \rightarrow \lnot \textsf{intersection}(u_1,x_1,y,z))$\\
\textbf{X3}: $\forall y, z~ ( \textsf{F}(y,z) \land \textsf{intersection}(u_1,v_1,y,z) \rightarrow \lnot \textsf{inside}(u_1,x_1,w_1,y))$\\
\textbf{X4}: $\forall y, z~ (\textsf{F}(y,z) \land \textsf{intersection}(u_1,v_1,y,z) \rightarrow \lnot \textsf{inside}(u_1,x_1,w_1,z))$\\
\\
{\bf Describing the properties of the convex hull.} The vertices $i$ and ${\sf next}(i)$ are vertices on the finest path of the convex hull of the triangle $u_1p_1w_1$.
The finest path is a path, such that there is no other vertex between $i$ and its successor ${\sf next}(i)$.
Every vertex $y$ different from $i$ and ${\sf next}(i)$ in the area left of $u_1v_1$ and left of $x_1w_1$ is left of $i$ and ${\sf next}(i)$ or $i$ is located between $y$ and ${\sf next}(i)$\\[2ex]
\textbf{Y0}: $\textsf{next}(w_1)=\textsf{nil}$
\\
\textbf{Y1}: $\forall y, i ~(\textsf{V}(y) \land \textsf{left}(u_1,v_1,y) \land \textsf{left}(x_1,w_1,y) \rightarrow  i=\textsf{nil} \lor \textsf{next}(i)=\textsf{nil} \lor  \\
~~~~~~~~~~~~~~~~y=i \lor  y=\textsf{next}(i) \lor \textsf{left}(i,\textsf{next}(i),y) \lor \textsf{left}(y,\textsf{next}(i),i))$\\
\textbf{Y2}: $\forall y, i ~ (\textsf{V}(y) \land \textsf{left}(u_1,v_1,y) \land \textsf{left}(x_1,w_1,y) \rightarrow  i=\textsf{nil} \lor \textsf{next}(i)=\textsf{nil} \lor \\
~~~~~~~~~~~~~~~~y=i \lor y=\textsf{next}(i) \lor \textsf{left}(i,\textsf{next}(i),y) \lor \textsf{left}(\textsf{next}(i),y,i))$\\
\textbf{Y3}: $\forall y, i ~ (\textsf{V}(y) \land \textsf{left}(u_1,v_1,y) \land \textsf{left}(x_1,w_1,y) \land \textsf{left}(\textsf{next}(i),i,y) \rightarrow \\
~~~~~~~~~~~~~~~~i=\textsf{nil} \lor \textsf{next}(i)=\textsf{nil} \lor y=i \lor y=\textsf{next}(i) \lor \textsf{left}(y,\textsf{next}(i),i))$\\
\textbf{Y4}: $\forall y, i ~ (\textsf{V}(y) \land \textsf{left}(u_1,v_1,y) \land \textsf{left}(x_1,w_1,y) \land \textsf{left}(\textsf{next}(i),i,y) \rightarrow  \\
~~~~~~~~~~~~~~~~i=\textsf{nil} \lor \textsf{next}(i)=\textsf{nil}  \lor y=i \lor y=\textsf{next}(i) \lor \textsf{left}(\textsf{next}(i),y,i))$\\
\\
The vertex ${\sf next}(i)$ is located inside the triangle $\Delta u_1v_1w_1$ or ${\sf next}(i)=w_1$. From the other axioms it follows that ${\sf next}(i)$ is also inside the triangle $\Delta u_1x_1w_1$ and therefore inside the triangle $u_1p_1w_1$ or equal to $w_1$.\\[2ex]
\textbf{Y11}: $\forall i ~(i=\textsf{nil} \lor \textsf{next}(i)=\textsf{nil} \lor \textsf{inside}(u_1,v_1,w_1,\textsf{next}(i)) \lor \textsf{next}(i)=w_1)$\\[2ex]
The vertex $(i)$ is located inside the triangle $\Delta u_1x_1w_1$ or $i=u_1$. From the other axioms it follows that $i$ is also inside the triangle $\Delta u_1v_1w_1$ and therefore inside the triangle $u_1p_1w_1$ or equal to $v_1$.\\
\textbf{Y12}: $\forall i ~(i=\textsf{nil} \lor \textsf{next}(i)=\textsf{nil} \lor \textsf{inside}(u_1,x_1,w_1,i) \lor i=u_1)$\\
There is an edge between the vertices of the convex hull.\\
\textbf{Y13}: $\forall i ~(\textsf{E}(i,\textsf{next}(i)) \lor i=\textsf{nil} \lor \textsf{next}(i)=\textsf{nil})$\\
\\
The vertex $i$ is located in the triangles $u_1v_1{\sf next}(i)$ and $u_1x_1{\sf next}(i)$ or equal to $u_1$\\
\textbf{Y14}: $\forall i ~(i=\textsf{nil} \lor \textsf{next}(i)=\textsf{nil} \lor \textsf{inside}(u_1,v_1,\textsf{next}(i),i) \lor i=u_1)$\\
\textbf{Y15}: $\forall i ~(i=\textsf{nil} \lor \textsf{next}(i)=\textsf{nil} \lor \textsf{inside}(u_1,x_1,\textsf{next}(i),i) \lor i=u_1)$\\
The vertex $\textsf{next}(i)$ is located in the triangles $\Delta v_1w_1i$ and $\Delta x_1
w_1i$ or equal to $w_1$\\
\textbf{Y16}: $\forall i ~(i=\textsf{nil} \lor \textsf{next}(i)=\textsf{nil} \lor \textsf{inside}(v_1,w_1,i,\textsf{next}(i)) \lor \textsf{next}(i)=w_1)$\\
\textbf{Y17}: $\forall i ~(i=\textsf{nil} \lor \textsf{next}(i)=\textsf{nil} \lor \textsf{inside}(x_1,w_1,i,\textsf{next}(i)) \lor \textsf{next}(i)=w_1)$

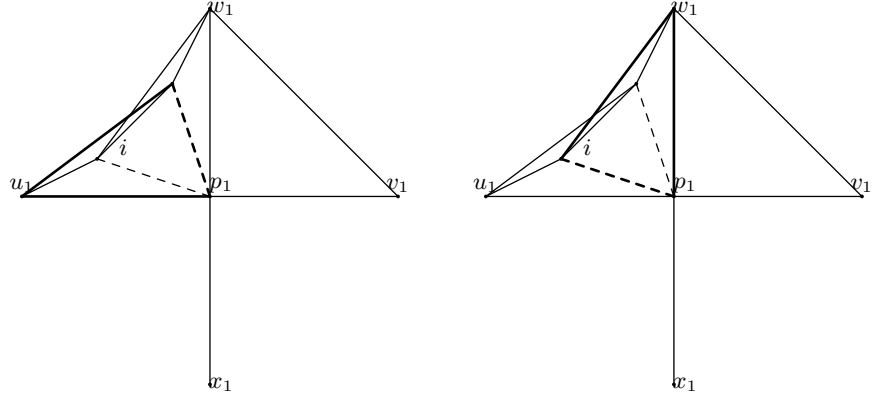
\begin{figure*}[h!]
    \centering
    \begin{subfigure}[t]{0.47\textwidth}
        \centering
\begin{tikzpicture}[line cap=round,line join=round,x=1cm,y=1cm,scale=0.5]
\draw [line width=0.5pt,color=black] (-5,0)-- (-5,0);
\draw [line width=1pt,color=black] (-5,0)-- (0,0);
\draw [line width=0.5pt,color=black] (0,0)-- (5,0);
\draw [line width=0.5pt,color=black] (5,0)-- (0,5);
\draw [line width=0.5pt,color=black] (0,5)-- (0,0);
\draw [line width=0.5pt,color=black] (0,-5)-- (0,0);
\draw [line width=0.5pt,color=black] (-5,0)-- (-3,1);
\draw [line width=0.5pt,color=black] (-3,1)-- (-1,3);
\draw [line width=0.5pt,color=black] (-1,3)-- (0,5);
\draw [line width=0.5pt,color=black] (-3,1)-- (0,5);
\draw [line width=1pt,color=black] (-1,3)-- (-5,0);
\draw [line width=0.5pt,color=black, dashed] (-3,1)-- (0,0);
\draw [line width=1pt,color=black, dashed] (-1,3)-- (0,0);
\draw [fill=black] (-5,0) circle (1pt);
\draw[color=black] (-5,0.3) node {$u_1$};
\draw [fill=black] (5,0) circle (1pt);
\draw[color=black] (5,0.3) node {$v_1$};
\draw [fill=black] (0,5) circle (1pt);
\draw[color=black] (0.3,5) node {$w_1$};
\draw [fill=black] (0,-5) circle (1pt);
\draw[color=black] (0.3,-5) node {$x_1$};
\draw [fill=black] (0,0) circle (1pt);
\draw[color=black] (0.3,0.3) node {$p_1$};
\draw [fill=black] (-3,1) circle (1pt);
\draw[color=black] (-2.3,1.3) node {$i$};
\draw [fill=black] (-1,3) circle (1pt);
\end{tikzpicture}

        \caption{The upper vertex of the highlighted triangle is $\textsf{next}(i)$. $i$ is located inside the triangle $\Delta u_1p_1\textsf{next}(i)$}\label{fig:convexhull_triangles}
    \end{subfigure}%
    ~~~ 
    \begin{subfigure}[t]{0.47\textwidth}
        \centering
\begin{tikzpicture}[line cap=round,line join=round,x=1cm,y=1cm,scale=0.5]
\draw [line width=0.5pt,color=black] (-5,0)-- (-5,0);
\draw [line width=0.5pt,color=black] (-5,0)-- (0,0);
\draw [line width=0.5pt,color=black] (0,0)-- (5,0);
\draw [line width=0.5pt,color=black] (5,0)-- (0,5);
\draw [line width=1pt,color=black] (0,5)-- (0,0);
\draw [line width=0.5pt,color=black] (0,-5)-- (0,0);
\draw [line width=0.5pt,color=black] (-5,0)-- (-3,1);
\draw [line width=0.5pt,color=black] (-3,1)-- (-1,3);
\draw [line width=0.5pt,color=black] (-1,3)-- (0,5);
\draw [line width=1pt,color=black] (-3,1)-- (0,5);
\draw [line width=0.5pt,color=black] (-1,3)-- (-5,0);
\draw [line width=1pt,color=black, dashed] (-3,1)-- (0,0);
\draw [line width=0.5pt,color=black, dashed] (-1,3)-- (0,0);
\draw [fill=black] (-5,0) circle (1pt);
\draw[color=black] (-5,0.3) node {$u_1$};
\draw [fill=black] (5,0) circle (1pt);
\draw[color=black] (5,0.3) node {$v_1$};
\draw [fill=black] (0,5) circle (1pt);
\draw[color=black] (0.3,5) node {$w_1$};
\draw [fill=black] (0,-5) circle (1pt);
\draw[color=black] (0.3,-5) node {$x_1$};
\draw [fill=black] (0,0) circle (1pt);
\draw[color=black] (0.3,0.3) node {$p_1$};
\draw [fill=black] (-3,1) circle (1pt);
\draw[color=black] (-2.3,1.3) node {$i$};
\draw [fill=black] (-1,3) circle (1pt);
\end{tikzpicture}

        \caption{$\textsf{next}(i)$ is located inside the triangle $\Delta p_1w_1i$}\label{fig:convexhull_Y}
    \end{subfigure}%
\caption{Axioms for the convex hull} \label{fig:convexhull}
\end{figure*}

\section{Proof tasks: Formulae}
\label{app:proofs-formulae}

We proved that ${\sf AxGeom} \cup {\sf AxGraphs} \cup \{ {\bf R1, C1} \}$ entails: 

\noindent {\bf Step 1:}
{\small \begin{align*}
    &\textsf{deleting}(u_1,v_1,w_1,x_1) & (e_0=u_1v_1, e_1=d_1=w_1x_1)\\ \land &\textsf{deleting}(w_1,x_1,w_2,x_2) & (e_1=w_1x_1, d_2=w_2x_2)\\ \rightarrow &\textsf{deleting}(u_1,v_1,w_2,x_1) & (e_2=w_2x_1 (Step1))\\ \lor &\textsf{intersection}(u_1,v_1,w_1,w_2) & (e_2=w_1w_2 (Step2)) 
\end{align*}} 

\noindent {\bf Step 2a:}
{\small 
\begin{align*}
    &\textsf{deleting}(u_1,v_1,w_1,x_1) & (e_0=u_1v_1, e_1=d_1=w_1x_1)\\ \land&
    \textsf{deleting}(w_1,x_1,w_2,x_2) & (e_1=w_1x_1, d_2=w_2x_2)\\ \land&
    \textsf{intersection}(u_1,v_1,w_1,w_2) & (e_2=w_1w_2)\\ \land& 
    \textsf{deleting}(w_1,w_2,w_3,x_3) & (d_3=w_3x_3)\\ 
    \rightarrow 
    &\textsf{deleting}(w_1,x_1,w_3,x_2)  & ( d_2=w_3x_2(Step1))\\
    \lor &\textsf{deleting}(w_1,x_1,w_3,x_3) & (d_2=w_3x_3(Step1))\\
    \lor &
    \textsf{deleting}(w_1,x_1,x_3,w_3)  & (d_2=x_3w_3 (Step1))\\
    \lor &\textsf{deleting}(w_1,x_1,x_3,x_2) & (d_2=x_3x_2(Step1))\\
    \lor &\textsf{deleting}(w_1,w_2,x_3,w_3) & (d_3=x_3w_3(Step2b))\\
    \lor &\textsf{intersection}(u_1,v_1,w_1,w_3) & (e_3=w_1w_3)\\ \lor &\textsf{intersection}(u_1,v_1,w_2,w_3) & (e_3=w_2w_3)
\end{align*}}

\noindent {\bf Step 2b:} 
{\small 
\begin{align*}
    &\textsf{deleting}(u_1,v_1,w_1,x_1) & (e_0=u_1v_1, e_1=d_1=w_1x_1)\\ \land&
    \textsf{deleting}(w_1,x_1,w_2,x_2) & (e_1=w_1x_1, d_2=w_2x_2)\\ \land&
    \textsf{intersection}(u_1,v_1,w_1,w_2) & (e_2=w_1w_2)\\ \land& 
    \textsf{deleting}(w_1,w_2,w_3,x_3) & (d_3=w_3x_3)\\ 
    \land &\textsf{deleting}(w_1,w_2,x_3,w_3) & (\text{or else } d_3=x_3w_3)\\
    \rightarrow 
     &\textsf{deleting}(w_1,x_1,w_3,x_2)  & (d_2=w_3x_2 (Step1))\\
    \lor &\textsf{deleting}(w_1,x_1,w_3,x_3) & (d_2=w_3x_3 (Step1))\\
    \lor &
    \textsf{deleting}(w_1,x_1,x_3,w_3)  & (d_2=x_3w_3 (Step1))\\
    \lor &\textsf{deleting}(w_1,x_1,x_3,x_2) & (d_2=x_3x_2 (Step1))\\
    \lor &\textsf{intersection}(u_1,v_1,w_1,w_3) & (d_3=w_3x_3, e_3=w_1w_3)\\ \lor &\textsf{intersection}(u_1,v_1,w_2,w_3) & (d_3=w_3x_3, e_3=w_2w_3)\\
    \lor &\textsf{intersection}(u_1,v_1,w_1,x_3) & (d_3=x_3w_3, e_3=w_1x_3)\\ \lor &\textsf{intersection}(u_1,v_1,w_2,x_3) & (d_3=x_3w_3, e_3=w_2x_3)
\end{align*}
}

\noindent {\bf Step 3:}
{\small \begin{align*}
    & \color{lightgray} \textsf{deleting}(u_1,v_1,w_1,x_1) & (e_0=u_1v_1, e_1=d_1=w_1x_1)\\\color{lightgray} \land&
   \color{black}
    \textsf{deleting}(w_1,x_1,w_2,x_2) & (e_1=w_1x_1, d_2=w_2x_2)\\ \color{lightgray}\land&
    \color{lightgray} \textsf{intersection}(u_1,v_1,w_1,w_2) & (e_2=w_1w_2)\\ \land& 
    \textsf{deleting}(w_1,w_2,w_3,x_3) & (d_3=w_3x_3)\\ 
    \land &\textsf{intersection}(w_1,w_3,w_2,x_2) \\
    \rightarrow 
     &\textsf{deleting}(w_1,x_1,w_3,x_2) 
     & (d_2=w_3x_2(Step1))\\
    \lor &\textsf{deleting}(w_1,x_1,w_3,x_3) & (d_2=w_3x_3(Step1))\\
    \lor &\textsf{deleting}(w_1,x_1,x_2,w_2) & (d_2=x_2w_2(Step1))\\ 
    \lor &
    \textsf{deleting}(w_1,x_1,x_3,w_3)  & (d_2=x_3w_3(Step1))\\
    \lor &\textsf{deleting}(w_1,x_1,x_3,x_2) & (d_2=x_3x_2(Step1))\\
\end{align*}}

\noindent {\bf Step 4:} 
{\small 
\begin{align*}
    & \color{lightgray} \textsf{deleting}(u_1,v_1,w_1,x_1) & (e_0=u_1v_1, e_1=d_1=w_1x_1)\\\color{lightgray} \land&
   \color{black}
    \textsf{deleting}(w_1,x_1,w_2,x_2) & (e_1=w_1x_1, d_2=w_2x_2)\\ \color{lightgray}\land&
    \color{lightgray} \textsf{intersection}(u_1,v_1,w_1,w_2) & (e_2=w_1w_2)\\ \land& 
    \textsf{deleting}(w_1,w_2,w_3,x_3) & (d_3=w_3x_3)\\ 
    \rightarrow 
      &\textsf{deleting}(w_1,x_1,w_3,x_2) & (d_2=w_3x_2(Step1))\\
    \lor &\textsf{deleting}(w_1,x_1,w_3,x_3) & (d_2=w_3x_3(Step1))\\
    \lor &\textsf{deleting}(w_1,x_1,x_3,w_3) & (d_2=x_3w_3(Step1))\\
    \lor &
    \textsf{deleting}(w_1,x_1,x_3,x_2)  & (d_2=x_3x_2(Step1))\\
    \lor &
    \textsf{inside}(w_1,x_1,w_2,w_3)  & (Step5)\\
    \lor &
    \textsf{inside}(w_1,x_1,w_2,x_3)  & (Step5)\\
    \lor &
    \textsf{inside}(w_1,w_2,x_1,w_3)  & (Step5)\\
    \lor &
    \textsf{inside}(w_1,w_2,x_1,x_3)  & (Step5)\\
    \lor &
    \textsf{F}(w_2,x_2) & (Step6)
\end{align*}}

\noindent {\bf Step 4a:} 
{\small 
\begin{align*}
    & \color{lightgray} \textsf{deleting}(u_1,v_1,w_1,x_1) & (e_0=u_1v_1, e_1=d_1=w_1x_1)\\\color{lightgray} \land&
   \color{black}
    \textsf{deleting}(w_1,x_1,w_2,x_2) & (e_1=w_1x_1, d_2=w_2x_2)\\ \color{lightgray}\land&
    \color{lightgray} \textsf{intersection}(u_1,v_1,w_1,w_2) & (e_2=w_1w_2)\\ \land& 
    \textsf{deleting}(w_1,w_2,x_1,x_3) & (d_3=x_1x_3)\\ 
    \rightarrow 
      &\textsf{deleting}(w_1,x_1,x_2,w_2) & (d_2=x_2w_2(Step1))\\
    \lor &\textsf{deleting}(w_1,x_1,x_3,x_2) & (d_2=x_3x_2(Step1))
\end{align*}}

\noindent {\bf Step 5a:} 
{\small \begin{align*}
    & \color{lightgray} \textsf{deleting}(u_1,v_1,w_1,x_1)   & (e_0=u_1v_1, e_1=d_1=w_1x_1)\\\color{lightgray}\land &
   \color{black}
    \textsf{deleting}(w_1,x_1,w_2,x_2) & (e_1=w_1x_1, d_2=w_2x_2)\\ \color{lightgray}\land&
    \color{lightgray} \textsf{intersection}(u_1,v_1,w_1,w_2) & (e_2=w_1w_2)\\ \color{lightgray} \land& \color{lightgray}
    \textsf{deleting}(w_1,w_2,y,y_2) & (d_3=yy_2, e_3=yw_2)\\
     \land&
   \color{black}
    \textsf{inside}(w_1,x_1,w_2,y)\\
     \land &\textsf{deleting}(w_2,y,w_3,x_3) &(d_4=w_3x_3)\\
    \rightarrow 
      &\textsf{deleting}(w_1,x_1,w_3,x_2) & (d_2=w_3x_2(Step1))\\
      \lor &\textsf{deleting}(w_1,x_1,w_3,x_3) & (d_2=w_3x_3(Step1))\\
    \lor &\textsf{deleting}(w_1,x_1,x_2,w_3) & (d_2=x_2w_3(Step1))\\
    \lor &\textsf{deleting}(w_1,x_1,x_2,x_3) & (d_2=x_2x_3 (Step1))\\
    \lor &
    \textsf{deleting}(w_1,x_1,x_3,w_3)  & (d_2=x_3w_3 (Step1))\\
    \lor &\textsf{deleting}(w_1,x_1,x_3,x_2) & (d_2=x_3x_2 (Step1))\\
     \lor &
    \textsf{deleting}(w_1,w_2,w_3,x_3)  & (d_3=w_3x_3 (Step2))\\
    \lor &
    \textsf{deleting}(w_1,w_2,x_3,w_3)  & (d_3=x_3w_3(Step2))\\
     \lor &
    \textsf{intersection}(w_1,y,w_2,x_2)  & (Step3)\\
     \lor &
    \textsf{inside}(w_1,x_1,w_2,w_3)  & (Step5)\\
    \lor &
    \textsf{inside}(w_1,x_1,w_2,x_3)  & (Step5)
\end{align*}
}

\noindent {\bf Step 5b:}  
{\footnotesize \begin{align*}
    & \color{lightgray} \textsf{deleting}(u_1,v_1,w_1,x_1) & (e_0=u_1v_1, e_1=d_1=w_1x_1)\\ \color{lightgray} \land&
   \color{black}
    \textsf{deleting}(w_1,x_1,w_2,x_2) & (e_1=w_1x_1, d_2=w_2x_2)\\ \color{lightgray}\land&
    \color{lightgray} \textsf{intersection}(u_1,v_1,w_1,w_2) & (e_2=w_1w_2)\\ \color{lightgray} \land& \color{lightgray}
    \textsf{deleting}(w_1,w_2,y,y_2) & (d_3=yy_2, e_3=yw_1)\\
     \land&
   \color{black}
    \textsf{inside}(w_1,x_1,w_2,y)\\
     \land &\textsf{deleting}(w_1,y,w_3,x_3) &(d_4=w_3x_3)\\
    \rightarrow 
      &\textsf{deleting}(w_1,x_1,w_3,x_2) & (d_2=w_3x_2(Step1))\\
      \lor &\textsf{deleting}(w_1,x_1,w_3,x_3) & (d_2=w_3x_3(Step1))\\
    \lor &\textsf{deleting}(w_1,x_1,x_2,w_3) & (d_2=x_2w_3(Step1))\\
    \lor &\textsf{deleting}(w_1,x_1,x_2,x_3) & (d_2=x_2x_3(Step1))\\
     \lor &\textsf{deleting}(w_1,x_1,x_3,w_3) & (d_2=x_3w_3(Step1))\\
    \lor &
    \textsf{deleting}(w_1,x_1,x_3,x_2)  & (d_2=x_3x_2(Step1))\\
     \lor &
    \textsf{deleting}(w_1,w_2,w_3,x_3)  & (d_3=w_3x_3(Step2))\\
    \lor &
    \textsf{deleting}(w_1,w_2,x_3,w_3)  & (d_3=x_3w_3(Step2))\\
     \lor &
    \textsf{intersection}(w_1,y,w_2,x_2)  & (Step3)\\
     \lor &
    \textsf{inside}(w_1,x_1,w_2,w_3)  & (Step5)\\
    \lor &
    \textsf{inside}(w_1,x_1,w_2,x_3)  & (Step5)
\end{align*}
} 

\newpage

\noindent {\bf Step 5c:} 

\vspace{-1cm}
{\small \begin{align*}
    & \color{lightgray} \textsf{deleting}(u_1,v_1,w_1,x_1) & (e_0=u_1v_1, e_1=d_1=w_1x_1)\\\color{lightgray} \land&
   \color{black}
    \textsf{deleting}(w_1,x_1,w_2,x_2) & (e_1=w_1x_1, d_2=w_2x_2)\\ \color{lightgray}\land&
    \color{lightgray} \textsf{intersection}(u_1,v_1,w_1,w_2) & (e_2=w_1w_2)\\ \color{lightgray} \land& \color{lightgray}
    \textsf{deleting}(w_1,w_2,y,y_2) & (d_3=yy_2, e_3=yw_1)\\
     \land
   \color{black}
    &\textsf{inside}(w_1,x_1,w_2,y)\\
     \color{lightgray}\land \color{lightgray} & \color{lightgray}\textsf{deleting}(w_1,y,z,z_2) &(d_4=zz_2, e_4=yz)\\
     \land &\textsf{inside}(w_1,x_1,w_2,z)\\
     \land &\textsf{deleting}(y,z,w_3,x_3) &(d_5=w_3x_3)\\
    \rightarrow 
      &\textsf{deleting}(w_1,x_1,w_3,x_2) & (d_2=w_3x_2(Step1))\\
      \lor &\textsf{deleting}(w_1,x_1,w_3,x_3) & (d_2=w_3x_3(Step1))\\
    \lor &\textsf{deleting}(w_1,x_1,x_2,w_3) & (d_2=x_2w_3(Step1))\\
    \lor &\textsf{deleting}(w_1,x_1,x_2,x_3) & (d_2=x_2x_3 (Step1))\\
     \lor &\textsf{deleting}(w_1,x_1,x_3,w_3)  & (d_2=x_3w_3 (Step1))\\
    \lor &\textsf{deleting}(w_1,x_1,x_3,x_2) & (d_2=x_3x_2 (Step1))\\
     \lor &
    \textsf{deleting}(w_1,w_2,w_3,x_3)  & (d_3=w_3x_3 (Step2))\\
    \lor &
    \textsf{deleting}(w_1,w_2,x_3,w_3)  & (d_3=x_3w_3 (Step2))\\
    \lor &
    \textsf{intersection}(w_1,z,w_2,x_2)  &(Step3)\\  
    \lor &
    \textsf{intersection}(w_1,y,w_2,x_2)  & (Step3)\\
     \lor &
    \textsf{inside}(w_1,x_1,w_2,w_3)  & (Step5)\\ 
    \lor &
    \textsf{inside}(w_1,x_1,w_2,x_3) & (Step5)
\end{align*}
} 

\noindent {\bf Step 5d:} 

\vspace{-1cm}
{\small \begin{align*}
    & \color{lightgray} \textsf{deleting}(u_1,v_1,w_1,x_1)  & (e_0=u_1v_1, e_1=d_1=w_1x_1)\\ 
    \color{lightgray}\land& \color{black}
    \textsf{deleting}(w_1,x_1,w_2,x_2) & (e_1=w_1x_1, d_2=w_2x_2)\\ \color{lightgray}\land&
    \color{lightgray} \textsf{intersection}(u_1,v_1,w_1,w_2) & (e_2=w_1w_2)\\ \color{lightgray} \land& \color{lightgray}
    \textsf{deleting}(w_1,w_2,y,y_2) & (d_3=yy_2, e_3=yw_2)\\
    \color{lightgray}\land&
    \color{lightgray}
    \textsf{intersection}(u_1,v_1,y,w_2) & (e_3=yw_2)\\ 
     \land&
   \color{black}
    \textsf{inside}(w_1,x_1,w_2,y)\\
     \land& \color{black}
    \textsf{deleting}(y,w_2,w_3,x_3) & (d_4=w_3x_3, e_4=yw_3)\\
     \land &\textsf{Intersection}(w_1,w_3,w_2,x_2)\\
    \rightarrow 
      &\textsf{deleting}(w_1,x_1,w_3,x_2) & (d_2=w_3x_2(Step1))\\
      \lor &\textsf{deleting}(w_1,x_1,w_3,x_3) & (d_2=w_3x_3(Step1))\\
    \lor &\textsf{deleting}(w_1,x_1,x_2,w_3) & (d_2=x_2w_3(Step1))\\
     \lor &\textsf{deleting}(w_1,x_1,x_2,x_3) & (d_2=x_2x_3 (Step1))\\
     \lor & \textsf{deleting}(w_1,x_1,x_3,w_3)  & (d_2=x_3w_3 (Step1))\\
    \lor &\textsf{deleting}(w_1,x_1,x_3,x_2) & (d_2=x_3x_2 (Step1))\\
     \lor &
    \textsf{deleting}(w_1,w_2,w_3,x_3)  & (d_3=w_3x_3 (Step2))\\
    \lor &
    \textsf{deleting}(w_1,w_2,x_3,w_3)  & (d_3=x_3w_3(Step2))\\
     \lor &
    \textsf{intersection}(w_1,y,w_2,x_2)  & (Step3)
\end{align*}
} 

\noindent {\bf Step 5e:} 
{\small \begin{align*}
    & \color{lightgray} \textsf{deleting}(u_1,v_1,w_1,x_1)  & (e_0=u_1v_1, e_1=d_1=w_1x_1)\\ 
    \color{lightgray}\land& \color{black}
    \textsf{deleting}(w_1,x_1,w_2,x_2) & (e_1=w_1x_1, d_2=w_2x_2)\\ \color{lightgray}\land&
    \color{lightgray} \textsf{intersection}(u_1,v_1,w_1,w_2) & (e_2=w_1w_2)\\ \color{lightgray} \land& \color{lightgray}
    \textsf{deleting}(w_1,w_2,y,y_2) & (d_3=yy_2, e_3=yw_1)\\
    \color{lightgray}\land&
    \color{lightgray}
    \textsf{intersection}(u_1,v_1,y,w_1) & (e_3=yw_1)\\ 
     \land&
   \color{black}
    \textsf{inside}(w_1,x_1,w_2,y)\\
     \land& \color{black}
    \textsf{deleting}(y,w_1,w_3,x_3) & (d_4=w_3x_3, e_4=yw_3)\\
     \land &\textsf{Intersection}(w_1,w_3,w_2,x_2)\\
    \rightarrow 
      &\textsf{deleting}(w_1,x_1,w_3,x_2) & (d_2=w_3x_2(Step1))\\
      \lor &\textsf{deleting}(w_1,x_1,w_3,x_3) & (d_2=w_3x_3(Step1))\\
    \lor &\textsf{deleting}(w_1,x_1,x_2,w_3) & (d_2=x_2w_3(Step1))\\
     \lor &\textsf{deleting}(w_1,x_1,x_2,x_3) & (d_2=x_2x_3 (Step1))\\
    \lor &
    \textsf{deleting}(w_1,x_1,x_3,w_3)  & (d_2=x_3w_3 (Step1))\\
    \lor &\textsf{deleting}(w_1,x_1,x_3,x_2) & (d_2=x_3x_2 (Step1))\\
     \lor &
    \textsf{deleting}(w_1,w_2,w_3,x_3)  & (d_3=w_3x_3 (Step2))\\
    \lor &
    \textsf{deleting}(w_1,w_2,x_3,w_3)  & (d_3=x_3w_3(Step2))\\
     \lor &
    \textsf{intersection}(w_1,y,w_2,x_2)  & (Step3)
\end{align*}
} 

\noindent {\bf Step 5f:} 
{\small \begin{align*}
    & \color{lightgray} \textsf{deleting}(u_1,v_1,w_1,x_1)  & (e_0=u_1v_1, e_1=d_1=w_1x_1)\\ 
    \color{lightgray}\land& \color{black}
    \textsf{deleting}(w_1,x_1,w_2,x_2) & (e_1=w_1x_1, d_2=w_2x_2)\\ \color{lightgray}\land&
    \color{lightgray}
    \textsf{deleting}(w_1,w_2,y,y_2) & (d_3=yy_2, e_3=yw_1)\\
     \land
   \color{black}
    &\textsf{inside}(w_1,x_1,w_2,y)\\
     \color{lightgray}\land \color{lightgray} & \color{lightgray}\textsf{deleting}(w_1,y,z,z_2) &(d_4=zz_2, e_4=yz)\\
     \land &\textsf{inside}(w_1,x_1,w_2,z)\\
     \land &\textsf{deleting}(y,z,w_3,x_3) &(d_5=w_3x_3)\\
     \land &\textsf{Intersection}(w_1,w_3,w_2,x_2)\\
    \rightarrow 
      &\textsf{deleting}(w_1,x_1,w_3,x_2) & (d_2=w_3x_2(Step1))\\
      \lor &\textsf{deleting}(w_1,x_1,w_3,x_3) & (d_2=w_3x_3(Step1))\\
    \lor &\textsf{deleting}(w_1,x_1,x_2,w_3) & (d_2=x_2w_3(Step1))\\
    \lor &\textsf{deleting}(w_1,x_1,x_2,x_3) & (d_2=x_2x_3 (Step1))\\
    \lor &
    \textsf{deleting}(w_1,x_1,x_3,w_3)  & (d_2=x_3w_3 (Step1))\\
    \lor &\textsf{deleting}(w_1,x_1,x_3,x_2) & (d_2=x_3x_2 (Step1))\\
     \lor &
    \textsf{deleting}(w_1,w_2,w_3,x_3)  & (d_3=w_3x_3 (Step2))\\
    \lor &
    \textsf{deleting}(w_1,w_2,x_3,w_3)  & (d_3=x_3w_3(Step2))\\
     \lor &
    \textsf{intersection}(w_1,y,w_2,x_2)  & (Step3)\\
    \lor &
    \textsf{intersection}(w_1,z,w_2,x_2)  & (Step3)
\end{align*}
} 

\ignore{

The test \color{blue}z3-test proof part2 step6g \color{black} proves 
\begin{align*}
    & \color{lightgray} \textsf{deleting}(u_1,v_1,w_1,x_1) (e_0=u_1v_1, e_1=d_1=w_1x_1)\\ \land&
   \color{black}
    \textsf{deleting}(w_1,x_1,w_2,x_2) (e_1=w_1x_1, d_2=w_2x_2)\\ \color{lightgray}\land&
    \color{lightgray} \textsf{intersection}(u_1,v_1,w_1,w_2) (e_2=w_1w_2)\\  \land&
    \textsf{deleting}(w_1,w_2,w_3,x_3) (d_3=w_3x_3)\\
     \land
   \color{black}
    &\textsf{inside}(w_1,x_1,w_2,w_3)\\
    \color{lightgray} \land& \color{lightgray} \textsf{intersection}(u_1,v_1,w_1,w_2) (e_3=w_1w_3)\\  
     \land  & \textsf{deleting}(w_1,w_3,w_4,x_4) (d_4=w_4x_4)\\
     \land &\textsf{inside}(w_1,x_1,w_2,w_4)\\
      \color{lightgray} \land& \color{lightgray} \textsf{intersection}(u_1,v_1,w_1,w_4) (e_4=w_1w_3)\\ 
     \land &\textsf{deleting}(w_1,w_4,w_5,x_5) (d_5=w_5x_5)\\
    \rightarrow 
      &\textsf{deleting}(w_1,x_1,w_3,x_2) (d_2=w_3x_2(Step1))\\
    \lor &\textsf{deleting}(w_1,x_1,w_4,x_4) (d_2=w_4x_4(Step1))\\
    \lor &\textsf{deleting}(w_1,w_2,w_4,x_4) (d_3=w_3x_3(Step2))\\
    \lor &
    \textsf{deleting}(w_1,w_3,w_5,x_4)  (d_4=w_5x_4(Step6g))\\
    \lor &\textsf{deleting}(w_1,w_3,w_5,x_5) (d_4=w_5x_5(Step6g))\\
    \lor &\textsf{deleting}(w_1,w_3,x_4,w_4) (d_4=x_4w4(Step6g))\\
     \lor &
    \textsf{deleting}(w_1,w_2,x_5,w_5)  (d_4=x_5w_5(Step6g))\\
    \lor &
    \textsf{deleting}(w_1,w_2,x_5,x_4)  (d_3=x_5x_4(Step6g))\\
    \lor &
    \textsf{intersection}(w_1,w_3,w_2,x_2)  (Step3)\\
    \lor &
    \textsf{intersection}(w_1,w_4,w_2,x_2)  (Step3)\\
     \lor &
     \textsf{intersection}(w_1,w_5,w_2,x_2)  (Step3)\\
     \lor &
    \textsf{inside}(w_1,w_3,w_4,w_5)\\
    \lor &
    \textsf{inside}(w_1,w_3,w_4,x_5)\\
    \lor &
    \textsf{inside}(w_1,w_4,w_3,w_5)\\
    \lor &
    \textsf{inside}(w_1,w_4,w_3,x_5) 
\end{align*}
}

\noindent {\bf Step 6a:} 
{\small \begin{align*}
    & \color{lightgray} \textsf{deleting}(u_1,v_1,w_1,x_1) & \vspace{-3cm}(e_0=u_1v_1, e_1=d_1=w_1x_1)\\ \color{lightgray} \land&
   \color{black}
    \textsf{deleting}(w_1,x_1,w_2,x_2) & \vspace{-3cm}(e_1=w_1x_1, d_2=w_2x_2)\\ \color{lightgray}\land&
    \color{lightgray} \textsf{intersection}(u_1,v_1,w_1,w_2) & \vspace{-3cm}(e_2=w_1w_2)\\  \land&
    \textsf{deleting}(w_1,w_2,w_3,x_3) & (d_3=w_3x_3)\\
    \color{lightgray} \land& \color{lightgray} \textsf{intersection}(u_1,v_1,w_1,w_3) & \vspace{-3cm}(e_3=w_1w_3)\\  
     \land  & \textsf{deleting}(w_1,w_3,w_4,x_4) & \vspace{-3cm}(d_4=w_4x_4)\\
    \rightarrow 
    &\textsf{deleting}(w_1,w_2,w_3,x_2) & \vspace{-3cm}(d_2=w_3x_2(Step1))\\
    \lor &\textsf{deleting}(w_1,x_1,w_3,x_3) & (d_2=w_3x_3(Step1))\\
    \lor &\textsf{deleting}(w_1,x_1,x_3,w_3) & \vspace{-3cm}(d_2=x_3w_3(Step1))\\
    \lor &\textsf{deleting}(w_1,w_2,x_3,x_2) & \vspace{-3cm}(d_2=x_3x_2(Step1))\\
    \lor &\textsf{deleting}(w_1,w_2,w_4,x_3)  & \vspace{-3cm}(d_3=w_4x_3(Step2))\\
    \lor &\textsf{deleting}(w_1,w_2,w_4,x_4)  & \vspace{-3cm}(d_3=w_4x_4(Step2))\\
    \lor &\textsf{deleting}(w_1,w_2,x_4,w_4)  & \vspace{-3cm}(d_3=x_4w_4(Step2))\\
    \lor &\textsf{deleting}(w_1,w_2,x_4,x_3)  & \vspace{-3cm}(d_3=x_4x_3(Step2))\\
     \lor &
    \textsf{inside}(w_1,x_1,w_2,w_3)\\
    \lor &
    \textsf{inside}(w_1,x_1,w_2,x_3)\\
    \lor & 
    \textsf{inside}(w_1,w_2,x_1,w_3)\\
    \lor &
    \textsf{inside}(w_1,w_2,x_1,x_3)\\
    \lor &
    \textsf{inside}(w_1,w_2,w_3,w_4)\\
    \lor &
    \textsf{inside}(w_1,w_2,w_3,x_4)\\
    \lor & 
    \textsf{inside}(w_1,w_3,w_2,w_4)\\
    \lor &
    \textsf{inside}(w_1,w_3,w_2,x_4) 
\end{align*}
} 

\noindent {\bf Step 6b:} 
{\small \begin{align*}
    & \color{lightgray} \textsf{deleting}(u_1,v_1,w_1,x_1) & \vspace{-3cm}(e_0=u_1v_1, e_1=d_1=w_1x_1)\\ \color{lightgray} \land&
   \color{black}
    \textsf{deleting}(w_1,x_1,w_2,x_2) & \vspace{-3cm}(e_1=w_1x_1, d_2=w_2x_2)\\ \color{lightgray}\land&
    \color{lightgray} \textsf{intersection}(u_1,v_1,w_1,w_2) & \vspace{-3cm}(e_2=w_1w_2)\\  \land&
    \textsf{deleting}(w_1,w_2,w_3,x_3) & (d_3=w_3x_3)\\
    \color{lightgray} \land& \color{lightgray} \textsf{intersection}(u_1,v_1,w_1,w_3) & \vspace{-3cm}(e_3=w_1w_3)\\  
     \land  & \textsf{deleting}(w_1,w_3,x_1,x_4) & \vspace{-3cm}(d_4=x_1x_4)\\
    \rightarrow 
    &\textsf{deleting}(w_1,x_1,w_3,x_2) & \vspace{-3cm}(d_2=w_3x_2(Step1))\\
    \lor &\textsf{deleting}(w_1,x_1,w_3,x_3) & (d_2=w_3x_3(Step1))\\
    \lor &\textsf{deleting}(w_1,x_1,x_2,w_2) & \vspace{-3cm}(d_2=x_2w_2(Step1))\\
    \lor &\textsf{deleting}(w_1,x_1,x_3,x_2) & \vspace{-3cm}(d_2=x_3x_2(Step1))\\
     \lor &
    \textsf{inside}(w_1,x_1,w_2,w_3)\\
    \lor &
    \textsf{inside}(w_1,x_1,w_2,x_3)\\
    \lor & 
    \textsf{inside}(w_1,w_2,x_1,w_3)\\
    \lor &
    \textsf{inside}(w_1,w_2,x_1,x_3)
\end{align*}
} 

\noindent {\bf Step 6c}: 
{\small \begin{align*}
    & \color{lightgray} \textsf{deleting}(u_1,v_1,w_1,x_1) & (e_0=u_1v_1, e_1=d_1=w_1x_1)\\\color{lightgray} \land&
   \color{black}
    \textsf{deleting}(w_1,x_1,w_2,x_2) & (e_1=w_1x_1, d_2=w_2x_2)\\ \color{lightgray}\land&
    \color{lightgray} \textsf{intersection}(u_1,v_1,w_1,w_2) & (e_2=w_1w_2)\\  \land&
    \textsf{deleting}(w_1,w_2,w_3,x_3) & (d_3=w_3x_3)\\
    \color{lightgray} \land& \color{lightgray} \textsf{intersection}(u_1,v_1,w_2,w_3) & (e_3=w_2w_3)\\  
     \land  & \textsf{deleting}(w_2,w_3,w_4,x_4) & (d_4=w_4x_4)\\
    \rightarrow 
    &\textsf{deleting}(w_1,w_2,w_3,x_2) & (d_2=w_3x_2 (Step1))\\
    \lor &\textsf{deleting}(w_1,x_1,w_3,x_3) & (d_2=w_3x_3 (Step1))\\
    \lor &\textsf{deleting}(w_1,x_1,x_3,w_3) & (d_2=x_3w_3 (Step1))\\
    \lor  &\textsf{deleting}(w_1,w_2,x_3,x_2) & (d_2=x_3x_2 (Step1))\\
    \lor &\textsf{deleting}(w_1,w_2,w_4,x_3)  & (d_3=w_4x_3 (Step2))\\
    \lor &\textsf{deleting}(w_1,w_2,w_4,x_4)  & (d_3=w_4x_4 (Step2))\\
    \lor &\textsf{deleting}(w_1,w_2,x_4,w_4)  & (d_3=x_4w_4 (Step2))\\
    \lor &\textsf{deleting}(w_1,w_2,x_4,x_3)  & (d_3=x_4x_3 (Step2))\\
     \lor &
    \textsf{inside}(w_1,x_1,w_2,w_3)\\
    \lor &
    \textsf{inside}(w_1,x_1,w_2,x_3)\\
    \lor &
    \textsf{inside}(w_1,w_2,x_1,w_3)\\
    \lor &
    \textsf{inside}(w_1,w_2,x_1,x_3)\\
    \lor &
    \textsf{inside}(w_1,w_2,w_3,w_4)\\
    \lor &
    \textsf{inside}(w_1,w_2,w_3,x_4)\\
    \lor &
    \textsf{inside}(w_1,w_3,w_2,w_4)\\
    \lor &
    \textsf{inside}(w_1,w_3,w_2,x_4) 
\end{align*}
} 

\noindent {\bf Step 6d:} 
{\small \begin{align*}
    & \color{lightgray} \textsf{deleting}(u_1,v_1,w_1,x_1) & \vspace{-3cm}(e_0=u_1v_1, e_1=d_1=w_1x_1)\\ \color{lightgray} \land&
   \color{black}
    \textsf{deleting}(w_1,x_1,w_2,x_2) & \vspace{-3cm}(e_1=w_1x_1, d_2=w_2x_2)\\ \color{lightgray}\land&
    \color{lightgray} \textsf{intersection}(u_1,v_1,w_1,w_2) & \vspace{-3cm}(e_2=w_1w_2)\\  \land&
    \textsf{deleting}(w_1,w_2,w_3,x_3) & (d_3=w_3x_3)\\
    \color{lightgray} \land& \color{lightgray} \textsf{intersection}(u_1,v_1,w_2,w_3) & \vspace{-3cm}(e_3=w_2w_3)\\  
     \land  & \textsf{deleting}(w_2,w_3,x_1,x_4) & \vspace{-3cm}(d_4=x_1x_4)\\
    \rightarrow 
       &\textsf{deleting}(w_1,x_1,w_3,x_2) & \vspace{-3cm}(d_2=w_3x_2(Step1))\\
      \lor &\textsf{deleting}(w_1,x_1,w_3,x_3) & (d_2=w_3x_3(Step1))\\
      \lor &\textsf{deleting}(w_1,x_1,x_2,w_2) & \vspace{-3cm}(d_2=x_2w_2(Step1))\\
    \lor &\textsf{deleting}(w_1,x_1,x_3,x_2) & \vspace{-3cm}(d_2=x_3x_2(Step1))\\
     \lor &
    \textsf{inside}(w_1,x_1,w_2,w_3)\\
    \lor &
    \textsf{inside}(w_1,x_1,w_2,x_3)\\
    \lor & 
    \textsf{inside}(w_1,w_2,x_1,w_3)\\
    \lor &
    \textsf{inside}(w_1,w_2,x_1,x_3)
\end{align*}
}

\noindent {\bf Step 6e:} 
{\small \begin{align*}
    & \color{lightgray} \textsf{deleting}(u_1,v_1,w_1,x_1) & \vspace{-3cm}(e_0=u_1v_1, e_1=d_1=w_1x_1)\\ \color{lightgray} \land&
   \color{black}
    \textsf{deleting}(w_1,x_1,w_2,x_2) & \vspace{-3cm}(e_1=w_1x_1, d_2=w_2x_2)\\ \color{lightgray}\land&
    \color{lightgray} \textsf{intersection}(u_1,v_1,w_1,w_2) & \vspace{-3cm}(e_2=w_1w_2)\\  \land&
    \textsf{deleting}(w_1,w_2,w_3,x_3) & (d_3=w_3x_3)\\
    \color{lightgray} \land& \color{lightgray} \textsf{intersection}(u_1,v_1,w_2,w_3) & \vspace{-3cm}(e_3=w_2w_3)\\  
     \land  & \textsf{deleting}(w_2,w_3,w_1,x_4) & \vspace{-3cm}(d_4=w_1x_4)\\
    \rightarrow 
       &\textsf{deleting}(w_1,x_1,w_3,x_2) & \vspace{-3cm}(d_2=w_3x_2(Step1))\\
      \lor &\textsf{deleting}(w_1,x_1,w_3,x_3) & (d_2=w_3x_3(Step1))\\
      \lor &\textsf{deleting}(w_1,x_1,x_2,w_2) & \vspace{-3cm}(d_2=x_2w_2(Step1))\\
    \lor &\textsf{deleting}(w_1,x_1,x_3,x_2) & \vspace{-3cm}(d_2=x_3x_2(Step1))\\
    \lor &\textsf{deleting}(w_1,w_2,x_3,w_3) & \vspace{-3cm}(d_3=x_3x_2(Step2))\\
    \lor &\textsf{deleting}(w_1,w_2,x_4,x_3) & \vspace{-3cm}(d_3=x_4x_3(Step2))\\
     \lor &
    \textsf{inside}(w_1,x_1,w_2,w_3)\\
    \lor &
    \textsf{inside}(w_1,x_1,w_2,x_3)\\
    \lor & 
    \textsf{inside}(w_1,w_2,x_1,w_3)\\
    \lor &
    \textsf{inside}(w_1,w_2,x_1,x_3)
\end{align*}
} 

\noindent 
{\bf Step 6g:} 
{\small\begin{align*}
    & \color{lightgray} \textsf{deleting}(u_1,v_1,w_1,x_1) & (e_0=u_1v_1, e_1=d_1=w_1x_1)\\ \color{lightgray} \land&
   \color{black}
    \textsf{deleting}(w_1,x_1,w_2,x_2) & (e_1=w_1x_1, d_2=w_2x_2)\\ \color{lightgray}\land&
    \color{lightgray} \textsf{intersection}(u_1,v_1,w_1,w_2) & (e_2=w_1w_2)\\  \land&
    \textsf{deleting}(w_1,w_2,w_3,x_3) & (d_3=w_3x_3)\\
     \land
   \color{black}
    &\textsf{inside}(w_1,x_1,w_2,w_3)\\
    \color{lightgray} \land& \color{lightgray} \textsf{intersection}(u_1,v_1,w_1,w_3) & (e_3=w_1w_3)\\  
     \land  & \textsf{deleting}(w_1,w_3,w_4,x_4) & (d_4=w_4x_4)\\
     \land &\textsf{inside}(w_1,x_1,w_2,w_4)\\
      \color{lightgray} \land& \color{lightgray} \textsf{intersection}(u_1,v_1,w_1,w_4) & (e_4=w_1w_4)\\ 
     \land &\textsf{deleting}(w_1,w_4,w_5,x_5) & (d_5=w_5x_5)\\
    \rightarrow 
      &\textsf{deleting}(w_1,x_1,w_3,x_2) & (d_2=w_3x_2(Step1))\\
    \lor &\textsf{deleting}(w_1,x_1,w_4,x_4) & (d_2=w_4x_4(Step1))\\
    \lor &\textsf{deleting}(w_1,w_2,w_4,x_4) & (d_3=w_4x_4(Step2))\\
 \lor &
    \textsf{deleting}(w_1,w_3,w_5,x_4) & (d_4=w_5x_4(Step6g))\\
    \lor &\textsf{deleting}(w_1,w_3,w_5,x_5) & (d_4=w_5x_5(Step6g))\\
    \lor &\textsf{deleting}(w_1,w_3,x_4,w_4) & (d_4=x_4w_4(Step6g))\\
     \lor &
    \textsf{deleting}(w_1,w_2,x_5,w_5)  & (d_4=x_5w_5(Step6g))\\
    \lor &
    \textsf{deleting}(w_1,w_2,x_5,x_4)  & (d_4=x_5x_4(Step6g))\\
       \lor &
    \textsf{intersection}(w_1,w_3,w_2,x_2)  & (Step3)\\
    \lor &
    \textsf{intersection}(w_1,w_4,w_2,x_2)  & (Step3)\\
     \lor &
     \textsf{intersection}(w_1,w_5,w_2,x_2)  & (Step3)\\
     \lor &
    \textsf{inside}(w_1,w_3,w_4,w_5)\\
    \lor &
    \textsf{inside}(w_1,w_3,w_4,x_5)\\
    \lor & 
    \textsf{inside}(w_1,w_4,w_3,w_5)\\
    \lor &
    \textsf{inside}(w_1,w_4,w_3,x_5) 
\end{align*}}

\noindent 
{\small{\bf Step 6h:} 
\begin{align*}
    & \color{lightgray} \textsf{deleting}(u_1,v_1,w_1,x_1) & (e_0=u_1v_1, e_1=d_1=w_1x_1)\\ \color{lightgray} \land&
   \color{black}
    \textsf{deleting}(w_1,x_1,w_2,x_2) & (e_1=w_1x_1, d_2=w_2x_2)\\ \color{lightgray}\land&
    \color{lightgray} \textsf{intersection}(u_1,v_1,w_1,w_2) & (e_2=w_1w_2)\\  \land&
    \textsf{deleting}(w_1,w_2,w_3,x_3) & (d_3=w_3x_3)\\
     \land
   \color{black}
    &\textsf{inside}(w_1,x_1,w_2,w_3)\\
    \color{lightgray} \land& \color{lightgray} \textsf{intersection}(u_1,v_1,w_1,w_3) & (e_3=w_1w_3)\\  
     \land  & \textsf{deleting}(w_1,w_3,w_4,x_4) & (d_4=w_4x_4)\\
     \land &\textsf{inside}(w_1,x_1,w_2,w_4)\\
      \color{lightgray} \land& \color{lightgray} \textsf{intersection}(u_1,v_1,w_3,w_4) & (e_4=w_3w_4)\\ 
     \land &\textsf{deleting}(w_3,w_4,w_5,x_5) & (d_5=w_5x_5)\\
    \rightarrow 
      &\textsf{deleting}(w_1,x_1,w_3,x_2) & (d_2=w_3x_2(Step1))\\
    \lor &\textsf{deleting}(w_1,x_1,w_4,x_4) & (d_2=w_4x_4(Step1))\\
    \lor &\textsf{deleting}(w_1,w_2,w_4,x_4) & (d_3=w_4x_4(Step2))\\
    \lor &
    \textsf{deleting}(w_1,w_3,w_5,x_4) & (d_4=w_5x_4(Step6h))\\
    \lor &\textsf{deleting}(w_1,w_3,w_5,x_5) & (d_4=w_5x_5(Step6h))\\
    \lor &\textsf{deleting}(w_1,w_3,x_4,w_4) & (d_4=x_4w_4(Step6h))\\
     \lor &
    \textsf{deleting}(w_1,w_2,x_5,w_5)  & (d_4=x_5w_5(Step6h))\\
    \lor &
    \textsf{deleting}(w_1,w_2,x_5,x_4)  & (d_4=x_5x_4(Step6h))\\
    \lor &
    \textsf{intersection}(w_1,w_3,w_2,x_2)  & (Step3)\\
    \lor &
    \textsf{intersection}(w_1,w_4,w_2,x_2)  & (Step3)\\
     \lor &
     \textsf{intersection}(w_1,w_5,w_2,x_2)  & (Step3)\\
     \lor &
    \textsf{inside}(w_1,w_3,w_4,w_5)\\
    \lor &
    \textsf{inside}(w_1,w_3,w_4,x_5)\\
    \lor & 
    \textsf{inside}(w_1,w_4,w_3,w_5)\\
    \lor &
    \textsf{inside}(w_1,w_4,w_3,x_5) 
\end{align*}}

\noindent 
{\bf Step 6i:} 
{\small\begin{align*}
    & \color{lightgray} \textsf{deleting}(u_1,v_1,w_1,x_1) & (e_0=u_1v_1, e_1=d_1=w_1x_1)\\\color{lightgray} \land&
   \color{black}
    \textsf{deleting}(w_1,x_1,w_2,x_2) & (e_1=w_1x_1, d_2=w_2x_2)\\ \color{lightgray}\land&
    \color{lightgray} \textsf{intersection}(u_1,v_1,w_1,w_2) & (e_2=w_1w_2)\\  \land&
    \textsf{deleting}(w_1,w_2,w_3,x_3) & (d_3=w_3x_3)\\
     \land
   \color{black}
    &\textsf{inside}(w_1,x_1,w_2,w_3)\\
    \color{lightgray} \land& \color{lightgray} \textsf{intersection}(u_1,v_1,w_2,w_3) & (e_3=w_2w_3)\\  
     \land  & \textsf{deleting}(w_2,w_3,w_4,x_4) & (d_4=w_4x_4)\\
     \land &\textsf{inside}(w_1,x_1,w_2,w_4)\\
      \color{lightgray} \land& \color{lightgray} \textsf{intersection}(u_1,v_1,w_2,w_4) & (e_4=w_2w_4)\\ 
     \land &\textsf{deleting}(w_2,w_4,w_5,x_5) & (d_5=w_5x_5)\\
    \rightarrow 
      &\textsf{deleting}(w_1,x_1,w_3,x_2) & (d_2=w_3x_2(Step1))\\
    \lor &\textsf{deleting}(w_1,x_1,w_4,x_4) & (d_2=w_4x_4(Step1))\\
    \lor &\textsf{deleting}(w_1,w_2,w_4,x_4) & (d_3=w_4x_4(Step2))\\
    \lor &
    \textsf{deleting}(w_2,w_3,w_5,x_4) & (d_4=w_5x_4(Step6i))\\
    \lor &\textsf{deleting}(w_2,w_3,w_5,x_5) & (d_4=w_5x_5(Step6i))\\
    \lor &\textsf{deleting}(w_2,w_3,x_4,w_4) & (d_4=x_4w_4(Step6i))\\
     \lor &
    \textsf{deleting}(w_1,w_2,x_5,w_5)  & (d_4=x_5w_5(Step6i))\\
    \lor &
    \textsf{deleting}(w_1,w_2,x_5,x_4)  & (d_4=x_5x_4(Step6i))\\
    \lor &
    \textsf{intersection}(w_1,w_3,w_2,x_2)  & (Step3)\\
    \lor &
    \textsf{intersection}(w_1,w_4,w_2,x_2)  & (Step3)\\
     \lor &
     \textsf{intersection}(w_1,w_5,w_2,x_2)  & (Step3)\\
     \lor &
    \textsf{inside}(w_2,w_3,w_4,w_5)\\
    \lor &
    \textsf{inside}(w_2,w_3,w_4,x_5)\\
    \lor & 
    \textsf{inside}(w_2,w_4,w_3,w_5)\\
    \lor &
    \textsf{inside}(w_2,w_4,w_3,x_5) 
\end{align*}}

\noindent 
{\bf Step 6j:} 
{\small\begin{align*}
    & \color{lightgray} \textsf{deleting}(u_1,v_1,w_1,x_1) & (e_0=u_1v_1, e_1=d_1=w_1x_1)\\ \color{lightgray} \land&
   \color{black}
    \textsf{deleting}(w_1,x_1,w_2,x_2) & (e_1=w_1x_1, d_2=w_2x_2)\\ \color{lightgray}\land&
    \color{lightgray} \textsf{intersection}(u_1,v_1,w_1,w_2) & (e_2=w_1w_2)\\  \land&
    \textsf{deleting}(w_1,w_2,w_3,x_3) & (d_3=w_3x_3)\\
     \land
   \color{black}
    &\textsf{inside}(w_1,x_1,w_2,w_3)\\
    \color{lightgray} \land& \color{lightgray} \textsf{intersection}(u_1,v_1,w_2,w_3) & (e_3=w_2w_3)\\  
      \land  & \textsf{deleting}(w_2,w_3,w_4,x_4) & (d_4=w_4x_4)\\
     \land &\textsf{inside}(w_1,x_1,w_2,w_4)\\
      \color{lightgray} \land& \color{lightgray} \textsf{intersection}(u_1,v_1,w_3,w_4) & (e_4=w_3w_4)\\ 
    \end{align*}
 \begin{align*}      \land &\textsf{deleting}(w_3,w_4,w_5,x_5) & (d_5=w_5x_5)\\
    \rightarrow 
      &\textsf{deleting}(w_1,x_1,w_3,x_2) & (d_2=w_3x_2(Step1))\\
    \lor &\textsf{deleting}(w_1,x_1,w_4,x_4) & (d_2=w_4x_4(Step1))\\
   \lor &\textsf{deleting}(w_1,w_2,w_4,x_4) & (d_3=w_4x_4(Step2))\\
   \lor &
    \textsf{deleting}(w_2,w_3,w_5,x_4) & (d_4=w_5x_4(Step6j))\\
    \lor &\textsf{deleting}(w_2,w_3,w_5,x_5) & (d_4=w_5x_5(Step6j))\\
    \lor &\textsf{deleting}(w_2,w_3,x_4,w_4) & (d_4=x_4w_4(Step6j))\\
     \lor &
    \textsf{deleting}(w_1,w_2,x_5,w_5)  & (d_4=x_5w_5(Step6j))\\
    \lor &
    \textsf{deleting}(w_1,w_2,x_5,x_4)  & (d_4=x_5x_4(Step6j))\\
    \lor &
    \textsf{intersection}(w_1,w_3,w_2,x_2)  & (Step3)\\
    \lor &
    \textsf{intersection}(w_1,w_4,w_2,x_2)  & (Step3)\\
     \lor &
     \textsf{intersection}(w_1,w_5,w_2,x_2)  & (Step3)\\
     \lor &
    \textsf{inside}(w_2,w_3,w_4,w_5)\\
    \lor &
    \textsf{inside}(w_2,w_3,w_4,x_5)\\
    \lor & 
    \textsf{inside}(w_2,w_4,w_3,w_5)\\
    \lor &
    \textsf{inside}(w_2,w_4,w_3,x_5) 
\end{align*}}

\noindent {\bf Step 7:}
{\small 
\begin{align*}
    &\textsf{deleting}(w_1,x_1,w_2,x_2) 
    \land 
    \textsf{deleting}(w_1,x_1,x_2,w_2)\\ 
    \land
    &\textsf{deleting}(w_1,w_2,x_1,w_3) \land 
    \textsf{deleting}(w_1,x_2,x_1,w_4)
    \rightarrow \bot
\end{align*}}

\noindent {\bf Step 8a:}

{\small 
\begin{align*}
    &\textsf{deleting}(u_1,v_1,w_1,x_1) \land
    \textsf{F}(w_1,x_1)\\ \land &
    \textsf{deleting}(v_1,w_1,w_2,x_2) \land
    \textsf{F}(w_2,x_2)\\ 
    \rightarrow& \textsf{inside}(u_1,v_1,w_1,w_2) \lor
    \textsf{inside}(v_1,u_1,w_1,w_2)\\ \lor  
    & \textsf{inside}(u_1,v_1,w_1,x_2) \lor
    \textsf{inside}(v_1,u_1,w_1,x_2)\\ \lor 
    &\textsf{deleting}(u_1,v_1,w_2,x_2)
    \lor
    \textsf{deleting}(u_1,v_1,x_2,w_2)
\end{align*}}

\noindent {\bf Step 8b:}

{\small 
\begin{align*}
&\textsf{deleting}(u_1,v_1,w_1,x_1) \land
    \textsf{F}(w_1,x_1)\\  \land
    &\textsf{deleting}(v_1,w_1,u_1,x_2)
    \land
    \textsf{F}(u_1,x_2)
    \rightarrow \bot \end{align*}}

\noindent {\bf Step 9a:} 

{\small \begin{align*}
    &\textsf{deleting}(u_1,v_1,w_1,x_1) \land
    \textsf{F}(w_1,x_1)\\ \land
&    \textsf{deleting}(v_1,w_1,w_2,x_2)
    \land \textsf{F}(w_2,x_2) \land 
    \textsf{inside}(u_1,v_1,w_1,w_2)\\ \land
    & \textsf{deleting}(v_1,w_2,w_3,x_3)
    \land
    \textsf{F}(w_3,x_3)\\
    \rightarrow& \textsf{inside}(u_1,v_1,w_1,x_1)\\ \lor
    &\textsf{inside}(v_1,w_1,w_2,w_3) \lor \textsf{inside}(v_1,w_1,w_2,x_3)
     \\ \lor
    &\textsf{inside}(w_1,v_1,w_2,w_3) \lor \textsf{inside}(w_1,v_1,w_2,x_3)\\
    \lor &
    \textsf{deleting}(u_1,v_1,w_3,x_3)
    \lor 
    \textsf{deleting}(u_1,v_1,x_3,w_3)\\
    \lor &
    \textsf{deleting}(v_1,w_1,w_3,x_3)
    \lor
    \textsf{deleting}(v_1,w_1,x_3,w_3)
\end{align*}}

\noindent {\bf Step 9b:}

{\small \begin{align*}
&\textsf{deleting}(u_1,v_1,w_1,x_1) \land
    \textsf{F}(w_1,x_1)\\ \land&
    \textsf{deleting}(v_1,w_1,w_2,x_2)
    \land
    \textsf{F}(w_2,x_2)
    \land 
    \textsf{inside}(u_1,v_1,w_1,w_2)\\
   \land &
   \textsf{deleting}(v_1,w_2,u_1,x_3)
    \land
    \textsf{F}(u_1,x_3)
    \rightarrow \bot
\end{align*}}

\end{document}